\documentclass[pdflatex,sn-nature]{sn-jnl}

\usepackage{graphicx}
\usepackage{amsmath,amssymb,amsfonts}
\usepackage{appendix}
\usepackage{booktabs}
\usepackage{subcaption}
\usepackage{siunitx}
\usepackage{url}
\usepackage{cleveref}
\usepackage{placeins}
\usepackage{caption}
\newcommand{\tabhead}[1]{\begin{tabular}[b]{@{}c@{}}#1\end{tabular}}
\usepackage{tabularx}
\usepackage[resetlabels]{multibib}
\makeatletter\def\@extra@binfo{@S}\makeatother
\newcites{S}{References}
\makeatletter\gdef\@extra@binfo{}\makeatother

\hypersetup{
    colorlinks=true,
    linkcolor=black,
    citecolor=blue,
    urlcolor=blue,
    bookmarksdepth=subsubsection
}
\crefname{figure}{Fig.}{Figs.}\Crefname{figure}{Fig.}{Figs.}
\crefname{table}{Table}{Tables}\Crefname{table}{Table}{Tables}
\crefname{equation}{Eq.}{Eqs.}\Crefname{equation}{Eq.}{Eqs.}
\crefname{section}{Section}{Sections}\Crefname{section}{Section}{Sections}
\crefname{subsection}{Section}{Sections}\Crefname{subsection}{Section}{Sections}
\crefname{subsubsection}{Section}{Sections}\Crefname{subsubsection}{Section}{Sections}
\crefname{appendix}{Supplementary Note}{Supplementary Notes}
\Crefname{appendix}{Supplementary Note}{Supplementary Notes}

\begin{document}

\title[Provenance-tracked, experimentally validated neural operator surrogates of three-dimensional laser melting with keyholes and free-form paths]{Provenance-tracked, experimentally validated neural operator surrogates of three-dimensional laser melting with keyholes and free-form paths}

\author*[1]{\fnm{Alix} \sur{Benoit}}\email{alix.benoit@alumni.polytechnique.org}

\author[1]{\fnm{Elia} \sur{Iseli}}

\affil[1]{\orgdiv{Light-Matter Dynamics Group, Laboratory for Multifunctional Materials and Interfaces}, \orgname{Swiss Federal Laboratories for Materials Science and Technology}, \orgaddress{\street{Feuerwerkerstrasse}, \city{Thun}, \postcode{3603}, \state{BE}, \country{Switzerland}}}

\abstract{%
A three-dimensional thermo-fluid-dynamic simulation of laser melting resolves the melt pool
and vapor depression but costs from one to thousands of core-hours per track. We train
FNO, RIGNO and GAOT neural operators that predict the temperature, melt-pool and
free-surface fields. Our datasets comprise 3D temperature and melt geometry fields covering lack-of-fusion to keyhole conditions under
varied laser power, scan speed, beam radius, beam shape and preheat, and include
free-form scan paths with speed changes and up to four beams. We generate them with
FLOW-3D WELD\textsuperscript{\textregistered}, using solver settings validated against
published measurements for four materials. We implement an AiiDA framework to store the provenance of every
simulation, model and evaluation, so each tabulated metric traces back to the solver and model
inputs behind it. 
On held-out simulations, the best architecture scores below \SI{2}{\percent} relative
temperature error on quasi-steady straight tracks, below \SI{4}{\percent} on
single-beam free-form paths and about \SI{5}{\percent} with two to four beams.
Inference takes under a second per predicted arbitrary time step on our single-GPU setup. We extend a
surrogate trained on Ti-6Al-4V to Inconel 718 by growing its training set
autonomously, and the extended models reach a relative temperature error of about \SI{3}{\percent} or less on both Inconel 718 and Ti-6Al-4V. At \num{20} Ti-6Al-4V and Inconel 718 conditions held out of training, the
two-alloy RIGNO predicts melt-pool and vapor-depression depths within \SI{30}{\percent} of
published measurements, and its depths differ from
FLOW-3D\textsuperscript{\textregistered} at the same conditions by \SI{2}{\percent} on
average.\\
Furthermore, these results provide a scalable framework which can readily be applied to computationally intensive simulations in fields other than laser melting.
}


\maketitle

\section{Introduction}\label{sec:intro}

Laser welding and laser-based additive manufacturing are governed by a tightly
coupled set of multi-physical mechanisms. Absorbed laser light acts as a heat input which then drives melting. Temperature gradients and evaporation lead to complex effects, such as Marangoni- and buoyancy-driven flows, which redistribute heat and mass within the melt pool. Once the vaporization threshold is exceeded, recoil pressure induces the formation of a vapor depression, whose evolving geometry further modulates laser absorption through multiple reflections~\cite{ye_energy_2019}, eventually leading to a vapor cavity or so-called \emph{keyhole}. The strong feedback among these mechanisms makes the process highly nonlinear and sensitive: small variations in operating conditions can produce substantial changes in melt-pool morphology, keyhole depth, and ultimately defect formation~\cite{gan_universal_2021}.

High-fidelity thermo-fluid solvers that combine volume-of-fluid free-surface
tracking with ray-traced absorption and recoil pressure reproduce much of this
behavior, and have been validated against melt-pool cross-sections and in-situ
X-ray imaging for a range of alloys and operating
conditions~\cite{cook_simulation_2020}.
Their computational cost is the main obstacle. A single three-dimensional track
at the resolutions such studies typically employ can occupy thousands of
core-hours~\cite{bayat_keyhole-induced_2019,katagiri_effect_2024}, which places large parameter sweeps,
inverse design and any use inside a control loop beyond routine reach. That cost is
paid on every query: a full computation integrates every intermediate step and
grows with the meshed volume even when the region of interest does not. A
surrogate that predicts a single time step directly escapes both constraints and enables almost instantaneous predictions for arbitrary time steps. 

Reduced descriptions trade physical completeness for speed. Analytical and
semi-analytical moving-source solutions~\cite{rosenthal_theory_1946} and thermal models that emulate
convective transport through an enhanced effective
conductivity~\cite{kamara_modelling_2011} are
inexpensive, are typically calibrated within a limited operating window, and
generally exclude free-surface deformation. Physics-informed
neural networks have been applied widely to additive manufacturing and in a few cases to
welding, with most
reported applications confined to the conduction
regime~\cite{ko_review_2024}; a standard
physics-informed network approximates the solution of a single
problem instance, so a change of power, speed or geometry usually requires
retraining~\cite{raissi_physics-informed_2019}.

Neural operators lift that constraint by approximating a mapping between
function spaces~\cite{kovachki_neural_2023}. Once trained across many solutions,
an operator evaluates a new one from its parameters in a single forward pass.
Architectures include the Fourier neural
operator (FNO)~\cite{li_fourier_2021} and its geometry-adapted
variant~\cite{li_fourier_2023}, branch-and-trunk constructions such as
DeepONet~\cite{lu_deeponet_2021}, convolutional
operators~\cite{raonic_convolutional_2023}, graph-based operators including the
region interaction graph neural operator (RIGNO)~\cite{mousavi_rigno_2025} and the harmonic-mapping
operator~\cite{yan_harmonic-mapping_2026}, and attention-based formulations such
as the geometry-aware operator transformer (GAOT)~\cite{wen_geometry_2025} and DoMINO~\cite{ranade_domino_2025}.
A detailed comparison of FNO and DeepONet finds that neither clearly
dominates across all tasks~\cite{lu_comprehensive_2022}. A graph-based operator has been trained to
predict part-scale thermo-mechanical distortion in metal additive
manufacturing~\cite{chen_graph_2026}, and to our knowledge the graph-based and attention-based
formulations have not yet been applied to laser melting at the scale of the melt pool.

Operator and deep-learning surrogates for laser processing are now
numerous~\cite{hemmasian_surrogate_2023,manav_meltpoolinr_2025,liu_deep_2024},
and most carry restrictions that limit their use as a general process
model. Several predict a two-dimensional field~\cite{manav_meltpoolinr_2025}, or a reduced set of melt-pool
descriptors~\cite{schlenger_fast_2025} instead of the full volume;
several omit fluid motion and free-surface
deformation~\cite{liu_deep_2024,yaseen_fast_2023}; and many are demonstrated in
the conduction regime, where the free surface stays close to
flat~\cite{schlenger_fast_2025}. Our own preceding
work~\cite{benoit_fast_2026} predicts
three-dimensional temperature fields and melt-pool boundaries across conduction
and keyhole regimes, for a single material on a straight constant-speed track,
by recasting the transient problem to quasi-steady in a frame translating with
the beam.

To extend a single-track model of this kind into a general laser-melting surrogate, the
simulation framework must scale to hundreds of simulations, each traceable to the solver settings
that produced it, and those settings must be validated against published measurements. The
surrogate must predict along arbitrary laser tracks and across process conditions, and transfer
to a new material without regenerating every dataset.

Provenance-tracked workflow managers are established in computational materials science at the
atomistic scale. AiiDA~\cite{huber_aiida_2020} and pyiron~\cite{janssen_pyiron_2019} run and record
electronic-structure and property workflows there, and pipelines for machine-learned interatomic
potentials record the chain from simulation to trained
model~\cite{bidoggia_aiida-trainspot_2026,menon_electrons_2024}. pyiron also includes continuum-scale
modules. The corpora behind neural-operator surrogates are distributed with generation scripts and
configuration metadata~\cite{takamoto_pdebench_2022,ohana_well_2025}, and melt-pool datasets are
usually described by their process window and simulation count. Neither links each simulation,
dataset, model and evaluation to the solver configuration that produced it, and to our knowledge no
such record exists for melt-pool simulation or for any continuum multiphysics solver used to
generate neural-operator training data. \Cref{sec:pipeline} describes our record.

FLOW-3D WELD\textsuperscript{\textregistered}~\cite{noauthor_flow-3d_2025}, the volume-of-fluid
solver we use, has been compared with measurements for melt-pool dimensions and
surface morphology~\cite{wu_parametric_2018,chen_numerical_2023}, melt-pool and porosity
evolution~\cite{ransenigo_evolution_2022}, keyhole-induced porosity~\cite{bayat_keyhole-induced_2019}
and shaped-beam response observed in situ~\cite{bayat_exploring_2024}. Each of these comparisons
holds for the settings its study chose, and a recoil setting matched at one spot size can miss
another~\cite{zenz_critical_2025}, so the settings that generate a training set need their own check.
We validate one set of settings against published measurements for four materials
(\cref{sec:expval_main}), and score the final surrogate the same way, which tests the surrogate and
the solver together (\cref{sec:closing}).

Data-driven thermal-history models for arbitrary toolpaths exist for
additive manufacturing~\cite{mozaffar_data-driven_2018,ren_thermal_2020}, and predict conduction-only
temperature, at part scale or in two dimensions, with no melt flow or free surface. A turning path
removes the quasi-steadiness that our preceding work relies on~\cite{benoit_fast_2026}, so the field depends on
earlier beam positions, which become model inputs. To our knowledge, no reported surrogate predicts the three-dimensional
temperature, melt-pool and free-surface fields of a keyholing process along a
free-form path, with variable scan speed and multiple beams, and across ranges of laser power, beam radius,
beam shape and preheat temperature.

Physics-derived inputs have been used before: signed-distance fields of the
geometry~\cite{guo_convolutional_2016},
invariants of the mean flow for turbulence closures~\cite{ling_reynolds_2016},
toolpath features~\cite{ren_thermal_2020},
and analytic solutions corrected by a
network~\cite{bock_hybrid_2021}. We supply the linear
heat-equation solution as an input field to the neural operator, together with
coordinates relative to the moving source and exposure histories that
approximate it (defined in \cref{sec:formulation}), so the network only has to
learn the departure from the linear solution. 

Choosing new training runs where a surrogate's error is largest is
established for cheap partial differential equation solvers that can be called during
training~\cite{cesar_learning_2026}, and Bayesian experimental design has been
applied to scalar models of melt-pool geometry~\cite{whalen_bayesian_2021}. No
reported loop lets a three-dimensional melt-pool surrogate choose and launch runs
of a commercial solver, with every run recorded in the provenance graph, and no
such surrogate has been extended from one alloy to another this way.

This work makes five main contributions: (1) a provenance-tracked pipeline and
simulation database that records every simulation, dataset, model and
evaluation of the study as one queryable graph, (2) validation of the solver used and of the final surrogate against published experimental measurements, (3) three-dimensional neural operator surrogates
that can generalize across free-form weld paths, varying laser power, scan speed,
beam radius, beam shape (Gaussian and uniform) and preheat temperature, from
conduction to keyhole, (4) engineered model input fields built in part from the linear
solution of the heat equation and computed from the process parameters
and (5) a closed
loop in which the new solver runs are chosen autonomously to minimize the surrogate's error and to cover parameter space efficiently, used to extend a Ti-6Al-4V model to Inconel 718 (IN718).

\FloatBarrier%
\section{Methodology}\label{sec:methods}
\begin{figure}[htbp]
\centering
\includegraphics[width=\linewidth]{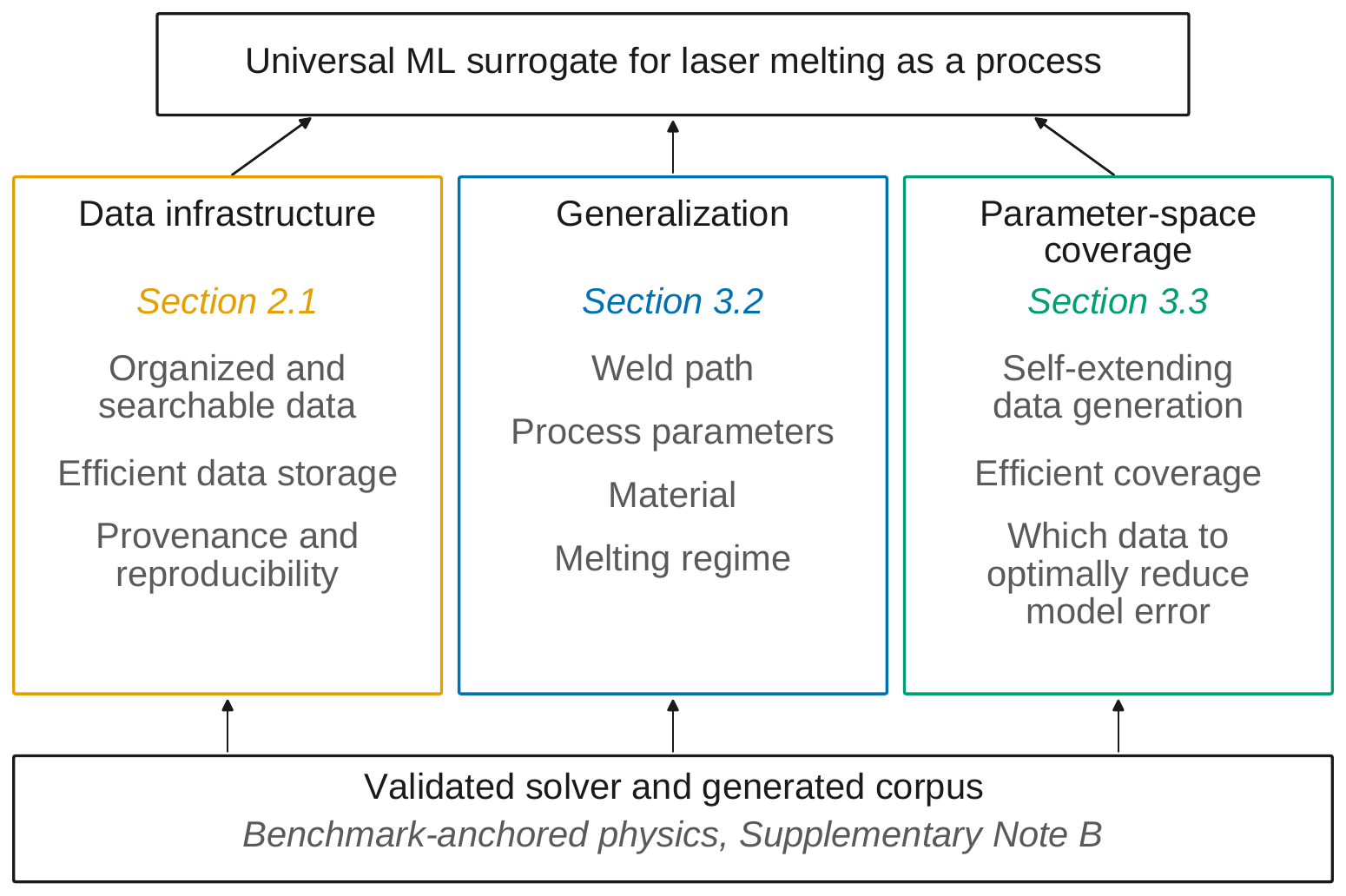}
\caption{We consider the problem of creating a universal surrogate for laser melting as a process to be split into three sub-problems: data infrastructure and handling, increasing surrogate generalization capabilities, and autonomous coverage and exploration of process parameter space for training data generation. In this work we mainly push the first two sub-problems, and briefly explore the third sub-problem through autonomous acquisition of new training data to extend the surrogate to a new alloy using incremental model evaluation results.}
\label{fig:pillars}
\end{figure}

We organize the problem of creating a universal welding surrogate model into
three sub-problems, summarized in \cref{fig:pillars}.

The first is \textbf{data infrastructure}. A surrogate spanning a wide process
space is only as trustworthy as the record of the simulations behind it. We
generate every simulation through a versioned, scriptable interface to FLOW-3D
WELD\textsuperscript{\textregistered} and catalog every simulation, dataset,
model and evaluation in a provenance database, so any reported number traces
back to, and can be re-derived from, the model configuration, solver version and
material data that produced it (\cref{sec:infra}).

The second is \textbf{generalization}, treated along three axes of the input
space. The weld path is the first: an industrial trajectory can turn, change speed during a weld, or involve multiple beams. The field
at a given instant depends on the history of beam positions, which a frame
translating with a straight track cannot represent. The process
parameters (laser power, scan speed, beam radius, preheat temperature and beam shape) are the second. The material
is the third, entering the governing equations as temperature-dependent
coefficients. The melting regime, spanning lack of
fusion (LoF), conduction melting and keyhole, is approximated by the parameters and the
material together through the normalized
enthalpy~\cite{king_observation_2014}, and
the datasets here span from LoF to keyhole regimes.

The third is \textbf{parameter-space coverage}. Every training simulation occupies a
licensed solver seat, for ten to forty minutes on ten cores and up to \num{90} minutes for the longest tracks, so we want to minimize the number of training simulations as much as possible and to choose which simulations to run in a well-thought-out way. Automating this process so that a model can autonomously improve itself by generating new data where it fails to generalize well is a key step to scalability and full process generalization.
We use a combination of two methods to autonomously choose new simulations, one which chooses simulations that are the most likely to reduce the surrogate's error, and a space-filling algorithm. We use it to
extend a Ti-6Al-4V model to IN718
(\cref{sec:acquisition,sec:al-results}).

Before generating any dataset we validate the solver against published benchmark
measurements of melt-pool and keyhole geometry, fix an acceptance criterion in
advance, and report both the operating range over which the validation holds and
the range where it does not (\cref{sec:calibration}). Recording which physics
produced which dataset lets the three axes be varied independently and compared
under one formulation and one set of metrics.

We restrict the scope in four respects. The geometry is a rectangular substrate,
so part-scale additive manufacturing geometry and domains of varying size are
future work. Training uses a single mesh resolution, leaving multi-fidelity
schemes outside the study. We run each architecture as its authors published it,
with physical information entering through the input fields of
\cref{sec:formulation} and no explicit constraints inside the network or the loss
function. In this work the predicted results are validated only against published measurement data
(\cref{sec:calibration} for the solver and \cref{sec:closing} for the surrogate).

\subsection{Simulation infrastructure and physical grounding}\label{sec:infra}

\subsubsection{Parametric simulation generation and provenance}\label{sec:pipeline}

Because we compare several architectures across several datasets and, as
\cref{sec:calibration} describes, across more than one generation of solver
parameter sets, every result had to stay tied to the exact data behind it for comparisons to make sense. And
because there are hundreds of simulations, they had to be generated
programmatically in a traceable and reproducible way. 

\paragraph{Solver.}
Simulations are computed with FLOW-3D WELD\textsuperscript{\textregistered}
2025R1~\cite{noauthor_flow-3d_2025}, which solves the Navier--Stokes and energy
equations with volume-of-fluid free-surface tracking while considering temperature-dependent
properties, phase change, surface tension with Marangoni convection, and recoil
pressure from evaporation. Laser absorption is ray-traced, so heat input varies
as the free surface deforms and rays reflect inside the vapor cavity.
We place $x$ and $y$ in the substrate surface at $z=0$, with $z$ pointing up into the gas layer
and the beam incident along $-z$. Straight tracks run along $+x$ at $y=0$, and free-form
paths lie in the $x$--$y$ plane. The top of the gas layer is an open boundary held at ambient
pressure, \SI{101.3}{\kilo\pascal}, through which gas can leave and enter. The side and bottom faces are solid walls that neither fluid nor heat can cross (impermeable and adiabatic). We check that the melt in each simulation stays relatively far from these walls, to avoid unwanted boundary effects.

\paragraph{Declarative simulation generation.}
The solver is driven through a wrapper we wrote for this work. A simulation is declared
as a JSON configuration (its \emph{declaration}) naming the process parameters, the geometry, the mesh and
a material, together with a material definition stored alongside the declaration. The
wrapper renders these into a complete solver input deck, writes the parallel
configuration, runs the pre-processor and solver, and exports every stored time
step to a single compressed npz array file, which is possible as we use a regular mesh. The fields shrink
losslessly by a factor of \numrange{16}{38}, from a few gigabytes of arrays to
a little over a hundred megabytes for a full time series, and to under a megabyte
for a simulation kept as one time-averaged steady frame. That file is what the
store holds, and with the further measures described in \cref{sec:provenance-supp}, it makes keeping
every simulation of this study on one machine practical.

Four properties of that layer matter for what follows. Derived quantities such as
the beam velocity and power against time along a curved path, the mesh bounds implied by the
substrate dimensions, and the temperature-dependent property tables are computed
from the declaration, so a sweep is specified by the parameters that vary across
it. A simulation is therefore described completely by files under version control, and
re-running the same declaration reproduces the same input deck. The wrapper holds
a run until disk and memory are available and serializes simulations to adapt to a single available license seat. The same entry point is always used, whether a simulation is submitted by hand, by a batch sweep, or by
an AiiDA~\cite{huber_aiida_2020} workflow layer (described next), so a simulation run in any of those ways is
identical.

\paragraph{Provenance store.}
Simulations are executed and cataloged through AiiDA, a
workflow engine that records provenance as a directed graph. Each simulation,
dataset, trained model and evaluation is a node in a PostgreSQL-backed store,
with the array payloads in an on-disk object repository, and each simulation, model and evaluation is linked to the inputs and the calculation that produced it. The graph is queryable: a
model node resolves to the dataset it trained on, that dataset to its member
simulations, and each simulation to the configuration, material definition and
solver version behind it.

\begin{figure}[htbp]
\centering
\includegraphics[width=\linewidth]{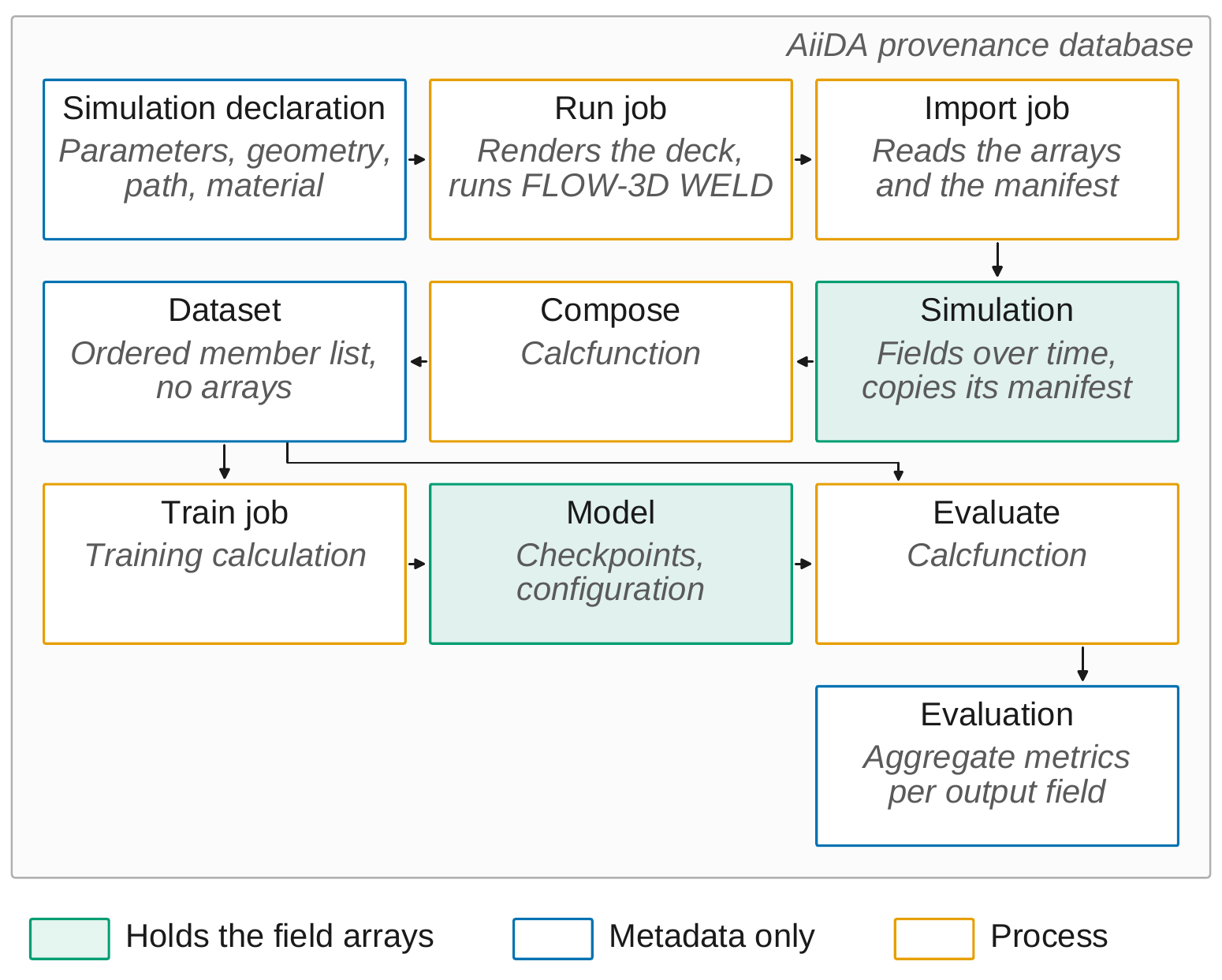}
\caption{How a simulation becomes a reported number. Every
artifact is a node and every arrow a stored link, so a reported metric resolves
backwards to the solver inputs that produced it. Green boxes hold the field
arrays, blue boxes carry metadata alone, and orange boxes are the processes that
create them. The run job writes a solver input deck from the simulation declaration and runs
FLOW-3D WELD\textsuperscript{\textregistered}, and the import job reads the resulting arrays and
their metadata file into a simulation node. A calcfunction is a function whose inputs and outputs
AiiDA records as nodes. One calcfunction assembles simulations into a dataset, the train job fits a
model to that dataset, and a second calcfunction evaluates the model on a dataset. Derived collections record their members as a stored list of identifiers, which
the store resolves the same way as links but which a graph export does not follow.}
\label{fig:prov-chain}
\end{figure}

\begin{figure}[htbp]
\centering
\includegraphics[width=\linewidth]{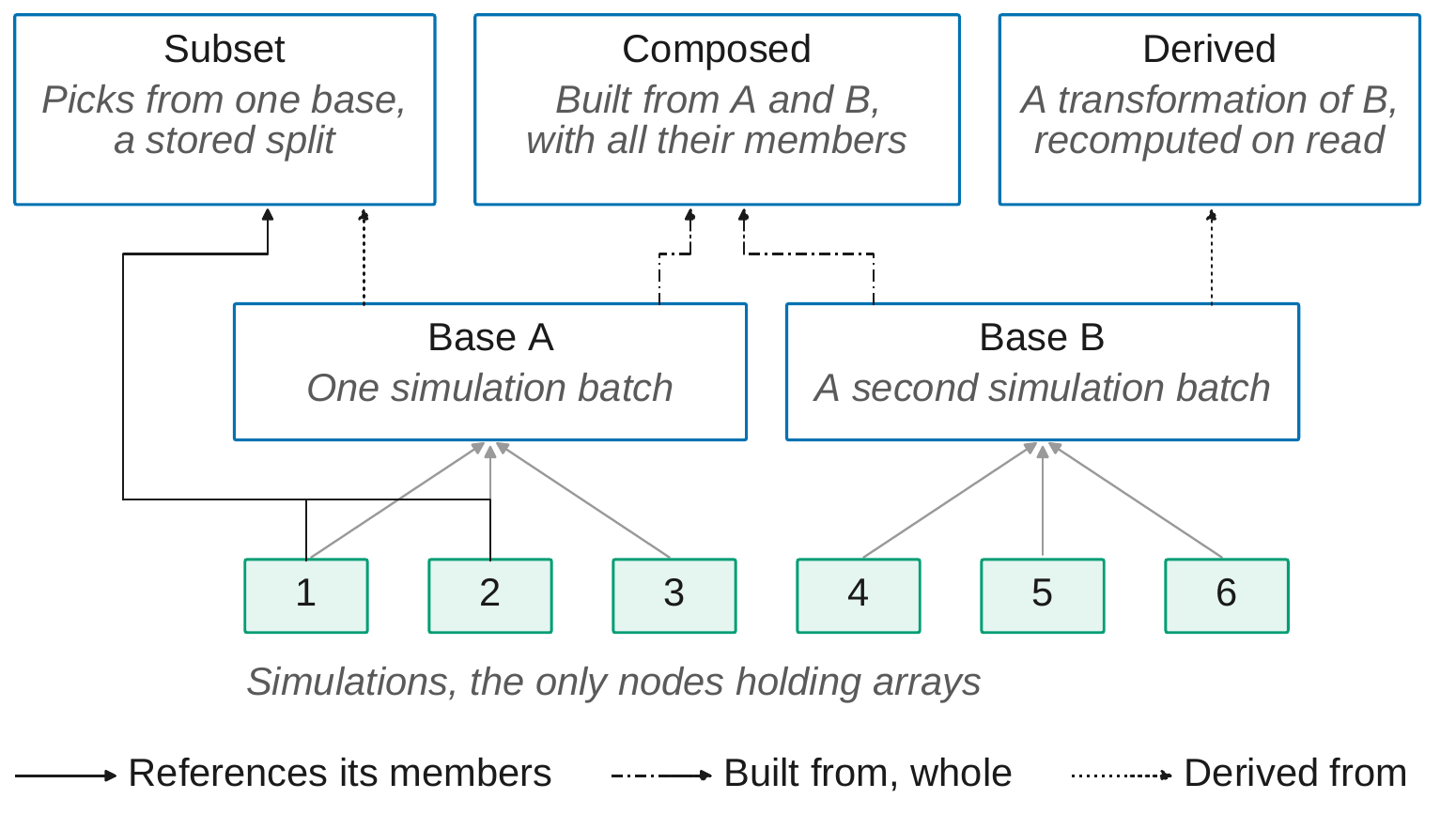}
\caption{Collection roles. Collections hold no arrays of
their own and reference their members, so a simulation is stored once and read by
many. A base collection owns its simulations, and a simulation belongs to no other
base. A subset picks simulations out of one base, which is how a split is
recorded. A composed collection is
built from whole datasets, and its members are their simulations. A derived collection stores a
transformation of its parent and recomputes the arrays on read.}
\label{fig:prov-roles}
\end{figure}

\paragraph{From a simulation to a reported number.}
\Cref{fig:prov-chain} shows the chain along which a simulation travels. The unit of
data is a \emph{simulation} node, holding one simulation as a compact array payload
with its process parameters, mesh description and full solver manifest.
Simulations are grouped into \emph{dataset collections}, which hold no arrays of
their own and reference their members by identifier in a fixed order, so a simulation
may appear in many collections without duplication. Collections carry a role
recording how they were built (\cref{fig:prov-roles}). A \emph{base} collection
is the direct product of one simulation batch and owns its members. A
\emph{subset} selects members of a base collection, which is how the train,
validation and test splits are represented; the rule used to draw them is stored
alongside, so a split can be audited and reused. A \emph{composed} collection
unions several collections, assembling an experiment that spans more than one
simulation batch. A \emph{derived} collection stores a transformation of a parent and
recomputes the arrays on demand; spatial subsampling and the moving-frame crop
used in previous work~\cite{benoit_fast_2026} are expressed this way.

Because membership is by reference, the datasets in \cref{tab:datasets} can
overlap without inflating storage or obscuring which simulations are shared, and a
comparison between two models can be checked for train and test leakage by
comparing member identifiers.

Every simulation and model result reported in this paper resolves to a node in this database
(\cref{fig:prov-chain}). A scoped export of the provenance graph covering the
cited nodes is supplementary material, so the chain from a reported metric back
to the solver input deck can be inspected without access to our machines.

\begin{table}[htbp]
\centering
\caption{Datasets used in this work, all solved on the validated solver
settings of \cref{sec:calibration}. $P$ is laser power, $V$ scan speed, $r_0$
the beam radius and $T_0$ the preheat temperature, a uniform initial temperature of the whole domain. The beam
radius is \SI{50}{\micro\metre} unless otherwise stated. Counts are the simulations the models are trained, validated and tested on. The two-alloy row counts the IN718 simulations of the acquisition study. The \num{171} Ti-6Al-4V training simulations of the two-alloy models are the training split of the
beam-radius sweep.}
\label{tab:datasets}
\small
\setlength{\tabcolsep}{5pt}%
\sisetup{range-phrase=--,range-units=single}%
\begin{tabularx}{\textwidth}{@{}ll>{\raggedright\arraybackslash}Xr@{}}
\toprule
Dataset & Path & Parameters varied, with ranges & Simulations \\
\midrule
Power and scan speed grid & straight & $P$ \SIrange{40}{215}{\watt}, $V$ \SIrange{0.10}{0.96}{\metre\per\second} & 123 \\
\addlinespace[3pt]
Variable paths      & random   & path and its length, \SIrange{0.11}{2.97}{\milli\metre}, $P$ \SIrange{68}{245}{\watt}, $V$ \SIrange{0.21}{0.99}{\metre\per\second} & 150 \\
\addlinespace[3pt]
Variable speed      & random   & path, \SIrange{0.20}{1.63}{\milli\metre} long, speed profile from 0 to \SI{2.0}{\metre\per\second} along it, $P$ \SIrange{53}{252}{\watt} & 140 \\
\addlinespace[3pt]
Preheat sweep        & straight & $T_0$ \SIrange{296}{761}{\kelvin}, $P$ \SIrange{55}{284}{\watt}, $V$ \SIrange{0.20}{1.19}{\metre\per\second}, $r_0$ \SIrange{35}{80}{\micro\metre}, Gaussian or uniform beam & {107} \\
\addlinespace[3pt]
Beam-radius sweep    & straight & $P$ \SIrange{90}{235}{\watt}, $V$ \SIrange{0.12}{1.25}{\metre\per\second}, $r_0$ \SIrange{42.5}{65}{\micro\metre} & {218} \\
\addlinespace[3pt]
Active-learning two-alloy & straight & material (Ti-6Al-4V, IN718), with IN718 at $P$ \SIrange{90}{350}{\watt}, $V$ \SIrange{0.15}{1.25}{\metre\per\second}, $r_0$ \SIrange{22}{65}{\micro\metre} & {272} \\
\addlinespace[3pt]
Multi-beam          & random   & 2 to 4 beams on their own paths, $P$ \SIrange{27}{155}{\watt} per beam, $V$ \SIrange{0.22}{0.98}{\metre\per\second}, $r_0$ \SI{47.5}{\micro\metre} & {79}  \\
\bottomrule
\end{tabularx}
\end{table}

\FloatBarrier%
\subsubsection{Validation against published measurements}
\label{sec:expval_main}
A surrogate trained on simulation inherits whatever the simulation gets wrong, so
we validated the FLOW-3D WELD\textsuperscript{\textregistered} solver and our
solver parameters against published measurements before any model was trained. We
perform no new physical experiments. Each comparison
is made per condition, by which we mean one published measurement at a single
material, laser power, beam size and scan speed or beam-on duration, scored on its
melt-pool or vapor-depression depth and, where reported, its melt-pool width. We consider an absolute error within
\SI{30}{\percent} on depth and width acceptable, a choice explained in
\cref{sec:calibration}. At a typical \SI{100}{\micro\metre} depth, that tolerance allows an error of three cells of our \SI{10}{\micro\metre} mesh. For process simulation with Ti-6Al-4V
we use in-situ synchrotron X-ray depression depths~\cite{gan_universal_2021,cunningham_keyhole_2019},
metallographic melt-pool depth and width across the conduction-to-keyhole
transition~\cite{galiegue_acoustic_2025}, and the NIST Additive Manufacturing Benchmark Series (AM-Bench) records, which
pair integrating-sphere absorptance with high-speed radiography of the same
shot~\cite{simonds_ability_2024,simonds_asynchronous_2022}. Stationary 304L spot
welds~\cite{hawk_laser_2022}, CoCr
cross-sections~\cite{soares_reliability_2026} and the IN718 conditions of the
AM-Bench 2022 challenge~\cite{weaver_am_2022} extend the validation to
three further materials. The full comparison appears
in \cref{sec:calibration}, and the validated solver settings are used for every
dataset generated in this study, unless marked otherwise.

\begin{figure}[htbp]
\centering
\begin{subfigure}[t]{0.49\textwidth}
  \includegraphics[width=\linewidth]{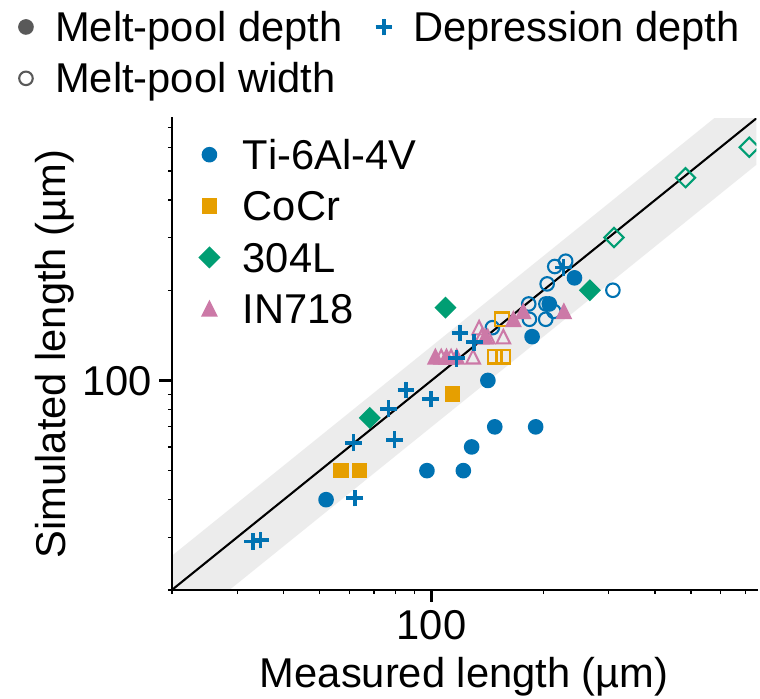}
  \subcaption{Simulated against measured length.}\label{fig:calib-materials-depth}
\end{subfigure}\hfill
\begin{subfigure}[t]{0.49\textwidth}
  \includegraphics[width=\linewidth]{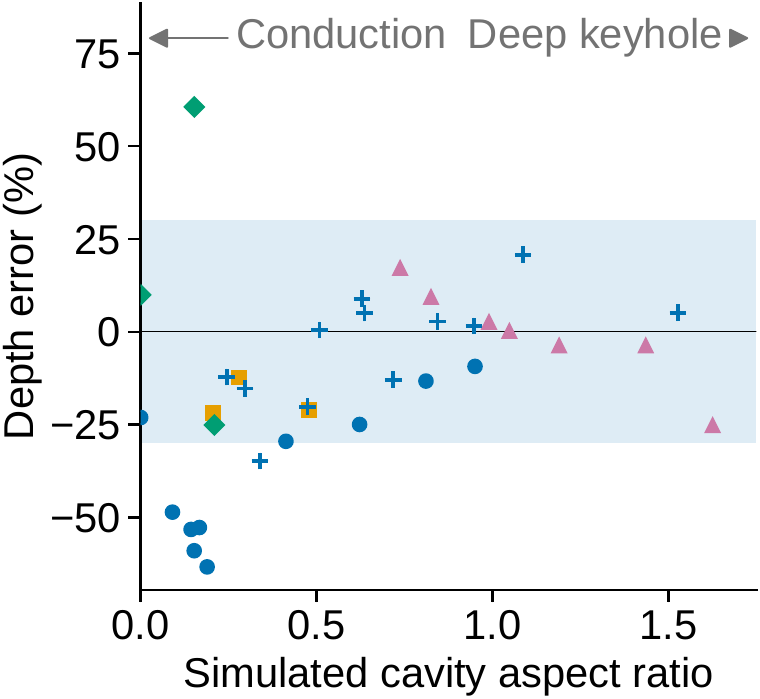}
  \subcaption{Depth error against cavity aspect ratio.}\label{fig:calib-materials-width}
\end{subfigure}
\caption{Every evaluated experimental condition, \num{35} across four
materials. (a) Simulated against measured length on logarithmic axes, with the
diagonal marking exact agreement and the band spanning $\pm\SI{30}{\percent}$;
filled markers are melt-pool depth, open markers melt-pool width, and crosses
vapor-depression depth. (b) The depth residuals plotted against simulated vapor cavity aspect ratio.}
\label{fig:calib-materials}
\end{figure}

\begin{figure}[htbp]
\centering
\begin{subfigure}[t]{0.49\textwidth}
  \includegraphics[width=\linewidth]{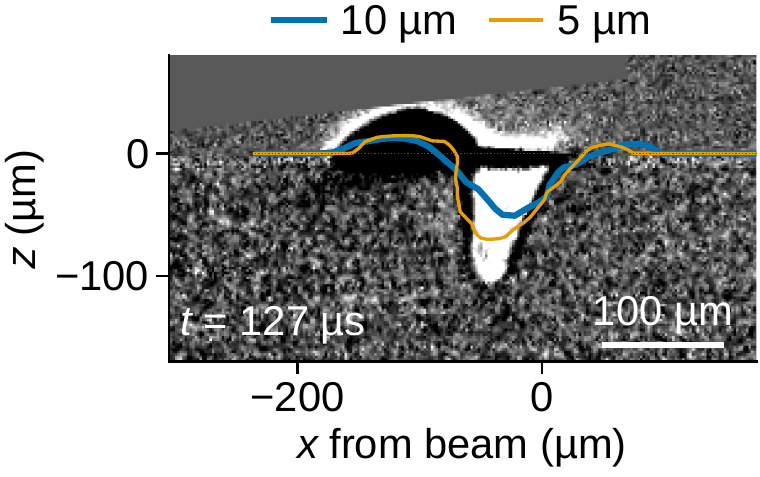}
  \subcaption{Scan, drilling transient.}\label{fig:calib-xray-transient}
\end{subfigure}\hfill
\begin{subfigure}[t]{0.49\textwidth}
  \includegraphics[width=\linewidth]{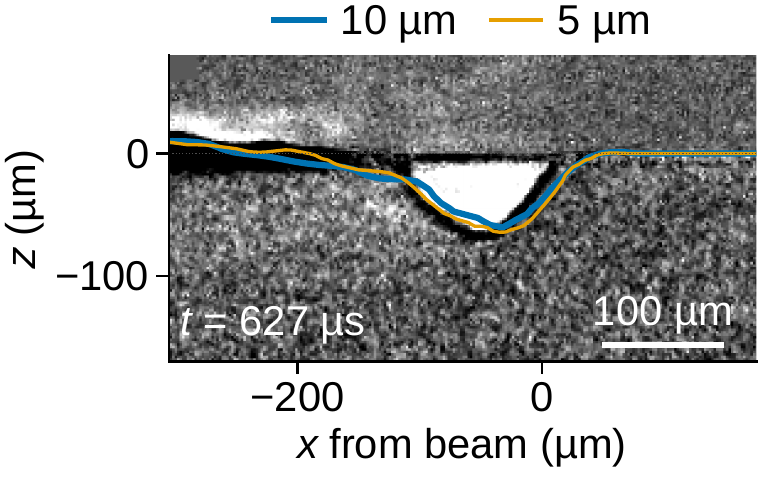}
  \subcaption{Scan, established cavity.}\label{fig:calib-xray-settled}
\end{subfigure}

\vspace{0.6em}
\begin{subfigure}[t]{0.49\textwidth}
  \includegraphics[width=\linewidth]{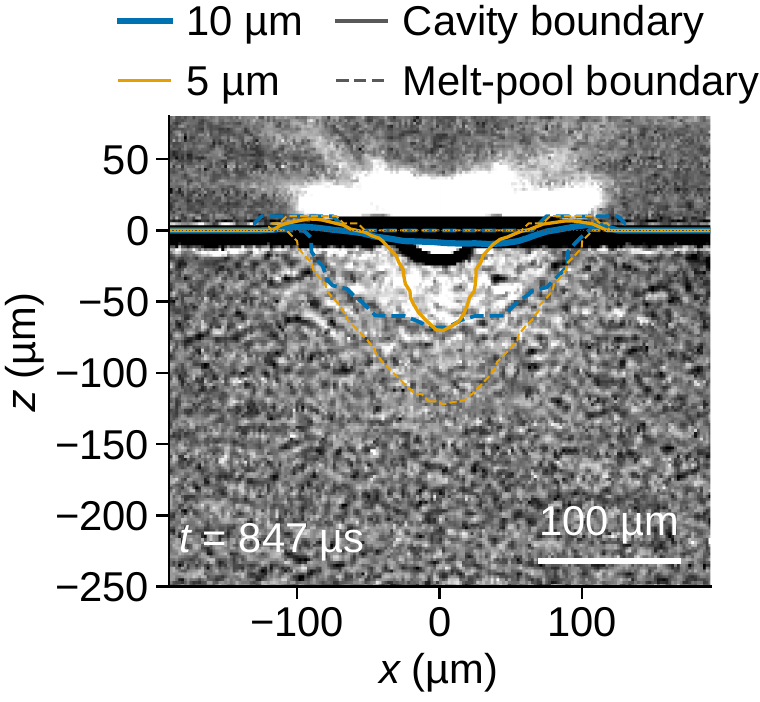}
  \subcaption{Stationary spot, frame before cavity onset at \SI{847}{\micro\second}, \SI{5}{\micro \metre} resolution simulation predicts early cavity onset.}\label{fig:calib-xray-spot-early}
\end{subfigure}\hfill
\begin{subfigure}[t]{0.49\textwidth}
  \includegraphics[width=\linewidth]{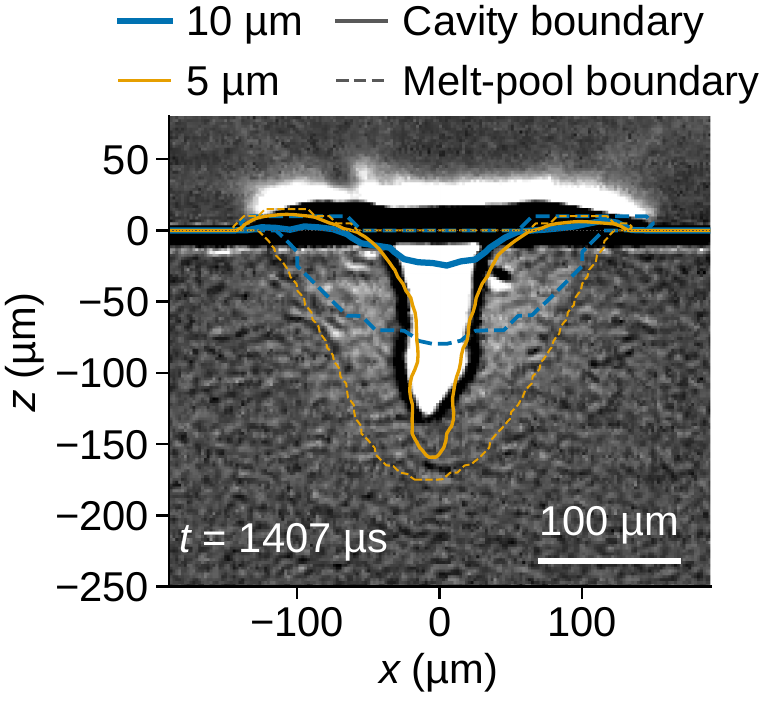}
  \subcaption{Stationary spot, during keyhole at \SI{1407}{\micro\second}, \SI{5}{\micro \metre} resolution simulation predicts keyhole well, \SI{10}{\micro\metre} one does not.}\label{fig:calib-xray-spot-late}
\end{subfigure}
\caption{Gas boundary from our FLOW-3D WELD\textsuperscript{\textregistered} simulations (lines) drawn on the radiograph at a common
scale, beam at $x=0$ and substrate surface at $z=0$~\cite{simonds_ability_2024,simonds_asynchronous_2022}.
Both conditions share $r_0=\SI{61.25}{\micro\metre}$, with simulations at \num{10} and \SI{5}{\micro\metre}
resolution. (a) is
inside the drilling transient and (b) the established cavity. (c, d) show a stationary spot with the beam turned on and held still. (c) \SI{847}{\micro\second}, the last frame before
the measured cavity opens. The \SI{5}{\micro\metre} simulation already shows a small vapor cavity while the radiograph still reads
\SI{15}{\micro\metre}. (d)
\SI{1407}{\micro\second}, past keyhole onset.}
\label{fig:calib-xray}
\end{figure}

\Cref{fig:calib-materials} shows every scored condition, \num{35} across
four materials. Aluminum is excluded because its coupling to the laser varies
too strongly between nominally identical
experiments~\cite{simonds_ability_2024}. Agreement on Ti-6Al-4V is good, with
deviations comparable to the variation the experiments themselves show between
repeats~\cite{simonds_ability_2024} and along a single
track~\cite{galiegue_acoustic_2025}. The exception is the
onset of keyholing, a stochastic and chaotic process that melt-pool solvers
struggle to reproduce~\cite{simonds_ability_2024}. \Cref{fig:calib-materials-width} plots the error in melt-pool or vapor-depression depth of each condition
against the aspect ratio of the simulated vapor cavity, its depth divided by its width (zero for no cavity, as in the conduction regime, and above one for a
keyhole deeper than it is wide). The five Ti-6Al-4V melt pools at keyhole onset, whose simulated cavity has only
begun to form (aspect ratios below \num{0.2}), are \SIrange{49}{63}{\percent} shallower than measured. At aspect
ratios between \num{0.2} and \num{0.5}, every simulated depth is \SIrange{12}{35}{\percent} too
shallow. Every condition above
\num{0.5} has a depth error within the \SI{30}{\percent} tolerance. \Cref{sec:calibration} examines the conditions of each melting regime in turn.
\Cref{fig:calib-xray,fig:calib-history} show keyhole onset in time against synchronized radiography
of the AM-Bench scan, where both simulations at 5 and \SI{10}{\micro \metre} resolutions understate the initial
drilling transient and then follow the settled keyhole well.
For the stationary spot, in \cref{fig:calib-xray-spot-early} the \SI{5}{\micro\metre} simulation predicts keyhole onset \SI{190}{\micro\second} early. And in \cref{fig:calib-xray-spot-late}, the simulation with
\SI{5}{\micro\metre} cells tracks the radiographed cavity boundary fairly well. The one with
\SI{10}{\micro\metre} cells forms no keyhole. In our segmentation of the radiograph, the cavity is
\SIrange{24}{42}{\micro\metre} wide, only two to four cells of that mesh. The cavity under a scanned
beam is broader, seven to nine cells across at the three \SI{0.40}{\metre\per\second} conditions analyzed
in \cref{sec:calibration}, and the \SI{10}{\micro\metre} cells of our datasets resolve it. This highlights the importance of using fine enough mesh resolution for simulating keyholes realistically. 

\FloatBarrier%
\subsection{Operator-learning formulation}\label{sec:formulation-pillar}

\subsubsection{Problem statement and model inputs}\label{sec:formulation}

\paragraph{What the neural operators predict.}

We learn an operator mapping the laser melting process parameters (power,
scan speed, scan path, beam radius, preheat temperature...) to the 3D temperature
and melt-pool geometry fields at a requested time. At each grid point in space, it outputs
the temperature $T$, the per-cell metal volume fraction $\alpha \in [0,1]$, which
locates the free surface, and the liquid fraction $f_l$, which marks the melt pool and is computed from $T$
through the simulated alloy's solidus and liquidus temperatures $T_s$ and $T_l$,
\begin{equation}
f_l(T)=\frac{1}{1+\exp\!\left[-2\beta\left(\varphi-\tfrac12\right)\right]},
\qquad
\varphi=\frac{T-T_s}{T_l-T_s},\qquad \beta=2.5,
\label{eq:fl}
\end{equation}
a logistic smoothing of the linear ramp between solidus and liquidus used
in~\cite{benoit_fast_2026}, with the same melt threshold $f_l=0.5$ at $(T_s+T_l)/2$.
Outside the metal the fields are masked: where $\alpha$ falls below
\num{0.5} the temperature is replaced by the material's vaporization temperature
and the remaining fields by zero through a smooth filter. This is done identically in the reference solver data and in the prediction, so that both carry the
same value in the gas phase and that the operator is trained and scored on the metal exclusively. The masking and the problem formulation are the same as in~\cite{benoit_fast_2026}, where further details can be found. The general
setting, which is a discretization-independent parametric map between function
spaces~\cite{kovachki_neural_2023}, covers the architectures used
here~(\cref{sec:models}).

For straight scanning paths the problem can be approximated as quasi-steady in a reference frame traveling with the laser beam,
averaged over time once the vapor cavity and melt pool are stable~\cite{benoit_fast_2026}. That gives an easier learning problem, better matched to $L_2$ losses, which regress high-frequency fluctuations to the mean anyway. The reduction works only for a straight track at constant speed, since a
curved trajectory has no single translation that removes the time dependence.

We therefore also consider a laboratory (stationary) frame of reference with explicit time
dependence. For free-form paths and further generalizations we supply physically
and spatially rich input channels that reduce the optimization leaps needed to
learn the underlying operator.

\paragraph{The linear solution as the learning scaffold.}
We want to give the models as much information about the problem as we can compute quickly from the
process parameters and the path. The linear conduction solution given below provides a good compromise between speed and information to give the model an initial starting point,
computed in about \SI{15}{\milli\second} per frame.
Ignoring detailed laser light-material interactions, convection, latent heat and any temperature change of the thermophysical properties, the
substrate obeys the linear heat equation, whose solution under a prescribed
heat source is a space-time convolution with the heat kernel:
\begin{equation}
\begin{aligned}
T_{\mathrm{lin}}(\mathbf{x},t)&=\int_{0}^{t}\!\!\int
  G(\mathbf{x}-\mathbf{x}',\,t-t')\,q(\mathbf{x}',t')\,
  \mathrm{d}^{3}\mathbf{x}'\,\mathrm{d}t',\\[4pt]
G(\mathbf{r},\lambda)&=\frac{\exp\!\left(-|\mathbf{r}|^{2}/4a\lambda\right)}{(4\pi a\lambda)^{3/2}},
  \qquad \lambda=t-t' ,
\end{aligned}
\label{eq:greens}
\end{equation}
with $a$ the thermal
diffusivity~\cite{carslaw_conduction_2011}. Here the source
$q$ is the beam shape on the surface, switched by the laser state
$\chi(t)\in\{0,1\}$ ($0$ off, $1$ on) and normalized to unit peak,
\begin{equation}
\begin{aligned}
q(\mathbf{x},t)&=\chi(t)\,f\!\left(|\mathbf{x}-\mathbf{x}_b(t)|\right),\\[4pt]
f(r)&=
\begin{cases}
\exp\!\left(-r^{2}/2\sigma^{2}\right) & \text{Gaussian beam},\\[4pt]
\tfrac12\left[1-\tanh\!\left(4\,(r-\sigma)/\sigma\right)\right] & \text{uniform beam}.
\end{cases}
\end{aligned}
\label{eq:footprint}
\end{equation}
with $\mathbf{x}_b(t)$ the beam position and $\sigma$ the configured beam
radius ($r_0$ in \cref{tab:datasets}). The second form is close to one inside radius $\sigma$, passes one half at $\sigma$, then falls to zero
over a transition about half a beam radius wide. Multiple beams
sum one shape each, weighted by their power share.

The trajectory input channels approximate \cref{eq:greens} in a sequence of
increasing fidelity, from the bare source to the linear field of that source, all
computable in milliseconds from the path before any simulation runs.

The simplest is the source itself, supplied as
the beam center coordinates and the footprint $f$ of \cref{eq:footprint}
evaluated around the current beam position, the $\lambda\rightarrow0$ limit of
\cref{eq:greens} where the kernel collapses onto the source. A network given
only the footprint must integrate the history itself.

\paragraph{Exposure history.}
The next set of channels approximates the time convolution. Writing
$f(\mathbf{x},t)$ for the footprint of \cref{eq:footprint} at the beam position
of time $t$, the exponentially weighted integrals
\begin{equation}
D_{\tau}(\mathbf{x},t)=\int_{0}^{t}e^{-(t-t')/\tau}\,f(\mathbf{x},t')\,\mathrm{d}t'
\label{eq:dose}
\end{equation}
are the convolution of the source with a single-exponential kernel of time
constant $\tau$. Each is a running total of exposure that forgets at a fixed
rate: a pass the beam made a time $\Delta t$ ago still contributes, but weighted
by $e^{-\Delta t/\tau}$, so a point's value says how much energy it has received
and how recently. It is computed as one recursion per frame, $D_\tau \leftarrow
D_\tau e^{-\Delta t/\tau} + f\,\Delta t$, which is the leaky integrator of signal
processing, the simplest first-order low-pass
filter.

\begin{figure}[htbp]
\centering
\begin{subfigure}[t]{0.32\textwidth}\includegraphics[width=\linewidth]{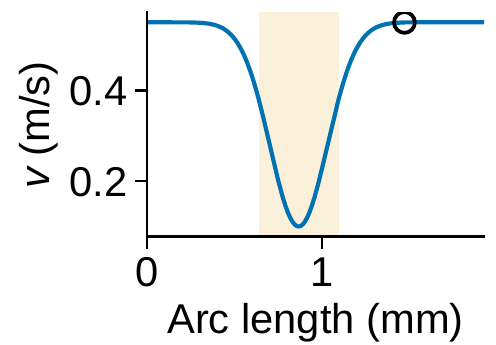}\subcaption{Scan speed along the path.}\label{fig:ch-speed}\end{subfigure}\hfill
\begin{subfigure}[t]{0.32\textwidth}\includegraphics[width=\linewidth]{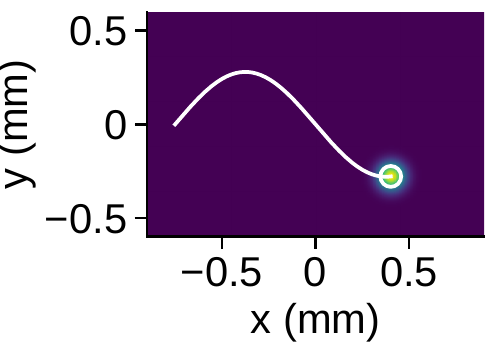}\subcaption{Beam footprint.}\label{fig:ch-footprint}\end{subfigure}\hfill
\begin{subfigure}[t]{0.32\textwidth}\includegraphics[width=\linewidth]{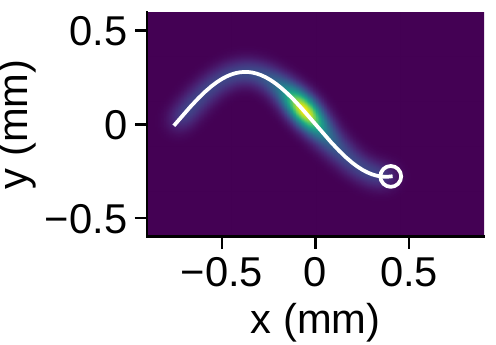}\subcaption{Dose, no decay.}\label{fig:ch-dose-inf}\end{subfigure}

\vspace{0.4em}
\begin{subfigure}[t]{0.32\textwidth}\includegraphics[width=\linewidth]{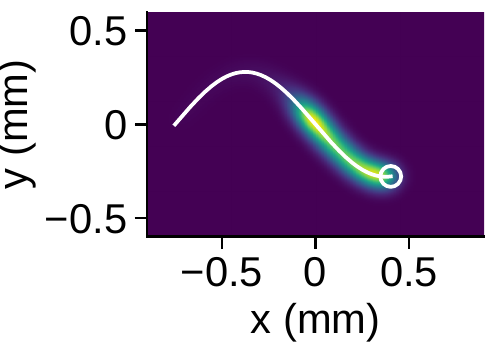}\subcaption{Dose, \SI{1}{\milli\second} decay.}\label{fig:ch-dose-1ms}\end{subfigure}\hfill
\begin{subfigure}[t]{0.32\textwidth}\includegraphics[width=\linewidth]{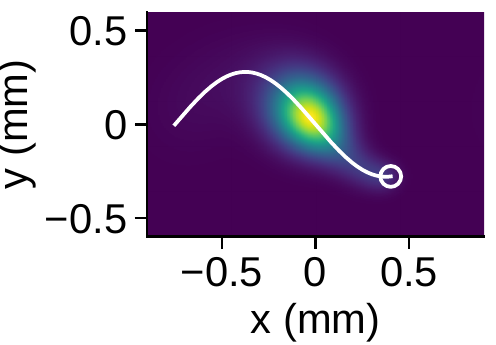}\subcaption{Conduction response.}\label{fig:ch-conduction}\end{subfigure}\hfill
\begin{minipage}[t]{0.32\textwidth}\includegraphics[width=0.4375\linewidth]{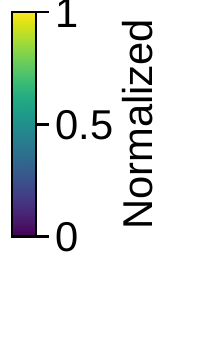}\end{minipage}

\vspace{0.4em}
\begin{subfigure}[t]{0.32\textwidth}\includegraphics[width=\linewidth]{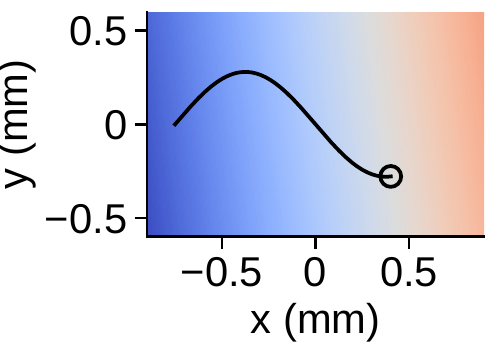}\subcaption{Along-track offset $s/\sigma$.}\label{fig:ch-s}\end{subfigure}\hfill
\begin{subfigure}[t]{0.32\textwidth}\includegraphics[width=\linewidth]{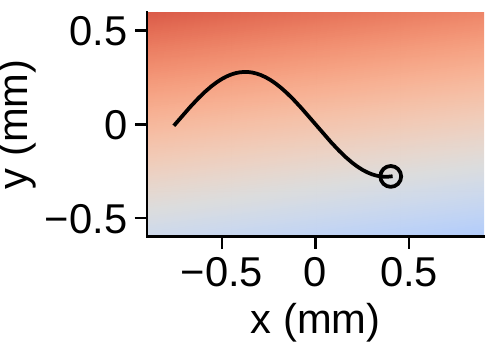}\subcaption{Cross-track offset $n/\sigma$.}\label{fig:ch-n}\end{subfigure}\hfill
\begin{minipage}[t]{0.32\textwidth}\includegraphics[width=0.4375\linewidth]{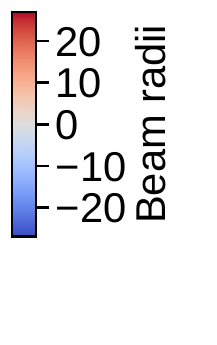}\end{minipage}
\caption{Model input channels computed from the laser trajectory, on a
synthetic variable-speed curved path at one instant shortly after the slow-down
shaded in the scan-speed profile \subref{fig:ch-speed}. The beam footprint \subref{fig:ch-footprint}, \cref{eq:footprint},
is integrated along the path into a dose with no decay \subref{fig:ch-dose-inf} and with the
\SI{1}{\milli\second} decay time of \cref{eq:dose} \subref{fig:ch-dose-1ms}, and the conduction
response \subref{fig:ch-conduction} is the temperature rise of a purely conducting substrate
(\cref{eq:greens}), each normalized to its own maximum. The along-track \subref{fig:ch-s} and cross-track \subref{fig:ch-n} offsets of
each point from the beam center are divided by the beam radius.}
\label{fig:channels}
\end{figure}

Two members are supplied to the models reported here. The cumulative dose is
the $\tau\rightarrow\infty$ member, with no temporal cooling, and one leaky integrator
at $\tau=\SI{1}{\milli\second}$ carries recency, a diffusion length
$\sqrt{a\tau}$ of about \SI{55}{\micro\metre} in Ti-6Al-4V
(\cref{fig:channels}). Both are useful because the cumulative dose alone cannot
distinguish the order in which energy arrived: a segment traversed fast and then
slow deposits the same total as the same segment traversed slow and then fast,
while leaving the substrate in a different state.
Where a path crosses or comes close to itself the field depends on how long
ago the beam last passed, which a single cumulative sum cannot express. Both exposure channels carry the laser power. Each is multiplied by
$P/P_{\mathrm{ref}}$ with $P_{\mathrm{ref}}=\SI{100}{\watt}$, and the cumulative
member is first divided by the total time the laser is on, so that it is the
time-averaged exposure of a point over the track.

\paragraph{The geometry of the moving source.}
The third set of channels provides the model with further geometric information about relative beam location. We supply $s/\sigma$ and $n/\sigma$ where $s$
is how far ahead of the beam a point lies and $n$ how far to its left, both
divided by the beam radius (\cref{fig:ch-s,fig:ch-n}). This helps the models learn the direction of movement, for example differentiating between the front of the beam and the back, which the radially symmetric footprint cannot do.

\paragraph{The linear field itself, and what is left to learn.}
The last channel evaluates \cref{eq:greens} on the simulation's grid
and at its own stored frames. Each spatial Fourier mode of the heat equation
decays independently of the others, so the field is carried as its Fourier coefficients
$\hat{T}$ and advanced one stored frame at a time. The coefficients $\hat{T}_{n}$ at frame $n$ follow from those
of the previous frame, the interval $\Delta t_{n}$ between the two frames and the beam position
$\mathbf{x}_{b,n}$ at frame $n$ as
\begin{equation}
\begin{aligned}
\hat{T}_{n}(\mathbf{k}) &= e^{-a|\mathbf{k}|^{2}\Delta t_{n}}\,\hat{T}_{n-1}(\mathbf{k})
 + \chi_{n}\,\Delta t_{n}\,\hat{g}(\mathbf{k})\,
   e^{-\mathrm{i}\,(k_{x}x_{b,n}+k_{y}y_{b,n})},\\[4pt]
\hat{g}(\mathbf{k}) &= (2\pi{\sigma_c}^{2})^{3/2}\,e^{-{\sigma_c}^{2}|\mathbf{k}|^{2}/2}.
\end{aligned}
\label{eq:conduction-update}
\end{equation}
The first term cools every mode over the interval and the second deposits
the beam's energy at its position in that frame. Here $\mathbf{k}$ is the
wavevector of the grid's discrete Fourier transform, in radians per meter, with
components $2\pi m/(N\Delta x)$ for integer $m$ and $N$ cells of size $\Delta x$ along each axis;
$\hat{g}$ is the closed-form transform of the heat source of \cref{eq:conduction-update}, a three-dimensional
Gaussian of standard deviation $\sigma_c$ centered on the free surface; and
$\chi_{n}\in\{0,1\}$ is the laser state of \cref{eq:footprint}. The
diffusivity $a$ is held constant at the room-temperature conductivity, density
and specific heat of the case's own material, \SI{2.9e-6}{\metre\squared\per\second}
for Ti-6Al-4V and \SI{2.5e-6}{\metre\squared\per\second} for IN718. The solver integrates the temperature-dependent curves.
The domain wraps in $x$ and $y$. In $z$ the
metal block is mirrored about its free surface, which makes that surface
adiabatic: the mirror image deposits the same heat at the same depth on the
other side, so no temperature gradient crosses the plane. The field is normalized per
simulation, after all frames have been built. 

Handing the operator
the linear part in closed form leaves it to predict the nonlinear remainder (latent heat,
convection in the melt pool, temperature-dependent properties, evaporation and
the free-surface depression, or keyhole). 

\Cref{fig:channels} shows the model input channels on a synthetic curved path, at one instant shortly
after the beam has slowed down and sped up again \subref{fig:ch-speed}. The laser power is
constant along this track. A power that varies along the path could enter the input channels in the
same way as the varying speed but is not considered here. The footprint \subref{fig:ch-footprint} marks only where the
beam is at that instant. The dose with no decay \subref{fig:ch-dose-inf} peaks where the beam
moved slowest, however long ago that was, while the dose with a \SI{1}{\milli\second} decay
\subref{fig:ch-dose-1ms} weights each point by how recently the beam passed. The conduction
response \subref{fig:ch-conduction}, defined above, also peaks at the slow-down, behind the beam,
and the offsets \subref{fig:ch-s} and \subref{fig:ch-n} give each point's position relative to
the beam, taking movement direction into account.

\begin{table}[h!]
\centering
\caption{Every input the operators take, grouped by what it supplies.
Process scalars and coordinates give ``raw'' process information; the trajectory groups
approximate the terms of \cref{eq:greens} and are computed from path geometry and
timing alone, before any simulation runs. $\sigma$ is the beam radius, $a$ the
thermal diffusivity, and a broadcast channel is constant over the domain at a
given instant.}
\label{tab:channels}
\small
\begin{tabular*}{\textwidth}{@{\extracolsep{\fill}}@{}>{\raggedright\arraybackslash}p{0.22\textwidth}>{\raggedright\arraybackslash}p{0.47\textwidth}>{\raggedright\arraybackslash}p{0.23\textwidth}@{}}
\toprule
Channel & What it is & Where it matters \\
\midrule
\multicolumn{3}{@{}l}{\emph{Broadcast scalars}} \\
$P$ & laser power & var. power \\
$V$ & laser scan speed at the current instant & var. scan speed \\
$H^*$ & normalized enthalpy & every dataset \\
$r_0$ & beam radius & var. radius \\
beam shape & boolean to differentiate two beam shapes & var. beam-shape \\
$T_0$ & uniform initial temperature & preheating \\
alloy label & 1 for IN718, 0 for Ti-6Al-4V & two-alloy models \\
\midrule
\multicolumn{3}{@{}l}{\emph{Coordinates}} \\
$x$, $y$, $z$ & cell position in the fixed domain & every model \\
\midrule
\multicolumn{3}{@{}l}{\emph{Source}} \\
$x_b$, $y_b$ (broadcast) & beam center & every path \\
footprint $f$ & \cref{eq:footprint} at the current beam position & every path \\
\midrule
\multicolumn{3}{@{}l}{\emph{Time convolution}} \\
recent dose $D_\tau$ & \cref{eq:dose} at $\tau = \SI{1}{\milli\second}$, $\sqrt{a\tau} \approx \SI{55}{\micro\metre}$ & free-form paths \\
cumulative dose & $D_\infty$, the footprint integrated along the path so far & free-form paths \\
\midrule
\multicolumn{3}{@{}l}{\emph{Geometry of the moving source}} \\
$s/\sigma$, $n/\sigma$ & signed offsets along and across the travel & curved paths \\
\midrule
\multicolumn{3}{@{}l}{\emph{The linear field}} \\
conduction response & \cref{eq:greens} evaluated along the path & every path \\
\bottomrule
\end{tabular*}
\end{table}

\paragraph{Case constants, and the channels the results use.}
We also supply laser power, scan speed and normalized enthalpy $H^*$~\cite{hann_simple_2011} as spatially uniform channels, and beam radius, beam shape and preheat
temperature likewise on datasets that vary them. We compute $H^*$ from laser power and scan speed alone, as $P/\sqrt{V}$ divided by its
maximum over the training set, and leave out the beam radius and the material constants of its
full definition in model inputs, as including them led to worse results in early experiments. The beam radius is a separate model input, and the material constants are fixed
within a single-alloy dataset. The two-alloy models keep the material constants in $H^*$ and divide $P/\sqrt{V}$ by $\rho\,[c_p(T_l - \SI{293}{\kelvin}) + L_f]\sqrt{\pi a}/A$, with the density $\rho$, specific heat $c_p$, liquidus $T_l$, latent heat of fusion $L_f$ and thermal diffusivity $a$ of each alloy and a nominal absorptivity $A$. With these constants, $H^*$ is about \SI{8}{\percent} lower for IN718 than for Ti-6Al-4V at equal $P$ and $V$. The two-alloy models also receive a binary alloy label as an input channel.
\Cref{tab:channels} summarizes the input channels.

Fields are normalized by reference scales before training,
and the gas is masked as described above.

\FloatBarrier
\subsubsection{Neural operator architectures}\label{sec:models}

We compare three neural operator architectures: FNO, RIGNO, and GAOT. All the architectures use the same general inputs and produce comparable outputs.
We provide a short introduction to each architecture below, and refer the reader to the original sources of each architecture for more details.

\paragraph{FNO.}
The FNO~\cite{li_fourier_2021} convolves the latent space in the Fourier
domain and keeps the lowest modes, making the learned operator independent
of the evaluation grid. It is the main architecture of~\cite{benoit_fast_2026} and is one of the most widely used neural operators. We use the implementation in NVIDIA
PhysicsNeMo~\cite{physicsnemo_contributors_nvidia_2023}, which also provides the
training loop and the distributed-execution utilities all three architectures are
run through. The main drawback of this architecture is that it is usually fixed to one geometry and to a regular mesh due to its use of the discrete Fourier transform.

\paragraph{RIGNO.}
RIGNO~\cite{mousavi_rigno_2025} is a graph neural operator which encodes the input field points onto progressively coarser
region nodes, passes messages on a multiscale graph over them, and decodes back
to the query points. It does not assume a regular grid, so it extends to
non-Cartesian domains and varying geometries. We repurpose RIGNO's conditioned normalization so that laser power, scan speed and $H^*$, and
in the two-alloy models the alloy label, set a scale and a shift applied to the internal features
of its processor, in the manner of
feature-wise linear modulation (FiLM)~\cite{perez_film_2018}. The same quantities, the alloy label included,
also enter RIGNO as input channels.
We ported RIGNO's reference implementation from JAX to PyTorch, the framework PhysicsNeMo is built
on.

\paragraph{GAOT.}
GAOT~\cite{wen_geometry_2025}, also designed for arbitrary domains, keeps the same encode, process and decode
structure as RIGNO but replaces message passing with a transformer over latent tokens,
with attentional encoders and an optional local-geometry embedding that we
disable, since a uniform Cartesian grid gives it nothing to describe. GAOT's
reference implementation is already in PyTorch, and we reuse its layers unchanged. Every sample
shares one grid, so we build the graphs linking grid points to latent tokens once and reuse them for
all samples.

\paragraph{Other possible architectures.}
Other possible architectures for a machine-learning laser processing surrogate such as this include U-Net~\cite{ronneberger_u-net_2015}, pointwise fully
connected neural networks and DeepONet~\cite{lu_deeponet_2021}. However, these were explored in~\cite{benoit_fast_2026} and were found to be less accurate 
or less efficient on steady melt-pool predictions varying power and scan speed. We therefore focus this study on the three architectures above.

\subsubsection{Time discretization and prediction mode}\label{sec:transient}

We predict each frame directly from the time and the trajectory history in one shot.
Field history reaches models only
through the trajectory channels described above.
While we could also train the model autoregressively, using previously predicted frames as input,
we cannot hope to learn chaotic fluctuations of keyholes~\cite{gan_universal_2021}
from a pointwise $L_2$ loss, which implicitly regresses the model to the mean field behavior.
By giving rich enough inputs to the surrogate, we can thus infer the slowly
varying part of the field, the part the process parameters determine, through
one-shot inference, which avoids error accumulation pathologies of autoregressive models. This regression to the mean means that in practice the models continue to
learn the quasi-steady state of the system even when trained in the transient
laboratory frame, so the quasi-steady formulation of~\cite{benoit_fast_2026}
remains the efficient choice for straight-path processes.
The regression to the mean also sets an error floor on what the models can actually learn.

\subsection{Autonomous acquisition under a fixed solver budget}\label{sec:acquisition}

Each FLOW-3D\textsuperscript{\textregistered} training simulation occupies a licensed solver seat 
for tens of minutes, up to an hour and a half, one at a time. Choosing which simulations to run thus becomes a crucial part of building the surrogate pipeline.

\paragraph{Fixed designs.}
Full-factorial grids, as used in~\cite{benoit_fast_2026}, cover parameter space well, but are prohibitively expensive for varying many parameters. 
We thus generate our fixed datasets using
Latin hypercube sampling (LHS)~\cite{mckay_comparison_1979}, which covers the parameter space by drawing one sample from each equally
probable interval of every parameter axis. The exception is the recreation of the straight path dataset varying power and speed from \cite{benoit_fast_2026}, for which we use the same grid and add more keyhole regime simulations using LHS.

\paragraph{Teaching a second alloy.}
We also test a scheme where, given a trained model and a corresponding dataset, the model error is used to expand the training set to a new parameter region.
More specifically, we take an FNO trained on Ti-6Al-4V simulations varying power, scan speed and beam radius, and we expand its training set to a new alloy, IN718, over five rounds of selection.
Once the dataset is created, other architectures can also be trained on it. Alloy information enters models through a binary label, and implicitly through the linear conduction response input and through $H^*$.

To choose new training simulations, we combine two methods. Each round chooses \num{50} new simulations, over five rounds. Candidate simulations are chosen within the following bounds:
power from \SIrange{90}{350}{\watt}, scan speed
\SIrange{0.15}{1.25}{\metre\per\second} and beam radius
\SIrange{22}{65}{\micro\metre}.
Half of those 50 are chosen by the maximin algorithm~\cite{gonzalez_clustering_1985}, a space-filling algorithm which greedily chooses the point furthest from the current points.

The other half is selected via a multi-armed bandit algorithm~\cite{lattimore_bandit_2020}. The parameter space is first divided into four normalized-enthalpy based regimes, lack of fusion, conduction, transition and keyhole. The bandit then picks its simulations one at a time. Each pick goes to the regime $r$ with the highest score $w_r\,(b_r E_r/E_{\max} + c)/(n_r+1)$, and inside it to the point chosen by maximin. Here $E_r$ is the beam-box mean absolute error (MAE, \cref{sec:metrics}) on the regime's four simulations of a \num{16}-simulation probe set, kept out of every training set and the test set. It is the median over three seeds of a model retrained each round on the Ti-6Al-4V set and the bandit's own selections, and $E_{\max}$ is its largest value across regimes. The exponent $b_r$ measures how fast that error falls as the regime gains simulations. It is fitted as $E_r(n)\sim a_r(n+1)^{-b_r}$ over past rounds and pulled toward the mean exponent across regimes. The count $n_r$ is the number of simulations the bandit has already chosen in the regime, $w_r$ is the regime's share of the candidate space and $c=0.5$ is an exploration constant. A regime thus keeps receiving simulations while its error is high and still falling.

The two simulation set halves are chosen independently, and the final models train on both
halves combined. The
test set is \num{24} held-out IN718 simulations, and the models of the loop are
trained at three different random seeds to average out run-to-run variability.

\subsection{Evaluation metrics}\label{sec:metrics}

\paragraph{Base metrics.}
For clarity, we report here only the metrics we find most informative about surrogate performance. Except for the beam-box MAE defined below, they are computed exactly in the same way as in \cite{benoit_fast_2026}. Every metric and prediction shown in this work is computed on simulations the models did not see during training.
We report mean absolute error (MAE) and root mean squared error (RMSE) of the temperature field in kelvin, and the relative error in percent, the T MAE divided by the mean true temperature of the substrate. 
We also report the intersection over union (IoU) of the thresholded melt pool defined by the liquid fraction $f_l$, the cells with $f_l > 0.5$.
This metric implicitly captures both the surrogate prediction accuracy of the vapor cavity interface and the mid liquidus--solidus isotherm in the substrate.
When comparing the surrogate against experiments, we report melt-pool and vapor cavity depth agreement. 

\paragraph{Beam-box MAE.}
Since we consider domains of varying sizes across different datasets, we introduce a new domain-agnostic metric we call \emph{beam-box MAE} (BB MAE). 
In laser processing, most of the complicated multi-physics behavior in the substrate happens around the beam center. Far-field predictions are easy, as they converge to the initial temperature; 
this means that increasing the domain of a simulation can artificially ``dilute'' a model's MAE and RMSE by giving lower average metrics and making roughly equivalent predictions look better.
We thus consider the MAE inside a ``box'' computed as the union of the cells located within two beam radii of the beam center.
For multiple beams we use the union of all the boxes. The BB MAE thus more precisely captures model accuracy in predicting close-to-beam substrate behavior.

\section{Results and Discussion}\label{sec:results}

\subsection{Straight path baseline dataset results}\label{sec:arch-straight}
As a baseline, we first consider the same dataset as in \cite{benoit_fast_2026} which varies laser scan speed and power along a straight path on Ti-6Al-4V across the LoF, conduction, and keyhole melting regimes. We recompute the dataset with our experimentally validated solver parameters, and add new simulations in the keyhole regime, as the new solver parameters used here lead to lower energy regimes for the same condition, partly because of the higher vaporization temperature $T_{v1}$ (\cref{sec:calibration}). The three architectures are also compared on the original dataset of~\cite{benoit_fast_2026} in \cref{sec:benchmark,tab:benchmark}. 

We score the three architectures on the dataset both in the quasi-steady formulation as in \cite{benoit_fast_2026} and as a transient process in the laboratory frame of reference. We note that the quasi-steady formulation is on a smaller domain, hence only beam-box MAE and melt-pool IoU are directly comparable across the two formulations (the beam box is defined in \cref{sec:metrics}). In the laboratory-frame comparison, we train all three architectures on every second cell along $x$ and $y$, drawing each frame at random from the whole training set. We have found that drawing frames uniformly in this way usually lowers the substrate T MAE compared with reading a few simulations at a time (coherent sampling). However, it needs the whole training set held in memory, which cannot fit on our hardware with full resolution training. All three models receive the same input channels.
We evaluate the models at full resolution, making use of the resolution invariance property of neural operators \cite{li_fourier_2021, mousavi_rigno_2025, wen_geometry_2025}.

The results are summarized in \cref{tab:arch-steady}. The results in this table and elsewhere use training hyperparameters and a training configuration found to be most advantageous to the architecture on the tested dataset. They were tuned iteratively, by a local search over the hyperparameters scored on the validation set. The full configuration of every model, from its architecture sizes to its optimizer, learning-rate schedule, batch size and loss, is stored with the model in the released provenance store (\cref{sec:provenance-supp}). The architecture rankings here should thus not be considered as final, as further tuning could likely switch the order especially when metrics are close to one another.

In the quasi-steady formulation, FNO dominates the other two architectures on all metrics except for IoU, with a relative temperature error of \SI{1.25}{\percent}, while RIGNO is most promising in the time-dependent laboratory frame, with the BB MAE closest to its quasi-steady counterpart: \SI{30.9}{\kelvin} for the laboratory frame versus \SI{23.6}{\kelvin} for quasi-steady.

In \cref{tab:cost} examples of the run times for training in both formulations and for inference of one frame on this dataset are given. Whatever architecture is used, inference is orders of magnitude faster than the equivalent simulation would be.
Among the surrogates, FNO is the fastest and cheapest to train and infer, followed by GAOT and then RIGNO, separated by factors of about 9 and 4 in inference time, and per training step by about 2 and 1.6 in the quasi-steady formulation and by 6 and 3 in the laboratory frame. The ordering matches the cost of the main operation of each architecture (\cref{sec:models}). FNO uses fast Fourier transforms on a regular grid, GAOT computes attention weights between a small set of coarse latent points, and RIGNO exchanges information along every edge of a multiscale graph. A laboratory-frame training step processes eight frames in a batch and takes 1.4 times as long as a quasi-steady step for FNO, 4.0 times for GAOT and 6.4 times for RIGNO.

Training data generation is the biggest bottleneck in terms of time of these neural operator surrogates, which is why the parameter space has to be covered as efficiently as possible. We note, however, that although data generation and training take a long time, they only have to be done once for a given parameter space (unless a self-improving active learning loop is considered). 

\begin{table}[htbp]
\centering
\caption{Accuracy on the straight-path power and scan speed (P-V) test split, \num{13}
held-out simulations of a dataset spanning \SIrange{40}{215}{\watt} and \SIrange{0.10}{0.96}{\metre\per\second}, in the quasi-steady and laboratory formulations. Mean and standard deviation over $3$ training
runs, best per block in bold.}
\label{tab:arch-steady}
\small
\begin{tabular*}{\textwidth}{@{\extracolsep{\fill}}lccccc}
\toprule
 & \tabhead{T MAE\\(\si{\kelvin})} & \tabhead{T RMSE\\(\si{\kelvin})} & \tabhead{BB MAE\\(\si{\kelvin})} & \tabhead{Rel. err.\\(\%)} & \tabhead{Melt-pool\\IoU} \\
\midrule
\multicolumn{6}{@{}l}{\emph{Quasi-steady formulation}} \\
FNO    & $\mathbf{7.71 \pm 0.05}$ & $\mathbf{29.6 \pm 0.8}$ & $\mathbf{15.0 \pm 1.1}$ & $\mathbf{1.25 \pm 0.01}$ & $0.920 \pm 0.001$ \\
RIGNO  & $11.36 \pm 1.69$ & $40.0 \pm 9.7$ & $23.6 \pm 2.3$ & $1.84 \pm 0.27$ & $\mathbf{0.926 \pm 0.004}$ \\
GAOT & $9.74 \pm 0.44$ & $32.3 \pm 0.9$ & $16.8 \pm 0.2$ & $1.58 \pm 0.07$ & $0.915 \pm 0.005$ \\
\midrule
\multicolumn{6}{@{}l}{\emph{Laboratory frame}} \\
FNO    & $10.01 \pm 0.10$ & $27.3 \pm 1.1$ & $49.1 \pm 2.7$ & $2.25 \pm 0.02$ & $0.863 \pm 0.002$ \\
RIGNO  & $\mathbf{6.65 \pm 0.46}$ & $\mathbf{21.4 \pm 0.5}$ & $\mathbf{30.9 \pm 0.5}$ & $\mathbf{1.49 \pm 0.10}$ & $\mathbf{0.926 \pm 0.001}$ \\
GAOT & $11.13 \pm 0.09$ & $41.7 \pm 5.3$ & $72.8 \pm 2.6$ & $2.50 \pm 0.02$ & $0.826 \pm 0.004$ \\
\bottomrule
\end{tabular*}
\end{table}

\begin{table}[htbp]
  \centering
  \caption{Indicative cost of training and inference of the models, timed on an NVIDIA V100
  GPU. The training time per step is the median over \num{100} steps after a warm-up, validation excluded, for
  one model of each formulation in \cref{tab:arch-steady}. Quasi-steady models train at batch size one, laboratory-frame models at batch size
  eight on the \numproduct{90x40x40} laboratory grid subsampled to every second cell in $x$ and $y$. 
  Inference is a batch-one forward
  pass of the laboratory-frame model on one test frame of the full grid.
  Timings here are to be taken as indicative,
  since hardware type and load, code optimizations, and hyperparameter choice can move them substantially.}
  \label{tab:cost}
  \small
  \setlength{\tabcolsep}{4pt}%
  \begin{tabular*}{\textwidth}{@{\extracolsep{\fill}}lrrrrr}
  \toprule
  & \multicolumn{2}{c}{Training, quasi-steady} & \multicolumn{2}{c}{Training, laboratory frame} & Inference \\
  \cmidrule(lr){2-3}\cmidrule(lr){4-5}
  & per step & \num{32000} steps & per step & \num{32000} steps & per frame \\
  \midrule
  FNO   & \SI{55}{\milli\second}  & \SI{29}{\minute}  & \SI{76}{\milli\second}   & \SI{40}{\minute}  & \SI{6.8}{\milli\second} \\
  GAOT  & \SI{118}{\milli\second} & \SI{1.1}{\hour}   & \SI{467}{\milli\second}  & \SI{4.2}{\hour}   & \SI{61.7}{\milli\second} \\
  RIGNO & \SI{188}{\milli\second} & \SI{1.7}{\hour}   & \SI{1210}{\milli\second} & \SI{10.8}{\hour}  & \SI{245}{\milli\second} \\
  \midrule
  \multicolumn{6}{@{}l}{\emph{Reference data generation}, FLOW-3D\textsuperscript{\textregistered}, \num{123} simulations, \SI{10}{\micro \meter} mesh, \num{10} CPU cores} \\
  & \multicolumn{5}{l}{\SI{28.1}{\hour} wall, \num{281} {core-hours}, \SI{13.7}{\minute} per simulation on average.} \\
  & \multicolumn{5}{l}{Takes longer on free-form datasets.} \\
  \bottomrule
  \end{tabular*}
  \end{table}

\FloatBarrier

\subsection{Axes of generalization}\label{sec:axes}

In this section we explore different ways to generalize the surrogates across different laser process conditions. We cover two of the three axes here: \cref{sec:paths} explores generalizing to arbitrary welding paths, multiple beams, and variable laser scan speed within the path{,} and \cref{sec:params} explores generalizing to new process parameter variations such as beam radius, beam shape, and preheat temperature. We look at generalization to a new material in \cref{sec:al-results}.

\subsubsection{Weld path variations}\label{sec:paths}
Many processes in laser melting, such as laser powder bed fusion (LPBF) and laser wobbling~\cite{hornik_laser_2022}{,} do not use straight laser trajectories. We thus test the abilities of the three neural operator architectures to infer 3D temperature and melt-pool geometry fields on free-form laser paths. 

To do so, we generate a dataset of \num{150} simulations with weld paths of varying lengths that can self-intersect, chosen by a randomized greedy space-filling algorithm in a simulation domain of \numproduct{1.5 x 1.0 x 0.4}~\si{\milli\metre} at \SI{10}{\micro\metre} resolution. We refer to this dataset as ``variable paths''. 

We generate a second similar dataset in which laser scan speed itself changes within a path. This dataset includes \num{140} simulations with short dwells and large speed changes. We refer to it as ``variable speed''. A third dataset includes simulations with two to four different beams of radius \SI{47.5}{\micro \metre} interacting with each other. This one only includes \num{79} simulations, since multi-beam simulations are more computationally expensive to run. It also has more simulations in the LoF and conduction regimes to avoid large keyholes that would not fit the domain when all lasers cross paths with one another. We refer to it as ``multi-beam''.

The simulations in all three datasets range from the LoF to keyhole melting regimes. More details on the datasets can be found in \cref{sec:datasets-detail}.

\begin{figure}[h]
\centering
\begin{subfigure}[t]{0.24\textwidth}\includegraphics[width=\linewidth]{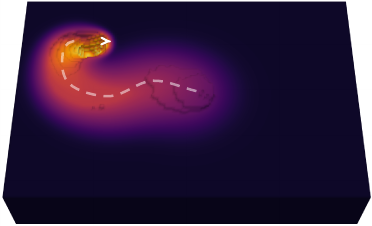}\subcaption{Ground truth.}\label{fig:rp-gt}\end{subfigure}\hfill
\begin{subfigure}[t]{0.24\textwidth}\includegraphics[width=\linewidth]{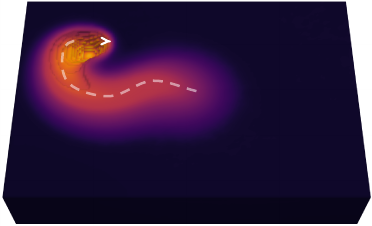}\subcaption{RIGNO, rel.~err.~\SI{6.5}{\percent}, BB MAE \SI{103}{\kelvin}, IoU \num{0.736}.}\label{fig:rp-rigno}\end{subfigure}\hfill
\begin{subfigure}[t]{0.24\textwidth}\includegraphics[width=\linewidth]{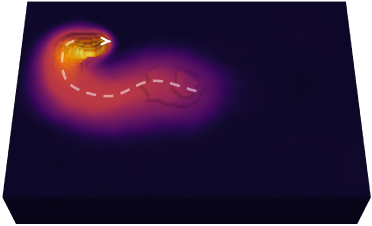}\subcaption{FNO, rel.~err.~\SI{3.1}{\percent}, BB MAE \SI{71}{\kelvin}, IoU \num{0.859}.}\label{fig:rp-fno}\end{subfigure}\hfill
\begin{subfigure}[t]{0.24\textwidth}\includegraphics[width=\linewidth]{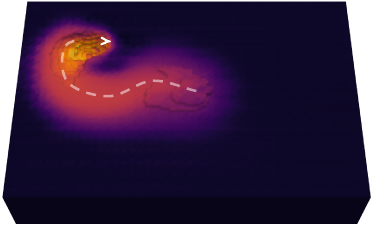}\subcaption{GAOT, rel.~err.~\SI{4.5}{\percent}, BB MAE \SI{113}{\kelvin}, IoU \num{0.736}.}\label{fig:rp-gaot}\end{subfigure}

\vspace{0.3em}
\includegraphics[width=0.5\textwidth]{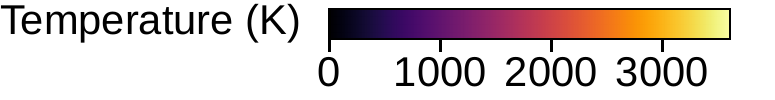}\hspace{1em}\includegraphics[width=0.2\textwidth]{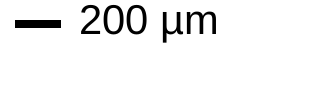}\hspace{1em}\includegraphics[width=0.22\textwidth]{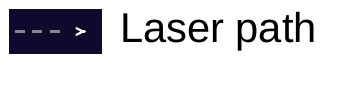}
\caption{One held-out simulation of the variable path dataset, at \SI{193}{\watt} and
\SI{0.74}{\metre\per\second} with a \SI{50}{\micro\metre} beam radius, a third of the way along its
\SI{3.9}{\milli\second} track with the laser on. Ground truth and the predictions of the three variable-path models of
\cref{tab:paths-accuracy} are rendered identically on one color scale.}
\label{fig:rp-artifact}
\end{figure}

Metrics for the three architectures on the three datasets can be found in \cref{tab:paths-accuracy}. All models in \cref{tab:paths-accuracy} are scored on the full mesh, and all are trained on every second cell along $x$ and $y$. Training frames are again drawn uniformly from the whole training set (\cref{sec:arch-straight}). On the variable path dataset, GAOT scores best on all five metrics, and RIGNO is second on all but BB MAE. The metrics are worse than for the straight path dataset from the previous section, as this is an intrinsically harder problem to infer. We have an average relative error of around \SI{4}{\percent} for the best model. A visualization of temperature field inference across the three architectures for one keyhole regime condition in the test set at \SI{193}{\watt} power and \SI{0.74}{\metre\per\second} scan speed can be seen in \cref{fig:rp-artifact}. Visually, the three predictions are similar, although RIGNO does not predict the small solidified mound of the ground truth, and GAOT shows some slight oscillation artifacts in the bulk substrate, which can be smoothed out in post-processing (\cref{sec:gaot-smoothing}).

\begin{figure}[h]
\centering
\begin{subfigure}[t]{0.24\textwidth}\includegraphics[width=\linewidth]{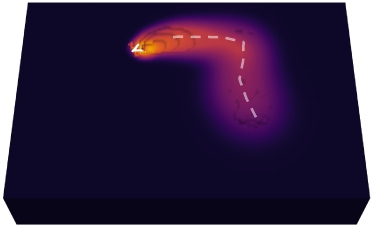}\subcaption{Ground truth, fast, \SI{178}{\watt}, \SI{0.782}{\milli\second}.}\label{fig:vspd-gt-fast}\end{subfigure}\hfill
\begin{subfigure}[t]{0.24\textwidth}\includegraphics[width=\linewidth]{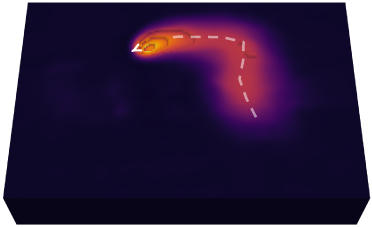}\subcaption{Prediction, fast, \SI{0.782}{\milli\second}, rel.~err.~\SI{3.7}{\percent}, BB MAE \SI{60}{\kelvin}, IoU \num{0.469}.}\label{fig:vspd-pr-fast}\end{subfigure}\hfill
\begin{subfigure}[t]{0.24\textwidth}\includegraphics[width=\linewidth]{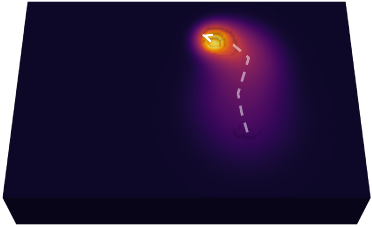}\subcaption{Ground truth, dwell, \SI{99}{\watt}, \SI{1.742}{\milli\second}.}\label{fig:vspd-gt-dwell}\end{subfigure}\hfill
\begin{subfigure}[t]{0.24\textwidth}\includegraphics[width=\linewidth]{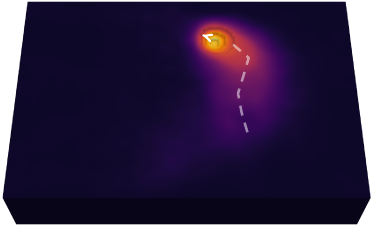}\subcaption{Prediction, dwell, \SI{1.742}{\milli\second}, rel.~err.~\SI{4.8}{\percent}, BB MAE \SI{133}{\kelvin}, IoU \num{0.752}.}\label{fig:vspd-pr-dwell}\end{subfigure}

\vspace{0.4em}
\begin{subfigure}[t]{0.42\textwidth}\includegraphics[width=\linewidth]{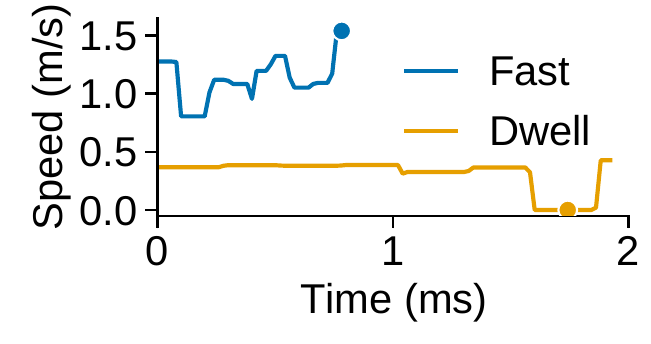}\subcaption{Beam speed along each track.}\label{fig:vspd-speed}\end{subfigure}\hspace{1em}%
\begin{minipage}[t]{0.15\textwidth}\includegraphics[width=\linewidth]{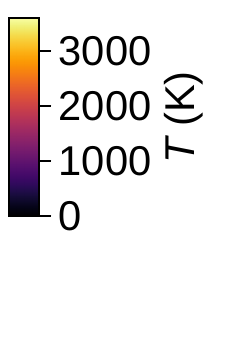}\end{minipage}\hspace{1em}%
\begin{minipage}[b]{0.22\textwidth}\includegraphics[width=\linewidth]{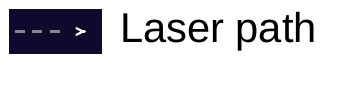}\\\includegraphics[width=0.909\linewidth]{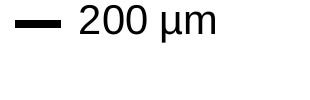}\end{minipage}
\caption{FNO temperature field predictions against ground truth on two held-out simulations varying laser scanning speed on a free-form scan path. Panels \subref{fig:vspd-gt-fast} and \subref{fig:vspd-pr-fast} show the instant of peak scan speed, \SI{1.54}{\meter\per\second}, in one condition, and panels
\subref{fig:vspd-gt-dwell} and \subref{fig:vspd-pr-dwell} show a condition mid-dwell after scanning with speed hovering around \SI{0.4}{\metre \per \second}. Both simulations use a \SI{50}{\micro\metre} beam radius.}
\label{fig:varspeed-pred}
\end{figure}

On the variable speed dataset, FNO and GAOT are within \SI{10}{\percent} of each other on the four temperature metrics (T MAE, T RMSE, BB MAE and relative error), and both have lower errors than RIGNO. GAOT has the highest melt-pool IoU, followed by RIGNO, then FNO. Overall metrics are surprisingly better on average than the variable path dataset ones for the three architectures. We attribute the lower errors to the variable path test simulations running for longer periods of time on average. \Cref{fig:varspeed-pred} shows two sample FNO temperature field predictions{:} one condition with a fast laser scan speed, varying from \SIrange{0.8}{1.54}{\metre\per\second} at \SI{178}{\watt} power; and another condition featuring a short laser dwell going from around \SI{0.4}{\metre \per \second} to stationary at \SI{99}{\watt} power. For both conditions the FNO prediction is visually very similar to the ground truth, differing only by one cell across some cavity details, and with some slight smearing artifacts at the edges of the cooling wake, although faint spurious low-temperature patches also appear in the substrate away from the tracks. This pathology of FNO can usually be fixed with more training data.

\begin{figure}[h]
\centering
\begin{subfigure}[t]{0.24\textwidth}\includegraphics[width=\linewidth]{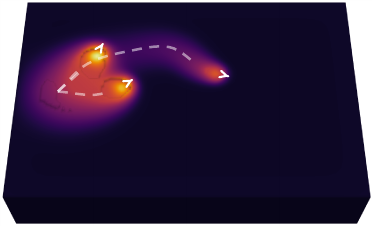}\subcaption{Ground truth, three beams at \SIlist[list-units=single]{63;71;36}{\watt}, \SI{0.925}{\milli\second}.}\label{fig:mb-gt-3}\end{subfigure}\hfill
\begin{subfigure}[t]{0.24\textwidth}\includegraphics[width=\linewidth]{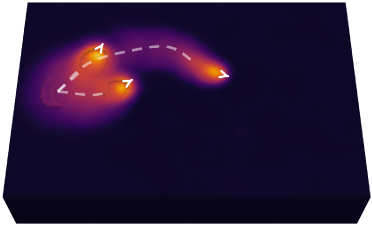}\subcaption{Prediction, three beams, \SI{0.925}{\milli\second}, rel.~err.~\SI{4.1}{\percent}, BB MAE \SI{120}{\kelvin}, IoU \num{0.535}.}\label{fig:mb-pr-3}\end{subfigure}\hfill
\begin{subfigure}[t]{0.24\textwidth}\includegraphics[width=\linewidth]{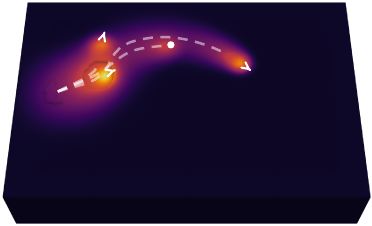}\subcaption{Ground truth, four beams at \SIlist[list-units=single]{28;50;31;54}{\watt}, \SI{1.025}{\milli\second}.}\label{fig:mb-gt-4}\end{subfigure}\hfill
\begin{subfigure}[t]{0.24\textwidth}\includegraphics[width=\linewidth]{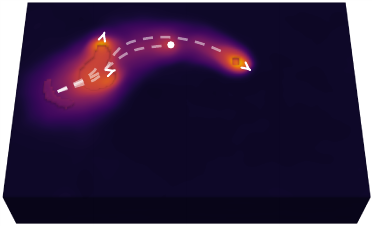}\subcaption{Prediction, four beams, \SI{1.025}{\milli\second}, rel.~err.~\SI{2.6}{\percent}, BB MAE \SI{111}{\kelvin}, IoU \num{0.596}.}\label{fig:mb-pr-4}\end{subfigure}

\vspace{0.3em}
\includegraphics[width=0.5\textwidth]{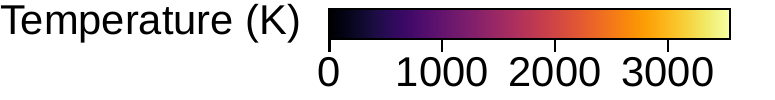}\hspace{1em}\includegraphics[width=0.16\textwidth]{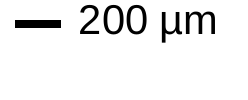}\hspace{1em}\includegraphics[width=0.26\textwidth]{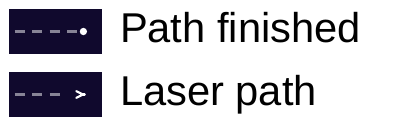}

\caption{FNO prediction against ground truth on two held-out multi-beam simulations, with three beams in
\subref{fig:mb-gt-3} and \subref{fig:mb-pr-3} and four in \subref{fig:mb-gt-4} and
\subref{fig:mb-pr-4}. Each beam has a \SI{47.5}{\micro\metre} radius and follows its own free-form path, at
speeds of \SIrange{0.28}{0.90}{\metre\per\second} in the three-beam simulation and
\SIrange{0.22}{0.94}{\metre\per\second} in the four-beam one.}
\label{fig:multibeam-pred}
\end{figure}

Lastly, on the multi-beam dataset, FNO and GAOT are within \SI{4}{\percent} of each other on the four temperature metrics, and both have lower errors than RIGNO. As on the variable speed dataset, GAOT has the highest melt-pool IoU, and FNO and RIGNO are about equal on that metric. Melt-pool IoU is lower than on the variable path and variable speed datasets, likely due in part to the dataset containing smaller melt pools, inflating small segmentation errors. The lower proportion of deep keyhole cases, however, may also make model predictions somewhat easier. These models are fine-tuned. Each starts from a variable-path model of the same architecture, and all of its weights are then trained for \num{6000} further iterations on the \num{61} multi-beam training simulations, with the same input fields as its starting model. \Cref{fig:multibeam-pred} shows FNO predictions against ground truth for a condition with three beams and another with four. In both cases the FNO inference is visually very similar to the ground truth.

\begin{table}[htbp]
\centering
\caption{Test-split metrics of one model per architecture on each free-form path dataset. All models are trained on
frames drawn uniformly from the training set (\cref{sec:arch-straight}), with the inputs given in \cref{sec:formulation}. Across the three datasets, power spans \SIrange{27}{252}{\watt} per beam, scan speed reaches \SI{2.0}{\metre\per\second} along the variable-speed paths, and the multi-beam paths carry up to four beams. Best metric per block is in bold.}
\label{tab:paths-accuracy}
\small
\sisetup{range-phrase=--,range-units=single,per-mode=symbol}%
\begin{tabular*}{\textwidth}{@{\extracolsep{\fill}}lccccc}
\toprule
& \tabhead{T MAE\\(\si{\kelvin})} & \tabhead{T RMSE\\(\si{\kelvin})} & \tabhead{BB MAE\\(\si{\kelvin})} & \tabhead{Rel. err.\\(\%)} & \tabhead{Melt-pool\\IoU} \\
\midrule
\multicolumn{6}{@{}l}{\emph{Variable-length free-form paths (\SIrange{0.11}{2.97}{\milli\metre} long, $P$ \SIrange{68}{245}{\watt}, $V$ \SIrange{0.21}{0.99}{\metre\per\second})}} \\
FNO   & {18.45} & {57.92} & {106.9} & {4.87} & {0.723} \\
RIGNO & {16.48} & {45.94} & {118.2} & {4.35} & {0.733} \\
GAOT  & \textbf{14.10} & \textbf{40.92} & \textbf{100.9} & \textbf{3.72} & \textbf{0.772} \\
\midrule
\multicolumn{6}{@{}l}{\emph{Variable speed along free-form paths (\SIrange{0.20}{1.63}{\milli\metre} long, $V$ \SIrange{0}{2.0}{\metre\per\second}, $P$ \SIrange{53}{252}{\watt})}} \\
FNO   & \textbf{7.35} & {29.60} & \textbf{95.8} & \textbf{2.20} & {0.741} \\
RIGNO & {11.44} & {41.89} & {164.5} & {3.42} & {0.751} \\
GAOT  & {7.56} & \textbf{27.37} & {96.1} & {2.26} & \textbf{0.773} \\
\midrule
\multicolumn{6}{@{}l}{\emph{Multiple beams, free-form paths (2 to 4 beams, $P$ \SIrange{27}{155}{\watt} per beam, $V$ \SIrange{0.22}{0.98}{\metre\per\second})}} \\
FNO   & \textbf{17.31} & {61.31} & {97.5} & \textbf{5.12} & {0.471} \\
RIGNO & 22.76 & 71.30 & 111.5 & 6.74 & 0.469 \\
GAOT  & {17.65} & \textbf{59.95} & \textbf{94.6} & {5.22} & \textbf{0.508} \\
\bottomrule
\end{tabular*}
\end{table}

\FloatBarrier
\subsubsection{Process parameters}\label{sec:params}

Having seen that neural operators can generalize successfully to free-form paths, we explore generalizations to new types of process parameter variations on a straight track, namely: varying beam radius, beam shape, and preheat temperature. We predict the quasi-steady state to make neural operator learning as easy as possible, in order to isolate the generalization capabilities of the models to new process variations and avoid confounds from the difficulties of transient laboratory-frame predictions.

\begin{table}[h!]
\centering
\caption{Metrics for FNO, RIGNO, and GAOT architectures across a dataset varying beam radius, laser power and velocity across LoF to keyhole regimes, and another varying those in addition to preheat temperature and beam shape (uniform or Gaussian).
Best metric result in bold.}
\label{tab:preheat}
\small
\sisetup{range-phrase=--,range-units=single,per-mode=symbol}%

\begin{tabular*}{\textwidth}{@{\extracolsep{\fill}}lccccc}
\toprule
& \tabhead{T MAE\\(\si{\kelvin})} & \tabhead{T RMSE\\(\si{\kelvin})} & \tabhead{BB MAE\\(\si{\kelvin})} & \tabhead{Rel. err.\\(\%)} & \tabhead{Melt-pool\\IoU} \\
\midrule
\multicolumn{6}{@{}l}{\emph{Beam-radius variation ($P$ \SIrange{129}{225}{\watt}, $V$ \SIrange{0.40}{1.10}{\metre\per\second}, $r_0$ \SIrange{47.5}{60}{\micro\metre})}} \\
FNO   & 8.5 & {30.9} & 32.6 & 2.05 & 0.923 \\
RIGNO & 9.4 & {42.4} & {30.9} & 2.26 & \textbf{0.932} \\
GAOT & \textbf{6.7} & \textbf{26.6} & \textbf{27.8} & \textbf{1.62} & 0.930 \\
\midrule
\multicolumn{6}{@{}l}{\emph{Preheat and beam shape ($T_0$ \SIrange{296}{761}{\kelvin}, $r_0$ \SIrange{35}{80}{\micro\metre}, $P$ \SIrange{55}{284}{\watt}, $V$ \SIrange{0.20}{1.19}{\metre\per\second})}} \\
FNO   & {21.0} & {47.9} & {38.2} & {3.18} & 0.823 \\
RIGNO & \textbf{11.4} & \textbf{37.0} & \textbf{26.9} & \textbf{1.72} & \textbf{0.867} \\
GAOT  & {41.9} & {63.4} & {44.0} & {6.35} & {0.837} \\
\bottomrule
\end{tabular*}
\end{table}

\begin{figure}[h!]
\centering
\begin{subfigure}[t]{0.32\textwidth}\includegraphics[width=\linewidth,trim=25 46 22 36,clip]{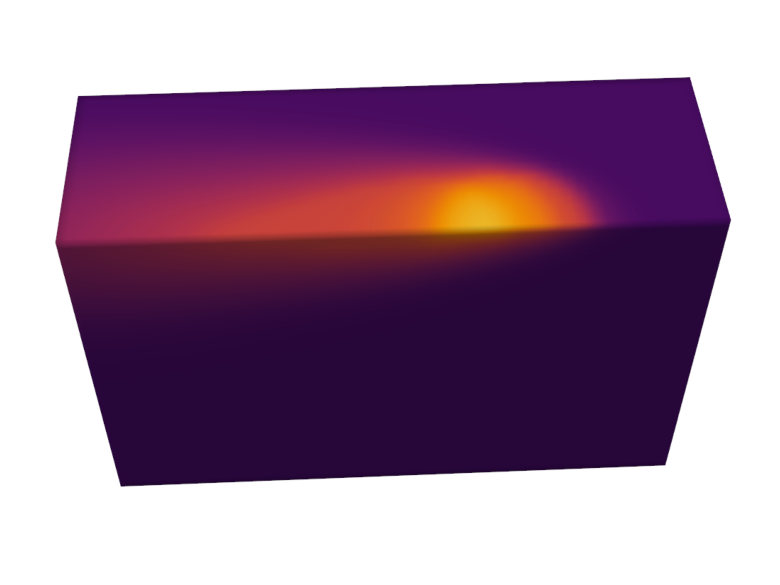}\subcaption{Ground truth, \SI{754}{\kelvin} preheat, \SI{116}{\watt}, \SI{1.15}{\metre\per\second}, \SI{72}{\micro\metre} beam.}\label{fig:ph3d-hot-gt}\end{subfigure}\hfill
\begin{subfigure}[t]{0.32\textwidth}\includegraphics[width=\linewidth,trim=25 46 22 36,clip]{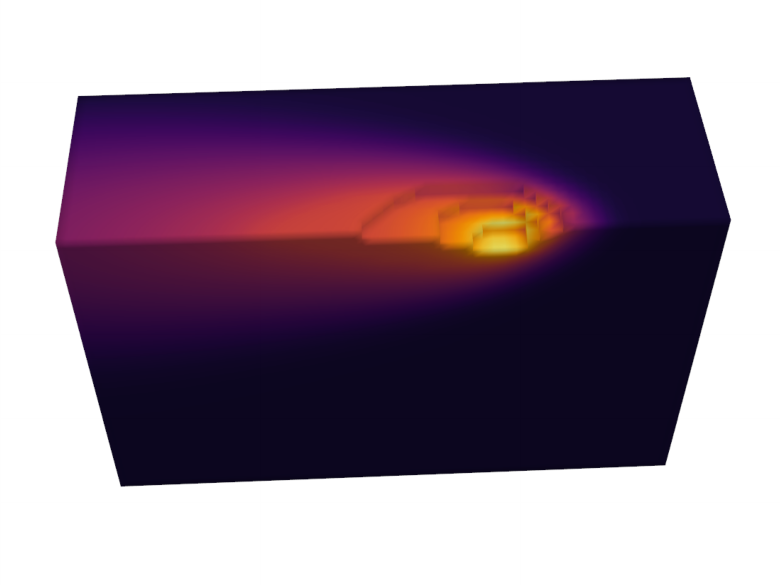}\subcaption{Ground truth, \SI{35}{\micro\metre} beam, \SI{108}{\watt}, \SI{0.65}{\metre\per\second}, \SI{337}{\kelvin} preheat.}\label{fig:ph3d-small-gt}\end{subfigure}\hfill
\begin{subfigure}[t]{0.32\textwidth}\includegraphics[width=\linewidth,trim=25 46 22 36,clip]{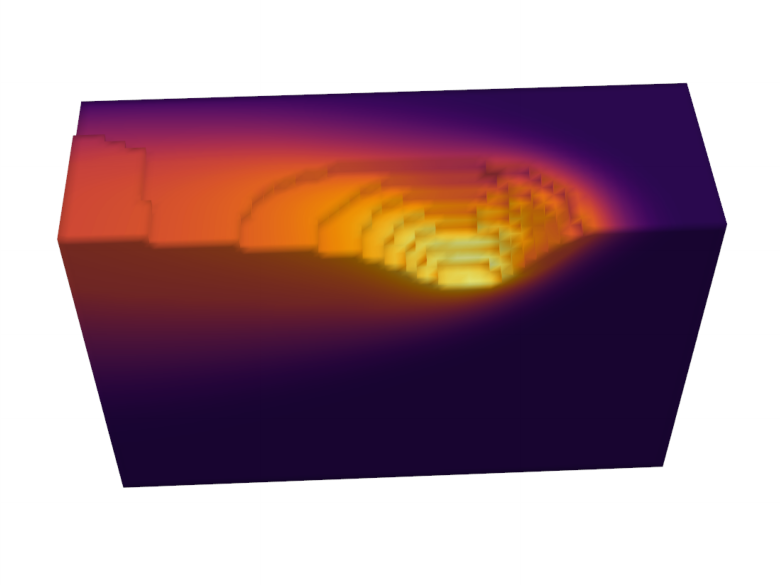}\subcaption{Ground truth, \SI{73}{\micro\metre} beam, \SI{231}{\watt}, \SI{0.60}{\metre\per\second}, \SI{514}{\kelvin} preheat.}\label{fig:ph3d-big-gt}\end{subfigure}

\vspace{0.3em}
\begin{subfigure}[t]{0.32\textwidth}\includegraphics[width=\linewidth,trim=25 46 22 36,clip]{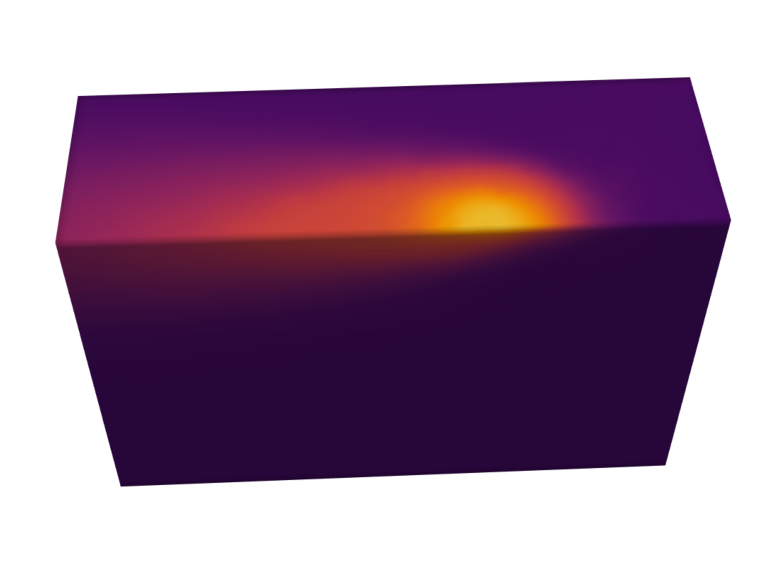}\subcaption{RIGNO prediction, \SI{754}{\kelvin} preheat, rel.~err.~\SI{1.2}{\percent}, BB MAE \SI{16}{\kelvin}, IoU \num{0.856}.}\label{fig:ph3d-hot-pred}\end{subfigure}\hfill
\begin{subfigure}[t]{0.32\textwidth}\includegraphics[width=\linewidth,trim=25 46 22 36,clip]{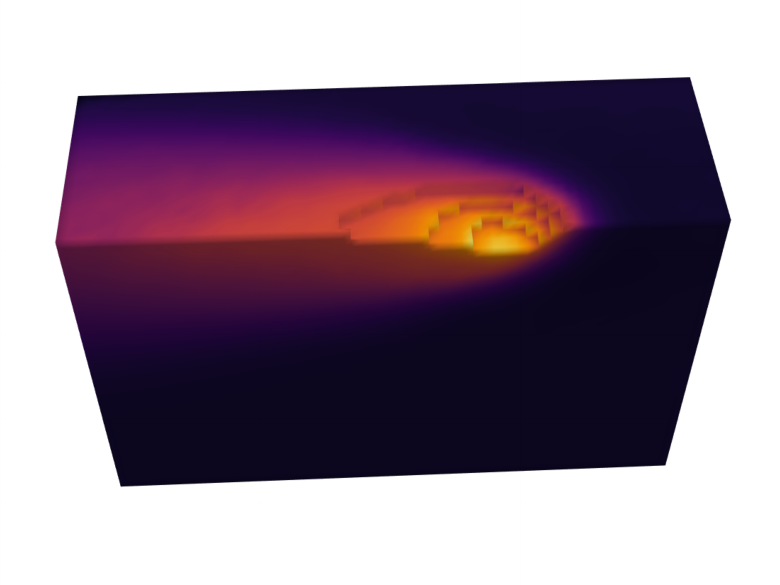}\subcaption{RIGNO prediction, \SI{35}{\micro\metre} beam, rel.~err.~\SI{2.3}{\percent}, BB MAE \SI{102}{\kelvin}, IoU \num{0.811}.}\label{fig:ph3d-small-pred}\end{subfigure}\hfill
\begin{subfigure}[t]{0.32\textwidth}\includegraphics[width=\linewidth,trim=25 46 22 36,clip]{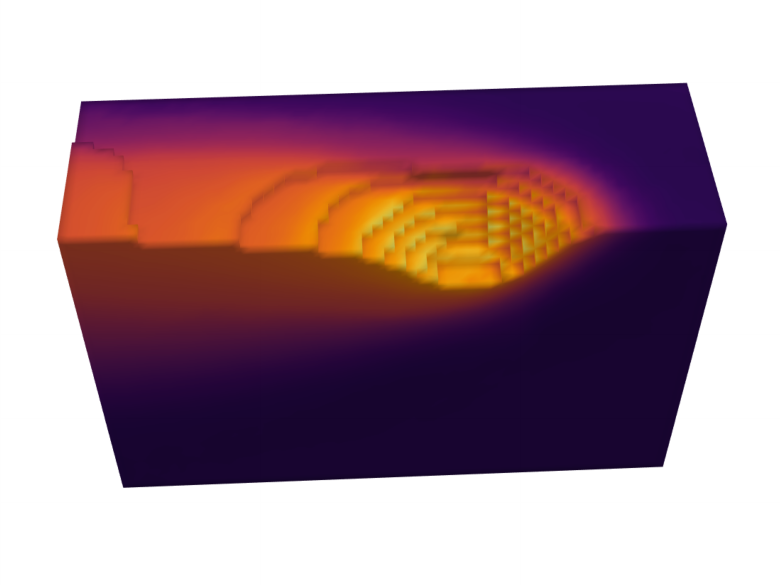}\subcaption{RIGNO prediction, \SI{73}{\micro\metre} beam, rel.~err.~\SI{1.7}{\percent}, BB MAE \SI{35}{\kelvin}, IoU \num{0.942}.}\label{fig:ph3d-big-pred}\end{subfigure}

\vspace{0.3em}
\includegraphics[width=0.5\textwidth]{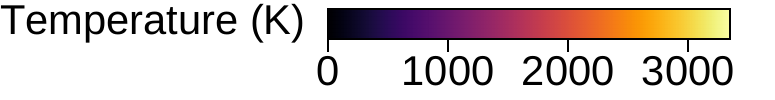}\hspace{1em}\includegraphics[width=0.2\textwidth]{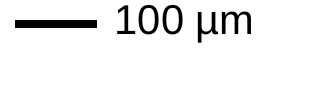}
\caption{RIGNO on three test set simulations from the preheat variation dataset. Each panel is the metal region of the time-averaged
quasi-steady state, cut at the beam plane and colored by temperature, with the beam
traveling rightward in the exposed cut face. The top row is ground truth and the bottom row
the prediction. The scale bar gives lengths along the scan direction in the cut face.}
\label{fig:preheat-pred}
\end{figure}

{\Cref{tab:preheat} gives} metrics across architectures on a dataset varying beam radius alongside the usual beam power and scan speed variations from LoF to keyhole melting regimes, and on another additionally varying beam shape between Gaussian and uniform, and preheat temperature. Accuracy is fairly good across both datasets on all architectures, with GAOT performing best on the beam radius dataset at \SI{1.62}{\percent} relative error and RIGNO performing best on the preheat set with \SI{1.72}{\percent} relative temperature error. We note that MAE gets comparatively inflated in the preheat set due to the higher bulk temperature, and that BB MAE is similar across both sets, although melt-pool IoU is slightly worse in the preheat set than in the beam radius set. \Cref{fig:preheat-pred} shows sample test set predictions of RIGNO on the preheat set across an LoF case with \SI{754}{\kelvin} preheat temperature, a case with small \SI{35}{\micro \metre} beam radius in conduction, and one with a larger \SI{73}{\micro \metre} beam radius at the keyhole transition regime. RIGNO predictions are slightly too cold in the keyhole surface for the large beam radius case, but otherwise there is good agreement, in accordance with the good metrics. The BB MAE of the small-beam case (\cref{fig:ph3d-small-pred}) looks high in part because its small beam radius narrows the beam box (which extends two beam radii around the beam) further to the hottest and most volatile section of the simulation, and BB MAE is not normalized by temperature.

\FloatBarrier
\subsection{Extension to a new alloy through autonomous data acquisition}\label{sec:al-results}

\begin{figure}[h!]
\centering
\includegraphics[width=\textwidth]{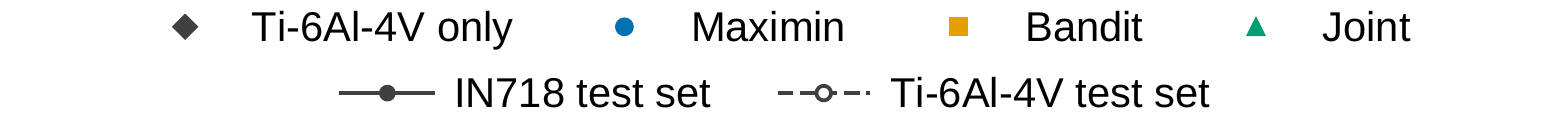}

\vspace{0.3em}
\begin{subfigure}[t]{0.49\textwidth}\includegraphics[width=\linewidth]{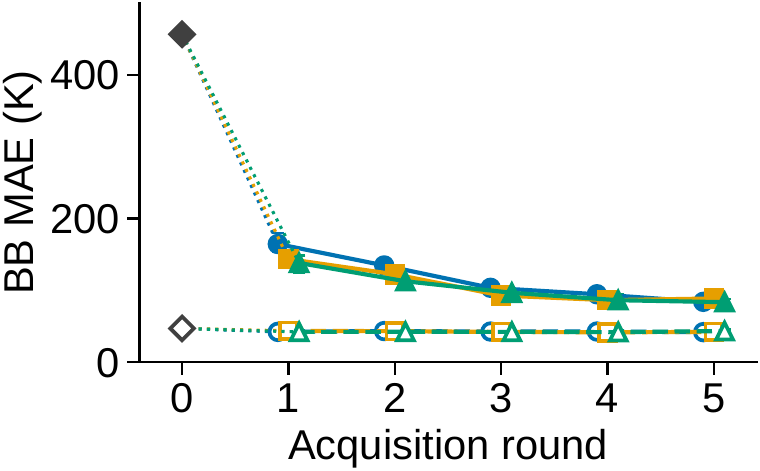}\subcaption{Beam-box MAE.}\label{fig:al-rounds-bbmae}\end{subfigure}\hfill
\begin{subfigure}[t]{0.49\textwidth}\includegraphics[width=\linewidth]{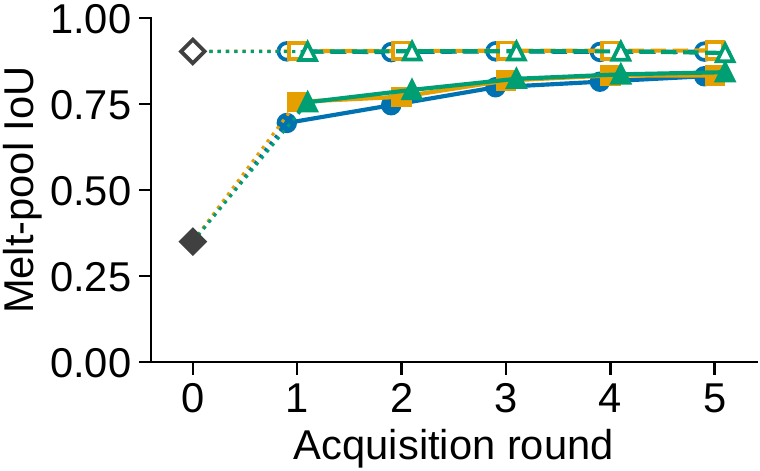}\subcaption{Melt-pool IoU.}\label{fig:al-rounds-iou}\end{subfigure}
\caption{FNO models after each acquisition round on both test sets of
\cref{tab:twoalloy}: the joint model and one model per half of the combined
selection, retrained from scratch every round with three seeds. Markers give the mean over the
seeds. The seed variation is at most about the size of the plotted markers. Round 0 is the
Ti-6Al-4V-only model, a single run.}
\label{fig:al-rounds}
\end{figure}

We now look at the results of the autonomous acquisition experiment described in \cref{sec:acquisition} to generalize the Ti-6Al-4V surrogates to IN718. \Cref{fig:al-rounds} indicates how much data is needed and how the model metrics improve per round. On IN718, the three FNO series (maximin half, bandit half and joint) improve over the rounds.
The bandit half leads the maximin half in the first four rounds, which we expected,
since the bandit sends its picks to the regimes where its own models' beam-box error is high
(\cref{sec:acquisition}). The bandit half stops improving after round 4, and by round 5 the two
halves are at the same level. The joint model has twice the IN718 simulations of either half, and in
every round its mean is either better than both halves or within the seed range of the more accurate
one. On the Ti-6Al-4V test set, the FNO models keep their accuracy as IN718 data are added. Over all rounds their mean beam-box MAE stays below the \SI{46.8}{\kelvin} of the Ti-6Al-4V-only FNO, and their mean melt-pool IoU within \num{0.004} of that model's \num{0.903} (\cref{fig:al-rounds}). 

\begin{figure}[h!]
\centering
\begin{subfigure}[t]{0.32\textwidth}\includegraphics[width=\linewidth,trim=39 19 36 34,clip]{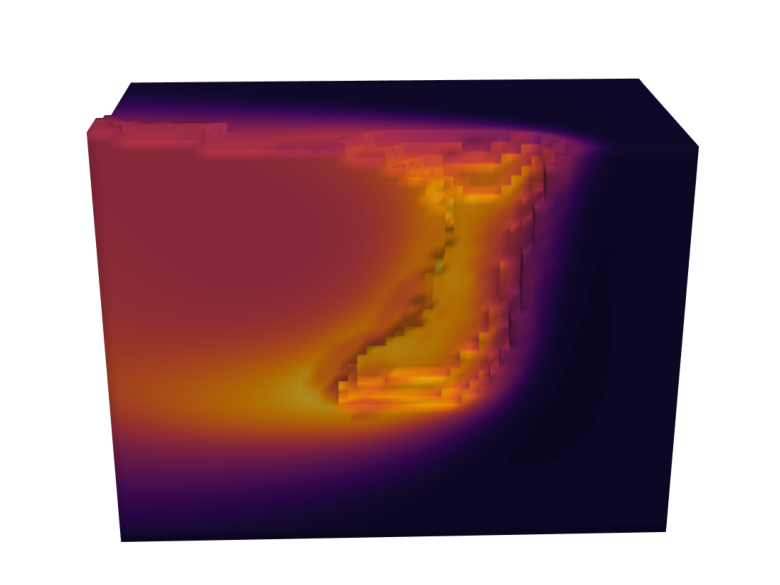}\subcaption{IN718, ground truth.}\label{fig:ta3d-in718-gt}\end{subfigure}\hfill
\begin{subfigure}[t]{0.32\textwidth}\includegraphics[width=\linewidth,trim=39 19 36 34,clip]{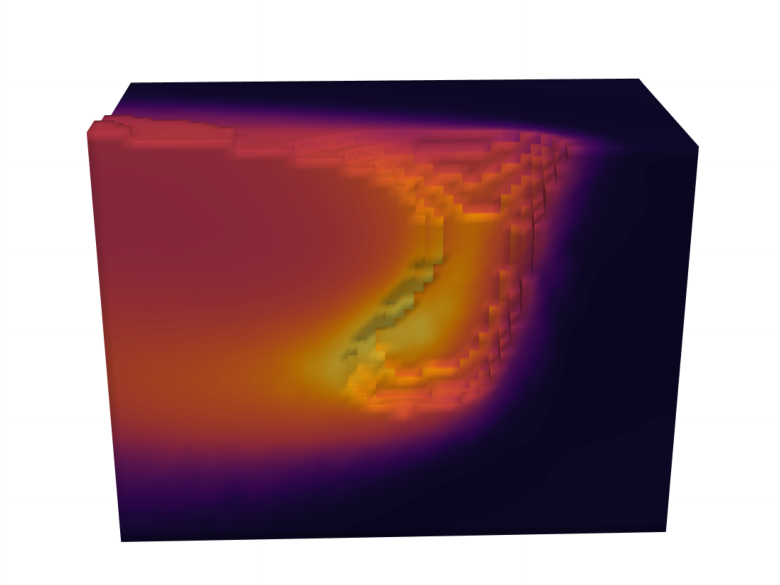}\subcaption{IN718, joint RIGNO model, rel.~err.~\SI{6.0}{\percent}, BB MAE \SI{95}{\kelvin}, IoU \num{0.714}.}\label{fig:ta3d-in718-pred}\end{subfigure}\hfill
\begin{minipage}[t]{0.32\textwidth}\includegraphics[width=0.4375\linewidth]{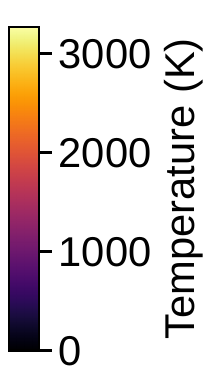}\hfill\includegraphics[width=0.5\linewidth]{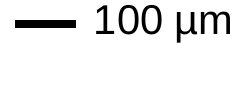}\end{minipage}

\vspace{0.4em}
\begin{subfigure}[t]{0.32\textwidth}\includegraphics[width=\linewidth,trim=39 19 36 34,clip]{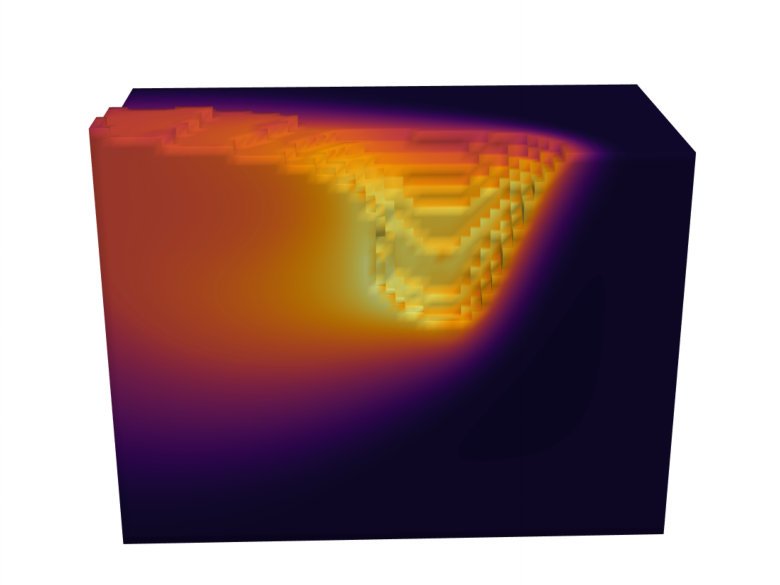}\subcaption{Ti-6Al-4V, ground truth.}\label{fig:ta3d-ti64-gt}\end{subfigure}\hfill
\begin{subfigure}[t]{0.32\textwidth}\includegraphics[width=\linewidth,trim=39 19 36 34,clip]{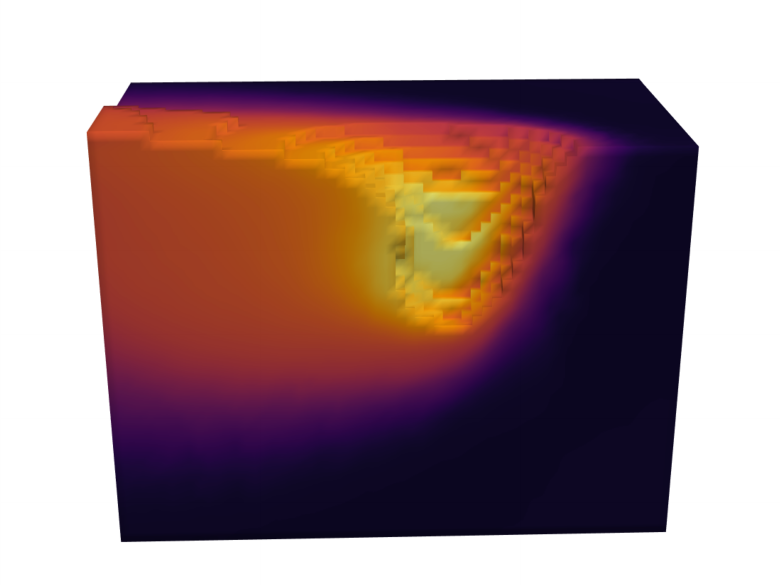}\subcaption{Ti-6Al-4V, joint RIGNO model, rel.~err.~\SI{5.6}{\percent}, BB MAE \SI{90}{\kelvin}, IoU \num{0.860}.}\label{fig:ta3d-ti64-pred}\end{subfigure}\hfill
\begin{subfigure}[t]{0.32\textwidth}\includegraphics[width=\linewidth,trim=39 19 36 34,clip]{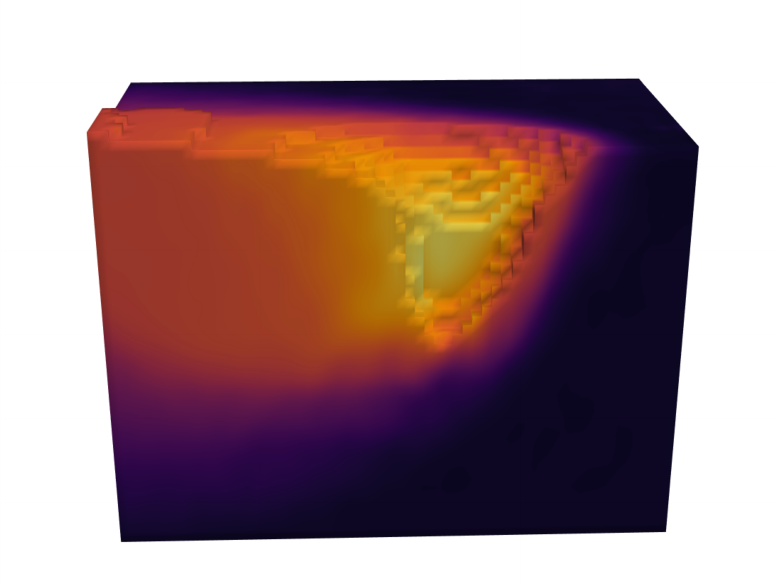}\subcaption{Ti-6Al-4V, Ti-6Al-4V-only FNO, rel.~err.~\SI{6.8}{\percent}, BB MAE \SI{111}{\kelvin}, IoU \num{0.833}.}\label{fig:ta3d-ti64-only}\end{subfigure}
\caption{The joint two-alloy RIGNO model on one held-out simulation of
each alloy. The IN718
simulation is the deepest keyhole of the IN718 test set, at \SI{347}{\watt},
\SI{0.48}{\metre\per\second} and a \SI{40}{\micro\metre} beam, and the
Ti-6Al-4V one is at \SI{225}{\watt}, \SI{0.40}{\metre\per\second} and a
\SI{60}{\micro\metre} beam, with a shallower keyhole. Panel \subref{fig:ta3d-ti64-only} shows the FNO trained on Ti-6Al-4V alone, the
reference model of \cref{tab:twoalloy}, on the same simulation. It shows slight FNO artifacts, faint ripples in the cool substrate below the melt pool.
}
\label{fig:twoalloy-pred}
\end{figure}

{\Cref{tab:twoalloy} gives} all metrics across architectures for the two alloys, for models trained on the maximin half, the bandit half and both halves together.
Every model in \cref{tab:twoalloy} whose training set includes IN718 has a relative temperature error of
about \SI{3}{\percent} or less on both alloys. The FNO trained on
Ti-6Al-4V alone performs badly as expected with \SI{30}{\percent} error on IN718. On IN718 the three architectures trained on the joint set are
close in relative error and melt-pool IoU and differ most in beam-box MAE, where RIGNO is lowest. On Ti-6Al-4V the joint FNO has
the lowest temperature and beam-box errors of the three joint models. \Cref{fig:twoalloy-pred} shows predictions of the RIGNO model
trained on both alloys for one held-out simulation of each alloy. For the Ti-6Al-4V simulation, the joint RIGNO is slightly closer to the ground truth than the
Ti-6Al-4V-only FNO (\cref{fig:ta3d-ti64-only}). The IN718 prediction is slightly less accurate than the joint model's Ti-6Al-4V one but gets the general melt-pool and field shapes right. The keyholes in the IN718 test set are also deeper on average (median vapor-depression depth \SI{108}{\micro\metre} against \SI{80}{\micro\metre} on the Ti-6Al-4V test set), which may partly explain why model metrics are worse on IN718 than those on Ti-6Al-4V.

\begin{table}[h!]
\centering
\caption{Quasi-steady models extended from Ti-6Al-4V to IN718 by the acquisition loop of
\cref{sec:acquisition}, scored on the IN718 test set and on the Ti-6Al-4V test split of the
beam-radius sweep. 
The joint models train on the simulations chosen by the maximin and bandit algorithms combined and on the original Ti-6Al-4V simulations, and the
Ti-6Al-4V-only FNO and the FNOs trained on both the original Ti-6Al-4V simulations and simulations from one algorithm alone are given for reference. FNO rows with
IN718 data are means over three training runs, and the other rows are single runs.}
\label{tab:twoalloy}
\small
\sisetup{range-phrase=--,range-units=single,per-mode=symbol}%
\begin{tabular*}{\textwidth}{@{\extracolsep{\fill}}lccccc}
\toprule
& \tabhead{T MAE\\(\si{\kelvin})} & \tabhead{T RMSE\\(\si{\kelvin})} & \tabhead{BB MAE\\(\si{\kelvin})} & \tabhead{Rel. err.\\(\%)} & \tabhead{Melt-pool\\IoU} \\
\midrule
\multicolumn{6}{@{}l}{\emph{IN718 only test set ($P$ \SIrange{90}{350}{\watt}, $V$ \SIrange{0.15}{1.25}{\metre\per\second}, $r_0$ \SIrange{23}{64}{\micro\metre})}} \\
FNO, Ti-6Al-4V only & {124.38} & {284.1} & {456} & {30.02} & {0.350} \\
FNO, maximin half & {13.64} & {41.2} & {84} & {3.31} & {0.831} \\
FNO, bandit half  & {14.00} & {44.2} & {88} & {3.40} & {0.833} \\
FNO, joint        & \textbf{12.79} & {40.8} & {83} & \textbf{3.10} & \textbf{0.843} \\
RIGNO, joint      & {13.55} & \textbf{40.8} & \textbf{78} & {3.29} & {0.835} \\
GAOT, joint & {12.86} & {42.2} & {92} & {3.12} & {0.837} \\
\midrule
\multicolumn{6}{@{}l}{\emph{Ti-6Al-4V only test set ($P$ \SIrange{100}{225}{\watt}, $V$ \SIrange{0.19}{1.10}{\metre\per\second}, $r_0$ \SIrange{42.5}{65}{\micro\metre})}} \\
FNO, Ti-6Al-4V only & {10.97} & {38.7} & {47} & {2.41} & {0.903} \\
FNO, maximin half & \textbf{9.89} & \textbf{35.0} & \textbf{42} & \textbf{2.17} & {0.903} \\
FNO, bandit half  & {10.08} & {35.3} & {42} & {2.21} & \textbf{0.906} \\
FNO, joint        & {10.47} & {35.9} & {43} & {2.30} & {0.899} \\
RIGNO, joint      & {14.40} & {38.0} & {51} & {3.16} & {0.902} \\
GAOT, joint & {11.90} & {39.8} & {55} & {2.62} & {0.887} \\
\bottomrule
\end{tabular*}
\end{table}

\FloatBarrier
\subsection{Surrogate against experiment, end to end}\label{sec:closing}

\begin{figure}[htbp]
\centering
\begin{subfigure}[t]{0.49\textwidth}\includegraphics[width=\linewidth]{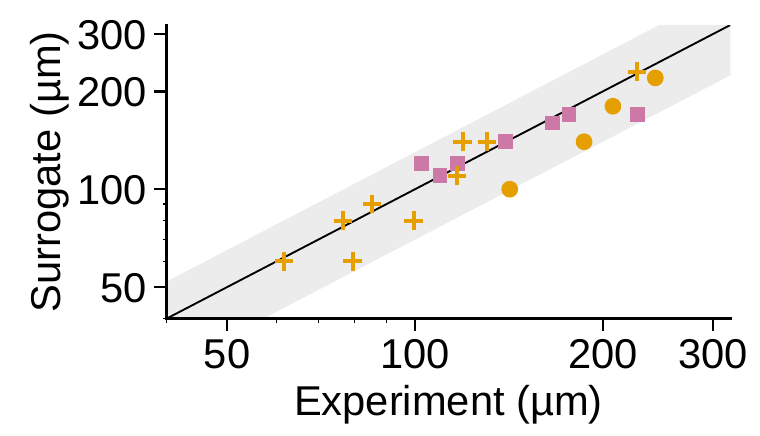}\subcaption{Surrogate vs. experiment.}\label{fig:closing-a}\end{subfigure}\hfill
\begin{subfigure}[t]{0.49\textwidth}\includegraphics[width=\linewidth]{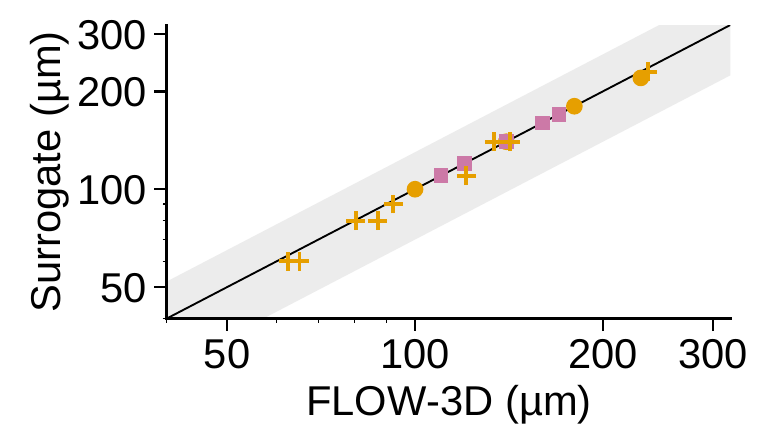}\subcaption{Surrogate vs. FLOW-3D\textsuperscript{\textregistered}.}\label{fig:closing-b}\end{subfigure}

\vspace{0.5em}
\begin{minipage}[t]{0.49\textwidth}\includegraphics[width=\linewidth]{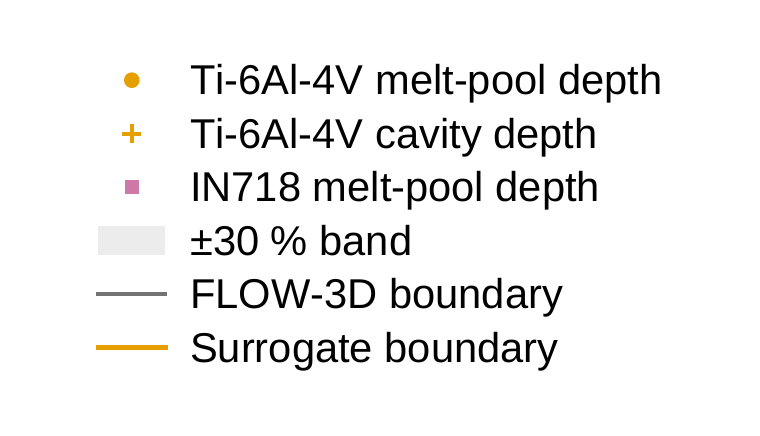}\end{minipage}\hfill
\begin{subfigure}[t]{0.49\textwidth}\includegraphics[width=\linewidth]{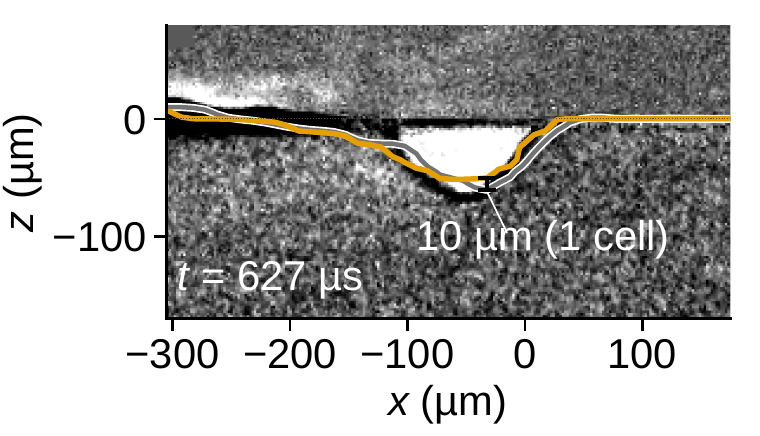}\subcaption{Experimental radiograph.}\label{fig:closing-xray}\end{subfigure}
\caption{Surrogate, FLOW-3D\textsuperscript{\textregistered} and experiment at published Ti-6Al-4V and
IN718 conditions, where experiment is the published measurement,
FLOW-3D\textsuperscript{\textregistered} the solver result at the same condition and \emph{surrogate} the
neural operator's prediction (in panels (a) and (b) one RIGNO model for both alloys). Panels (a) and (b) compare every scored depth on thirteen Ti-6Al-4V
and seven IN718 conditions, with the band
spanning $\pm\SI{30}{\percent}$. Melt-pool depths were measured on cross-sections after solidification for
Ti-6Al-4V~\cite{galiegue_acoustic_2025} and IN718~\cite{weaver_am_2022}, and Ti-6Al-4V vapor-depression depths by
in-situ X-ray imaging during the scan~\cite{gan_universal_2021,cunningham_keyhole_2019}. Surrogate depths are read from the predicted
quasi-steady frame. \Cref{tab:closing-conditions} lists every condition. (c) The laboratory-frame FNO's and FLOW-3D\textsuperscript{\textregistered}'s gas boundaries on
one radiograph of the scanned track from~\cite{simonds_ability_2024,simonds_asynchronous_2022}.}
\label{fig:closing}
\end{figure}

Finally we look at surrogate performance against published experimental data as done for the FLOW-3D\textsuperscript{\textregistered} solver in \cref{sec:expval_main}. We restrict the comparison to the published conditions of \cref{sec:expval_main} reproduced by test simulations of our datasets, thirteen Ti-6Al-4V conditions of the beam-radius sweep and the seven IN718 conditions of \cref{sec:calibration}. We leave out a fourteenth Ti-6Al-4V condition, at \SI{100}{\watt}, whose tabulated depth lies well above the trend of its own source (\cref{sec:calibration}), and which neither FLOW-3D nor the surrogate reproduces. At all of these conditions the reference FLOW-3D\textsuperscript{\textregistered} solver is within \SI{30}{\percent} of the measurement. The surrogate is trained to reproduce the solver, so its error against experiment includes the solver's own error, which is currently a fundamental limitation of surrogates of this type. Of the three architectures trained on both alloys with the alloy label
(\cref{sec:acquisition}), RIGNO predicts all 20 measured depths to within \SI{30}{\percent}, and the
FNO and GAOT 19 each.
Results
are summarized in \cref{fig:closing}.

\Cref{fig:closing-a} shows that, for melt-pool and cavity depths on both Ti-6Al-4V and IN718, the surrogate is accurate within \SI{30}{\percent}. As \cref{fig:closing-b} shows, surrogate predictions stay close to the reference FLOW-3D simulations, with every depth within about one \SI{10}{\micro\metre} cell of them and an average gap of
about a quarter of a cell (\SI{2}{\percent} of the depth), a cell being about \SI{10}{\percent} of a typical depth here. A radiograph records a single instant, so for \cref{fig:closing-xray} we use an FNO trained on the same beam-radius sweep in the laboratory frame, which predicts the field at a given time. \Cref{fig:closing-xray} compares its vapor-cavity prediction with the radiograph of \cref{fig:calib-xray-settled}. Measured from its own free surface, its cavity bottom lies one cell above that of FLOW-3D\textsuperscript{\textregistered}, whose depth matches the one we segment from the radiograph.

The comparison with published measurements can also serve as a final benchmark for a surrogate retrained over rounds of autonomous acquisition (\cref{sec:acquisition}), since every other test of the surrogate in this work takes the solver as the reference. Incorporating experimentally measured depths in training, alongside the simulations, with some conditions held out for this benchmark, could perhaps bring the surrogate's depth error against experiment below that of the solver (\cref{sec:conclusion}).

\FloatBarrier
\section{Conclusion}\label{sec:conclusion}

The first decision when building a laser-melting surrogate is whether to pose the
problem in the quasi-steady formulation or in the laboratory frame. For straight tracks at constant speed, the quasi-steady formulation gives a lower beam-box MAE than the
laboratory frame for every architecture (\cref{tab:arch-steady}), and laboratory-frame predictions
averaged over time remain less accurate in BB MAE than native quasi-steady ones (\cref{tab:frame-difficulty}). For
RIGNO, the averaged laboratory-frame predictions come within the seed spread of the native quasi-steady model
(\cref{tab:arch-steady}) in T MAE and relative error and have the lower T RMSE and the higher melt-pool IoU, so on this
dataset the quasi-steady formulation does not seem to improve RIGNO model error. It remains cheaper to train, since it predicts a single time-averaged field per simulation. We therefore recommend it
wherever the process allows. Free-form paths, speed changes and transients require the
laboratory frame.

FNO is the cheapest architecture to train and evaluate (\cref{tab:cost}) and leads four of the five metrics in the quasi-steady formulation (\cref{tab:arch-steady}). We recommend it for fixed geometries on regular meshes. For free-form paths we recommend either FNO or GAOT, which cost less than RIGNO and are on average more accurate (\cref{tab:paths-accuracy}). We recommend GAOT where the domain geometry varies or uses non-regular meshes, since FNO normally requires fixed geometries and regular meshes to perform DFTs. RIGNO, the slowest of the three, is the most accurate on the straight-path set in the laboratory frame and on the preheat set, but trails GAOT on every metric of the three free-form path sets.

FNO and GAOT predictions occasionally show
small spurious oscillations in the bulk substrate, seen for example in the GAOT prediction of \cref{fig:rp-gaot} and the FNO predictions of \cref{fig:vspd-pr-dwell} and \cref{fig:ta3d-ti64-only}. RIGNO predictions are
usually smoother but sometimes under-predict the peak temperature of the melt pool, as FNO also does on the multi-beam simulations of \cref{fig:multibeam-pred} for example.

Overall, the choice of neural operator architecture does not seem crucial to building a successful laser processing neural operator. While some architectures performed better on some datasets, all three architectures considered here performed quite well on average at the different generalization tasks, and we believe that any existing architecture gap can likely be closed by further tuning and feature engineering. We also note that new and more powerful neural operator architectures come out almost every year, and thus believe that the generalization gaps seen in the datasets here will quickly close. We believe that given a few years, we could see laser processing surrogates performing with the same breadth and accuracy as current traditional solvers for a fraction of the cost. We are already partly there. 

We list some of the limits of these models below, each with the work that would address it.

\textbf{One large fully generalized model.} Apart from the two-alloy models, we trained a separate model
for each dataset, so no single model yet spans varied scan paths, process parameters and varying alloys
together. A single surrogate covering all of these would likely need larger training sets that vary
the axes together, and possibly larger models.

\textbf{Unseen alloys.} Our two-alloy models receive material properties through $H^*$, the
conduction response and a binary alloy label, which cannot describe an alloy outside
the two seen in training.
Replacing the label with further material properties and testing on an alloy held out of training would be a first step toward generalization to unseen materials. Material generalization is the biggest open gap toward full process generalization.

\textbf{Amount of training data.} We have not measured how accuracy grows with the number of
training simulations beyond the five IN718 rounds of \cref{fig:al-rounds}, where the joint FNO's
gains in IN718 beam-box MAE and melt-pool IoU shrank each round. Training each model on growing subsets of
its dataset would show where more simulations stop improving it, and how many solver runs a new
alloy or process axis needs to generalize. 

\textbf{Autonomous acquisition of new data.} Our acquisition experiment is a proof of concept, with one new
alloy, one selection scheme combining maximin and a bandit, and five rounds of \num{50} simulations over power, scan speed and beam
radius. A bigger experiment with other selection algorithms, batch sizes and budgets across larger generalizations remains to be tested.

\textbf{Stochastic keyhole fluctuations.} Our surrogates
cannot represent chaotic keyhole fluctuations (\cref{sec:transient}) and predict an ``averaged'' behavior. Engineering model loss, data and metrics to predict pulsations or realistic fluctuations and stochastic processes would be a natural future direction.

\textbf{Part geometry.} Each of our datasets uses a single rectangular substrate for all its
simulations. Because neither RIGNO nor GAOT assumes a regular grid (\cref{sec:models}), both could be
trained on substrates of other shapes and on domains that vary in size. This matters in particular for applications such as directed energy deposition (DED) and LPBF part{-}scale modeling{.}

\textbf{Solver physics and measured data.} The surrogates inherit the solver's error against
the measurements (\cref{sec:closing}), so improvements to the solver physics would carry over to surrogates
retrained on the new simulations. Measured experimental data such as melt-pool and vapor-depression depths could also enter
training directly, for example in a process control loop.

In summary, we built three-dimensional FNO, RIGNO and GAOT surrogates of laser melting, trained on
FLOW-3D WELD\textsuperscript{\textregistered} simulations whose settings we validated against published
measurements for four materials. With one model per dataset, they predict the temperature,
melt-pool and free-surface fields for varied process parameters, free-form scan paths and up to four
beams. On one GPU each network predicts a frame of the straight-path P-V dataset in a few to a few hundred milliseconds (\cref{tab:cost}).

Throughout our different experiments, we observe high accuracy of the models in straight-path quasi-steady prediction varying power, scan speed, beam radius, beam shape, and preheat temperature (below \SI{2}{\percent} relative error for the best models in \cref{tab:arch-steady,tab:preheat}). For models predicting transient free-form path fields, we observe slightly worse metrics as expected, but models still perform well. Paths themselves are predicted accurately, and varying the speed during the path does not seem to make inference harder for the best model. Indeed the best model on the free-form paths dataset with constant scan speed has an average relative error of around \SI{3.7}{\percent}, while on the dataset varying speed within the path the best model has an error of \SI{2.2}{\percent}. The discrepancy likely comes from total scan time and other path differences between the two datasets. Increasing the number of beams to predict modestly increased average relative error of the best model to \SI{5.1}{\percent}.

We have also shown that a surrogate can be extended to a second alloy with training simulations
chosen autonomously. Trained on the Ti-6Al-4V set together with these IN718 simulations, all three
architectures reach a relative temperature error of about \SI{3}{\percent} on IN718
(\cref{tab:twoalloy}), and the two-alloy RIGNO predicts all \num{20} measured Ti-6Al-4V and
IN718 depths to within \SI{30}{\percent} (\cref{sec:closing}).

We
release the models, evaluations and simulation configurations with the AiiDA provenance record that
traces each reported metric back to the solver runs and settings behind it.

\section*{Statements and Declarations}
\paragraph{Data availability.} The \num{45} models of
\cref{tab:arch-steady,tab:paths-accuracy,tab:preheat,tab:twoalloy}, the models they
were fine-tuned from, the models trained after each acquisition round
(\cref{fig:al-rounds}) and their evaluations are deposited on Zenodo under the
CC BY-NC-SA 4.0 license, with the training, validation and test data each model used,
in one open and one restricted-access record. The open record (DOI
\href{https://doi.org/10.5281/zenodo.23325086}{10.5281/zenodo.23325086}, about \SI{15.7}{\giga\byte}) holds the AiiDA archive
carrying the provenance graph from solver configuration to reported metric, the
model checkpoints, the evaluations and the configuration of every simulation. The
FLOW-3D WELD\textsuperscript{\textregistered} output, the simulated fields of every
simulation used to train and score the models, is shared on request for
non-commercial use, as agreed with the solver's vendor. It is in a
restricted-access record (DOI \href{https://doi.org/10.5281/zenodo.23325088}{10.5281/zenodo.23325088}, about
\SI{158}{\giga\byte}), in which the test-split simulations needed to re-run
evaluations take about \SI{23}{\giga\byte}. 

\paragraph{Code availability.} The scripts that set up and run the
FLOW-3D\textsuperscript{\textregistered} simulations (\texttt{flow3D\_scripts}), the
training and evaluation code (\texttt{lpno}) and the AiiDA plugin that
records provenance (\texttt{aiida-pino}) are released in one repository under the PolyForm Noncommercial License 1.0.0 at
\url{https://github.com/azbenoit/lpno_code.git}.

\paragraph{Competing interests.} The authors declare that they have no competing
interests.

\paragraph{Use of LLMs in this work.}
\label{sec:ai}
Large language models (LLMs), specifically Claude and the agentic coding interface
Claude Code, were used throughout this work under the authors' supervision to
code and debug parts of the simulation-generation harness, the provenance layer,
the training and evaluation pipeline and the analysis notebooks, and to organize
and rephrase ideas supplied by the authors. The scientific questions, the
experimental design, the interpretation of results and the final text are the
authors' own.

Because such assistance can introduce errors that are fluent enough to pass
casual reading, we adopted a set of practices intended to make them more visible.
Code was developed test-first, with a regression test accompanying each bug fix,
and results were accepted only after the underlying simulation or training run
had been checked against its own logs and stored outputs. Every simulation and model result
in this manuscript is traceable to a recorded run: the provenance
database of \cref{sec:pipeline} links each number to the configuration, solver
version and material data that produced it. Draft text and code were additionally passed
through repeated adversarial review, in which separate model instances were asked to
identify unsupported claims, internal inconsistencies and numbers that
disagreed with their sources. This process helped catch both human errors and LLM hallucinations when rephrasing.

The code repositories behind this work are organized around a small set of plain-text
files that persist across sessions and act as a shared memory between the
authors and the assistant. A \texttt{CHANGELOG.md} in each repository records
what was changed and why, including failed approaches, the command used to test
a change, and the outcome of individual simulation campaigns, so that earlier
work is neither silently repeated nor contradicted. A \texttt{potential\_bugs.md}
records behavior that is suspected to be incorrect or that could mislead a user
outside the conditions the code was written for, each entry naming the
assumption it violates. Entries are removed only once a fix has landed with a
regression test. A \texttt{future\_work.md} holds proposed improvements,
separated into those a human has approved and those awaiting review, which keeps
speculative suggestions from being implemented unprompted. A \texttt{CLAUDE.md}
carries the operating rules for the repository, including environment
constraints, validation requirements and conventions that would otherwise have
to be restated each session.

This way of working is inspired by the approach described for long-running agentic
scientific computing~\cite{mishra-sharma_long-running_2026} and demonstrated in a
differentiable cosmological Boltzmann solver~\cite{mishra-sharma_clax_2026}. An
independent single-case study of a physicist supervising the same
tooling~\cite{nguyen_physics_2026} names shared changelogs among the three
practices that caught errors its automated tests did not. We found the same here,
and the same limitation: an assistant will sometimes treat the disappearance of a
symptom as the resolution of its cause, and these records exist to catch
that failure mode.

\bibliography{references}

\clearpage
\makeatletter\immediate\write\@auxout{\string\gdef\string\lpnosuppstart{\the\value{page}}}\makeatother
\setcounter{page}{1}\renewcommand{\thepage}{S\arabic{page}}
\makeatletter%
\let\lpno@cite\cite
\protected\def\cite{\let\@citex\mb@@citex\expandafter\let\expandafter\@newciteauxhandle\csname @auxoutS\endcsname\lpno@cite}
\gdef\@extra@b@citeb{@S}%
\makeatother
\begin{center}
{\LARGE\bfseries {Supplementary Information}}\par\medskip
\makeatletter{\large \@title}\makeatother\par\bigskip
\end{center}

\renewcommand{\appendixpagename}{Supplementary Materials}%
\renewcommand{\appendixtocname}{Supplementary Materials}%
\renewcommand{\appendixname}{Supplementary Note}%
\begin{appendices}
\crefalias{section}{appendix}%
\counterwithin{figure}{section}%
\counterwithin{table}{section}
\renewcommand{\thefigure}{\thesection\arabic{figure}}%
\renewcommand{\thetable}{\thesection\arabic{table}}

\section{Datasets}\label{sec:datasets-detail}

\begin{figure}[htbp]
\centering
\includegraphics[width=\linewidth]{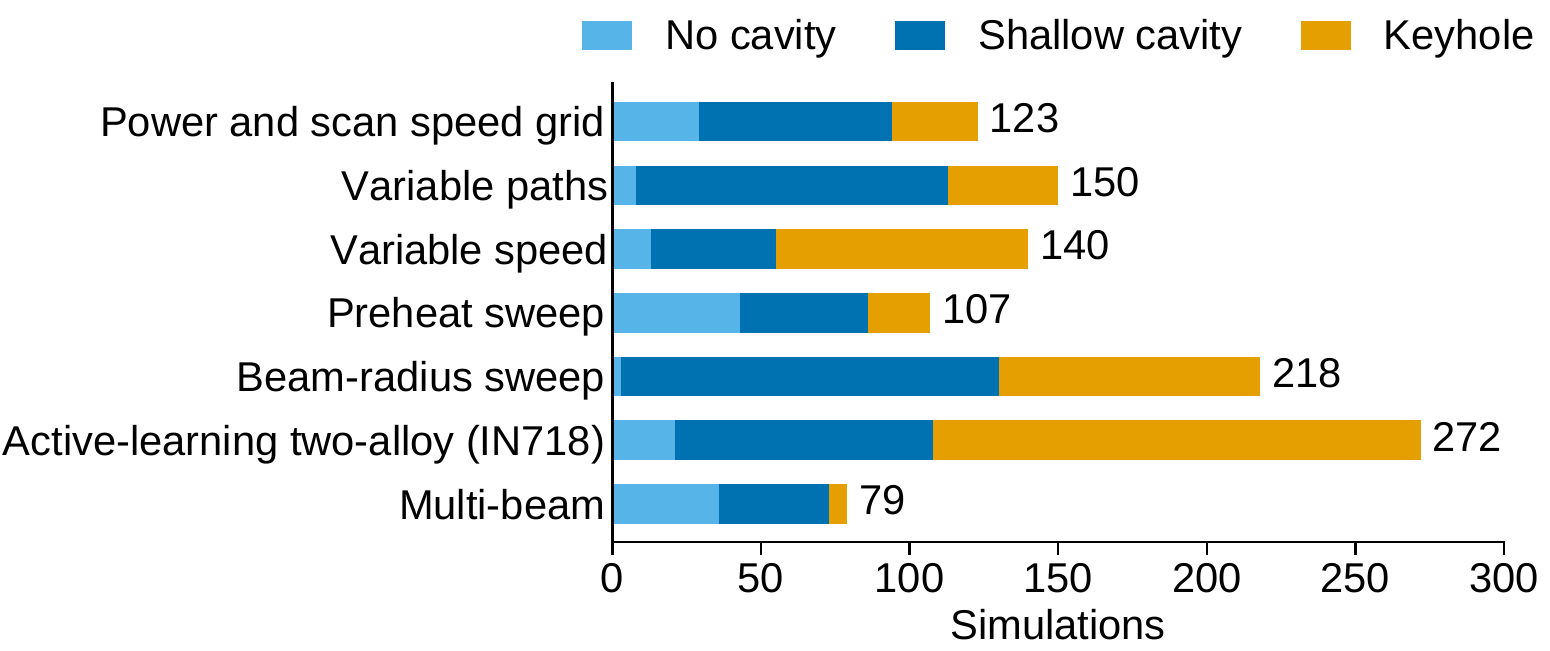}
\caption{The number of simulations in each dataset of \cref{tab:datasets} that the models are trained, validated and tested on, grouped by the kind of vapor cavity each develops. The \emph{no-cavity} class holds simulations whose cavity, if any, stays less than two cells deep at its largest. The \emph{keyhole} class holds simulations in which the aspect ratio, averaged over time, exceeds one and the cavity reaches a depth of at least \num{2.1} beam radii. Every other simulation is classified as \emph{shallow cavity}. The two-alloy bar counts only the IN718 simulations of the acquisition study.}
\label{fig:dataset-regimes}
\end{figure}

\Cref{fig:dataset-regimes} counts the simulations of each dataset of \cref{tab:datasets} by the kind of vapor cavity each develops. \Cref{tab:closing-conditions}
lists the published conditions of the comparison with experiment in \cref{sec:closing}.

\begin{table}[h!]
\centering
\caption{The published conditions of \cref{fig:closing}. Each row gives the depth its source measured, the
FLOW-3D WELD\textsuperscript{\textregistered} depth of the test simulation at that condition, obtained as in \cref{sec:calibration}, and the depth
read from the quasi-steady frame predicted by the joint RIGNO model of \cref{tab:twoalloy}. Simulated and predicted
depths are counted in whole \SI{10}{\micro\metre} cells, except the FLOW-3D\textsuperscript{\textregistered}
vapor-depression depths, which are averages over frames. On the predicted frame, the melt-pool depth is the depth of the deepest cell hotter than the midpoint of $T_s$ and $T_l$, and the vapor-depression depth that of the deepest cell with $\alpha<0.5$ below the original surface. The IN718 beam radius is
half the spot diameter of the AM-Bench record.}
\label{tab:closing-conditions}
\small
\begin{tabular*}{\textwidth}{@{\extracolsep{\fill}}cccccc}
\toprule
{\tabhead{$P$\\(\si{\watt})}} & \tabhead{$V$\\(\si{\metre\per\second})} & \tabhead{$r_0$\\(\si{\micro\metre})} & \tabhead{Measured\\(\si{\micro\metre})} & \tabhead{FLOW-3D\textsuperscript{\textregistered}\\(\si{\micro\metre})} & \tabhead{RIGNO\\(\si{\micro\metre})} \\
\midrule
\multicolumn{6}{@{}l}{\emph{Ti-6Al-4V, melt-pool depth on cross-sections after solidification~\cite{galiegue_acoustic_2025}}} \\
150 & 0.40 & 60 & 141.8 & 100 & 100 \\
175 & 0.40 & 60 & 186.6 & 140 & 140 \\
200 & 0.40 & 60 & 207.5 & 180 & 180 \\
225 & 0.40 & 60 & 242.6 & 230 & 220 \\
\midrule
\multicolumn{6}{@{}l}{\emph{Ti-6Al-4V, vapor-depression depth by in-situ X-ray imaging during the scan~\cite{gan_universal_2021}}} \\
129.4 & 0.40 & 47.5 & 85.3 & 92.3 & 90 \\
155.5 & 0.40 & 47.5 & 130.4 & 133.9 & 140 \\
206.4 & 0.40 & 47.5 & 226.8 & 235.9 & 230 \\
156 & 0.60 & 47.5 & 76.6 & 80.3 & 80 \\
208 & 0.60 & 47.5 & 119.3 & 142.1 & 140 \\
156 & 0.70 & 47.5 & 61.6 & 62.6 & 60 \\
208 & 0.70 & 47.5 & 116.8 & 120.7 & 110 \\
209 & 0.90 & 47.5 & 99.5 & 87.1 & 80 \\
209 & 1.10 & 47.5 & 79.5 & 65.3 & 60 \\
\midrule
\multicolumn{6}{@{}l}{\emph{IN718, melt-pool depth on cross-sections after solidification~\cite{weaver_am_2022}}} \\
245 & 0.96 & 33.5 & 116.9 & 120 & 120 \\
285 & 0.80 & 33.5 & 176.5 & 170 & 170 \\
285 & 0.96 & 24.5 & 227.2 & 170 & 170 \\
285 & 0.96 & 33.5 & 139.7 & 140 & 140 \\
285 & 0.96 & 41 & 102.4 & 120 & 120 \\
285 & 1.20 & 33.5 & 109.7 & 110 & 110 \\
325 & 0.96 & 33.5 & 166.1 & 160 & 160 \\
\bottomrule
\end{tabular*}
\end{table}

\FloatBarrier
\section{Solver validation against published measurements}\label{sec:calibration}

A surrogate inherits whatever its training simulations get wrong, so we validate FLOW-3D\textsuperscript{\textregistered} parameters against published benchmark measurements to ensure simulated melt-pool geometry is physically consistent. 

\paragraph{Chosen solver settings.}

We base simulations on the solver configuration previously developed in \cite{benoit_fast_2026}, which was designed to predict thermal and melt fields for operator-learning models. It was never validated against measurements, and we thus had to make some minor parameter changes to better match experimental data. The solver computes recoil pressure from a Clausius--Clapeyron saturation relation,
\begin{equation}
p_{\mathrm{sat}}(T) = p_{v1}\,
  \exp\!\left[\frac{1}{t_{vexp}}\left(\frac{1}{T_{v1}}-\frac{1}{T}\right)\right],
  \qquad t_{vexp} = \frac{R}{M L_v},
\label{eq:clausius}
\end{equation}
in which $T_{v1}$ and $p_{v1}$ fix one point on the pressure curve ($p_{v1} = \SI{101.3}{\kilo\pascal}$) and $t_{vexp}$ fixes how steeply it rises. With $R$ the gas constant, $M$ the molar mass and $L_v$ the latent heat of vaporization, $1/t_{vexp} = M L_v / R$ is the exponential's characteristic temperature, and Ti-6Al-4V ($M = \SI{0.0479}{\kilo\gram\per\mole}$, $L_v = \SI{9.83e6}{\joule\per\kilo\gram}$) gives $t_{vexp}=\SI{1.77e-5}{\per\kelvin}$.

The simulation model evaluates $t_{vexp} = R/(M L_v)$ per material, and changes four further settings, all selecting among options the solver already provides.

The first is the recoil pressure that escaping vapor exerts back on the liquid. The base configuration applied an analytic recoil pressure, $p_r = p_{r0}\exp[k_r(1 - T_{v1}/T)]$ with fixed coefficients $p_{r0} = \SI{50.65}{\kilo\pascal}$ and $k_r = \num{10.96}$, wherever the free surface was hotter than $T_{v1}$.
 The validated settings derive recoil from the net evaporation rate, driven by the excess of $p_{\mathrm{sat}}$ over the ambient pressure $p_\infty$ the vapor escapes into,
\begin{equation}
\dot{m}(T) \;\propto\; p_{\mathrm{sat}}(T) - p_\infty ,
\label{eq:recoil}
\end{equation}
so recoil grows from zero at the boiling point where $p_{\mathrm{sat}}(T) = p_\infty$, and $T_{v1}$ only sets the reference point of the saturation curve. The two differ most at the onset of evaporation, the regime this section finds hardest to reproduce.

The second is the vapor partial pressure assigned to the gas phase, $p_\infty$ in \cref{eq:recoil}. We raised it from \SI{50.65}{\kilo\pascal} to standard atmospheric pressure, \SI{101.3}{\kilo\pascal}, so that net evaporation begins at $T_{v1}$, the boiling point at atmospheric pressure.

The third is surface tension, read from a tabulated $\sigma(T)$. The base Ti-6Al-4V table ends at \SI{1881}{\kelvin}, below the liquidus, and the validated settings extend it from the liquidus at the measured liquid slope, $\mathrm{d}\sigma/\mathrm{d}T = \SI{-2.6e-4}{\newton\per\metre\per\kelvin}$~\cite{mills_ti_2002}, up to \SI{3123}{\kelvin}.

Lastly, the absorption model also differs from the base configuration, the solver's Fresnel treatment with the material's own optical constants in place of a constant absorptivity.

\paragraph{Vaporization anchor.}

We adjusted one parameter, $T_{v1}$, by comparing four possible values, from the base configuration's \SI{3123}{\kelvin} to \SI{3520}{\kelvin}, against the
\SI{129}{\watt} Ti-6Al-4V track of Gan et al.~\cite{gan_universal_2021}, whose
vapor-depression depth comes from in-situ synchrotron X-ray imaging. This parameter mostly affects vapor cavity depth for a given temperature. Values of
$T_{v1} = \SI{3420}{\kelvin}$ and $T_{v1} = \SI{3520}{\kelvin}$ both reproduced
that measurement to within $\pm 15\,\%$. We chose to use $T_{v1} = \SI{3520}{\kelvin}$, which agreed better across other conditions. This $T_{v1}$ value fits
inside the \SIrange{3315}{3560}{\kelvin} range published for this alloy, running
from the vaporization temperature used in laser powder-bed fusion
simulations~\cite{pantawane_computational_2020} to the boiling point of pure
titanium~\cite{haynes_crc_2017}.

\paragraph{Benchmark sources.}

We score against five published sources, each reporting enough process
detail to reproduce the process it measured. Throughout this section a
\emph{condition} is one set of process parameters, laser power, scan speed and
beam radius. Gan et
al.~\cite{gan_universal_2021} give vapor-depression depth for Ti-6Al-4V
across power and across scan speed, from in-situ X-ray imaging~\cite{cunningham_keyhole_2019}. Gali\`egue et
al.~\cite{galiegue_acoustic_2025} give metallographic melt-pool depth and width for
Ti-6Al-4V across power, speed and defocus. Hawk et al.~\cite{hawk_laser_2022}
give melt-pool depth for stationary 304L spot welds, and Soares et
al.~\cite{soares_reliability_2026} a cross-section series for CoCr.
The fifth, the Additive Manufacturing Benchmark Series (AM-Bench) run by NIST,
supplies three separate things. Its Ti-6Al-4V record
\texttt{mds2-2525} pairs integrating-sphere absorptance with high-speed
radiographs on the same shot, for a scanned track and for a stationary spot,
which is the only source here that resolves anything in time. Its 2022 challenge
series supplies the seven IN718 conditions scored below, the AMB2022-03
single-track cross-sections~\cite{weaver_am_2022}.

Besides the three lengths below, a fourth quantity is compared where a record supplies it. Absorptance is
the fraction of incident laser power that crosses the surface, measured by
integrating-sphere calorimetry~\cite{simonds_ability_2024,simonds_asynchronous_2022}. The solver reports
the share of incident power still held in the metal as thermal energy once
radiation, convection and evaporating material have carried some back out, and
that share is what we plot. FLOW-3D\textsuperscript{\textregistered} does not itemize radiation and convection to
the surrounding gas, so it stays a lower bound on the measured absorptance.

\paragraph{Experimental measurements used for validation.}

\begin{figure}[htbp]
\centering
\begin{subfigure}[t]{0.49\textwidth}
  \includegraphics[width=\linewidth]{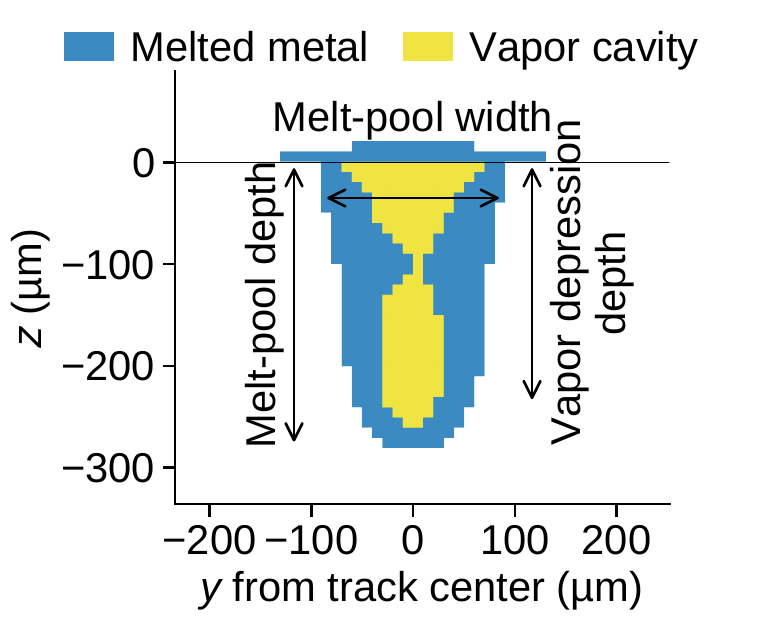}
  \subcaption{Section across the track.}\label{fig:calib-metrics-across}
\end{subfigure}\hfill
\begin{subfigure}[t]{0.49\textwidth}
  \includegraphics[width=\linewidth]{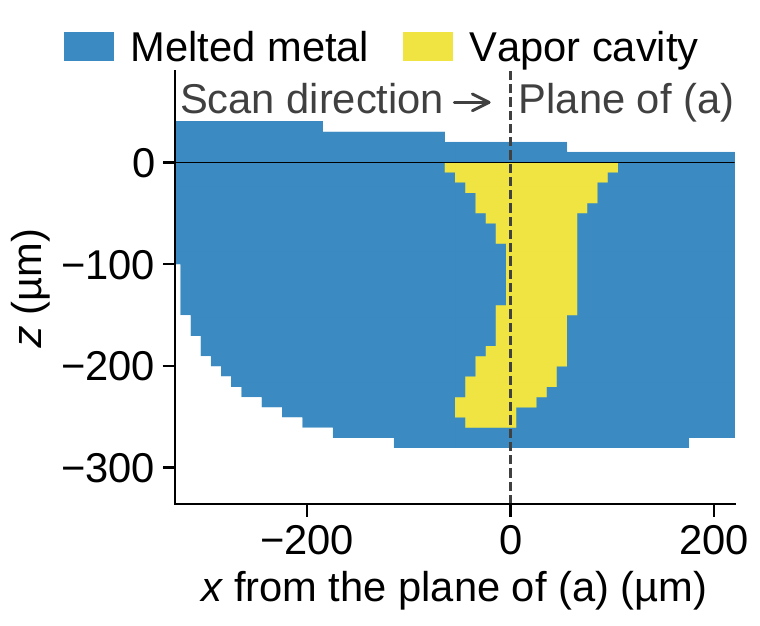}
  \subcaption{Section along the track.}\label{fig:calib-metrics-along}
\end{subfigure}
\caption{The three scored lengths on a simulated Ti-6Al-4V melt pool at
\SI{206}{\watt} and \SI{0.4}{\metre\per\second}, with the original substrate
surface at $z=0$. Blue is the union over time of melted cells, which is what a
metallographic section provides, and yellow the vapor cavity at the instant
shown. (a) is the plane across the track at the beam and (b) the longitudinal
section.}
\label{fig:calib-metrics}
\end{figure}

We compare the three lengths the sources report directly, namely melt-pool depth, melt-pool width and vapor-depression depth (\cref{fig:calib-metrics}). Metallography gives the melt-pool boundary, the outer limit of melted and resolidified material. And in-situ X-ray radiography gives the vapor depression zone. We consider beam radius $r_0$ as half the $D4\sigma$ diameter (four times the intensity profile's standard deviation, the ISO beam width), which equals the $1/e^2$ diameter for a Gaussian beam.

We read each length off the exported field grids in the same way as its instrument does. A metallographic section is cut after the track has solidified, so it records the largest extent the pool reached as it passed. We take one fixed plane across the track, through the beam's position at mid-track, and collect every cell melted there at any exported time step. A cell counts as melted when its liquid fraction exceeds one half ($f_l > 0.5$ in \cref{eq:fl}, equivalently a temperature above the midpoint of $T_s$ and $T_l$). Melt-pool depth runs from the original surface to the bottom of the deepest of these cells, and width is their extent across the track. Vapor-depression depth is the deepest cell where gas has displaced metal below the original surface, averaged over the last \SI{30}{\percent} of the laser-on interval because a keyhole oscillates and no single frame represents it. At the X-ray conditions we score, that depth has a standard deviation of \SIrange{7}{18}{\micro\metre} over the averaging interval (one to two cells), and up to \SI{35}{\micro\metre} in our deepest cavities. Each length is a whole number of \SI{10}{\micro\metre} cells in every frame, so it carries one cell of quantization before any averaging over frames, and where a feature spans only a few cells we give its size in cells.

\paragraph{What counts as agreement between solver and experiment.}

Neither the measurements nor the published simulations support a sharp
uncertainty threshold. We treat agreement within \SI{30}{\percent} on a length as
close agreement, and a larger deviation as a sign of physics the model does not
resolve or chaotic behavior. We choose this value based on multiple factors.
The measurements themselves have fairly large uncertainty. The metallographic series quotes along-track variation near \SI{10}{\percent} and
measures most conditions at a single location~\cite{galiegue_acoustic_2025}, and
depths from two independent groups differ by \SI{12}{\percent} at matched power
and speed~\cite{galiegue_acoustic_2025,vaglio_single_2020,totis_data_2021}. The process is
partly stochastic at the scale being measured, so exact replication is not
available even in principle. A keyhole's depth fluctuates by more than
\SI{40}{\percent} of its own mean within a single simulated
track~\cite{wang_mechanism_2022}, and in the AM-Bench absorptance experiments the
onset of keyholing repeats to \SI{30}{\micro\second} in Ti-6Al-4V but scatters
across \SIrange{190}{650}{\micro\second} in aluminum, which the authors
attribute to the sensitivity of coupling to surface
state~\cite{simonds_ability_2024}. Due to this extreme experimental
variability, we do not model aluminum. Published simulations of these benchmarks deviate comparably. A thermal-fluid model resolving its beam with \num{12.5} cells reports keyhole depths within \SIrange{2.5}{15}{\percent} of experiment across its speed range~\cite{wang_mechanism_2022}, and most 2022 AM-Bench challenge submissions over-predicted both melt-pool width and depth for the scanned laser~\cite{simonds_ability_2024}.

\paragraph{Agreement in the keyhole regime.}

\begin{figure}[htbp]
\centering
\begin{subfigure}[t]{0.49\textwidth}
  \includegraphics[width=\linewidth]{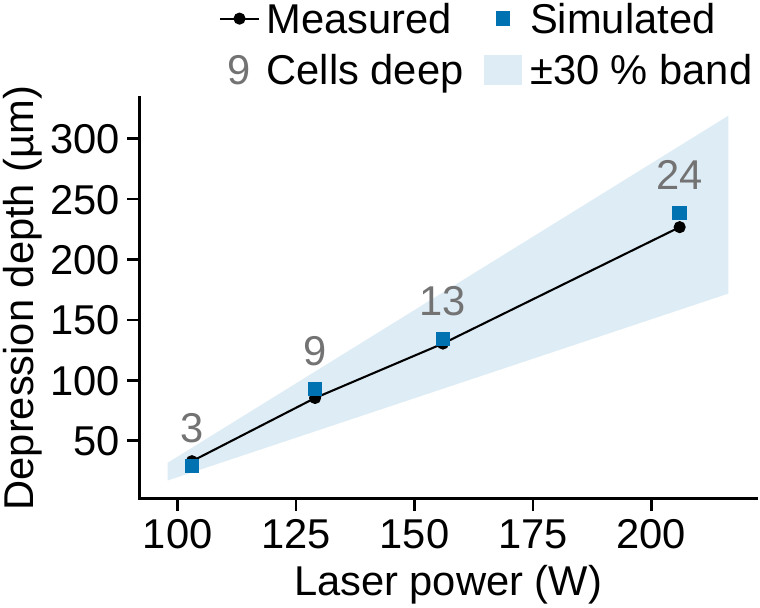}
  \subcaption{Depth against laser power.}\label{fig:calib-keyhole-power}
\end{subfigure}\hfill
\begin{subfigure}[t]{0.49\textwidth}
  \includegraphics[width=\linewidth]{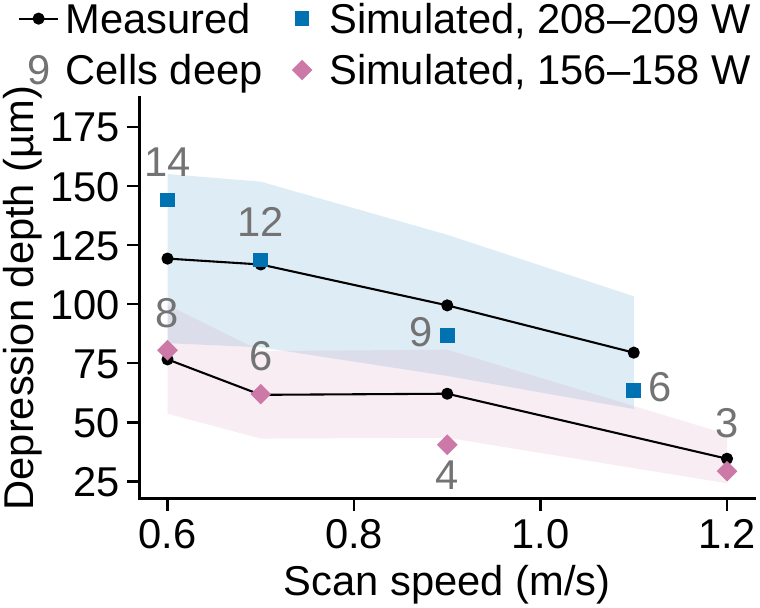}
  \subcaption{Depth against scan speed.}\label{fig:calib-keyhole-speed}
\end{subfigure}
\caption{Vapor depression depth in the keyhole regime, simulated against
measured by in-situ X-ray radiography~\cite{gan_universal_2021,cunningham_keyhole_2019}, for Ti-6Al-4V at
a beam radius of \SI{47.5}{\micro\metre}. (a) Against laser power at
\SI{0.4}{\metre\per\second}. (b) Against scan speed at \SIrange{208}{209}{\watt} and
at \SIrange{156}{158}{\watt}. Shading marks agreement within \SI{30}{\percent} of the measurement. Annotations give
the number of \SI{10}{\micro\metre} cells across the depth of each simulated
cavity, rounded to an integer because that depth is a mean over an oscillating
cavity and carries \numrange{0.2}{1.5} cells of scatter.}
\label{fig:calib-keyhole}
\end{figure}

Where a keyhole is established the validated model reproduces the measured cavity quite well (\cref{fig:calib-keyhole}). On Gan et al.'s power series~\cite{gan_universal_2021} at \SI{0.4}{\metre\per\second}, vapor depression against vapor depression, all four measured points agree to better than \SI{30}{\percent}, with a mean signed error of $+\SI{1.2}{\percent}$ (the shallowest, the \SI{103}{\watt} cavity, is under three cells deep) and a depth-versus-power slope within \SI{7}{\percent} of measurement. On the speed series at \SIrange{208}{209}{\watt} and at \SIrange{156}{158}{\watt}, seven of eight measured points agree. Absolute depth therefore transfers across both axes we can test, while its scaling in speed does not. Fitted as depth $\propto V^{n}$, our cavity shallows with speed \num{1.4} to \num{2} times faster than the measurement.

\paragraph{The sub-keyhole depth deficit.}

\begin{figure}[htbp]
\centering
\begin{subfigure}[t]{0.49\textwidth}
  \includegraphics[width=\linewidth]{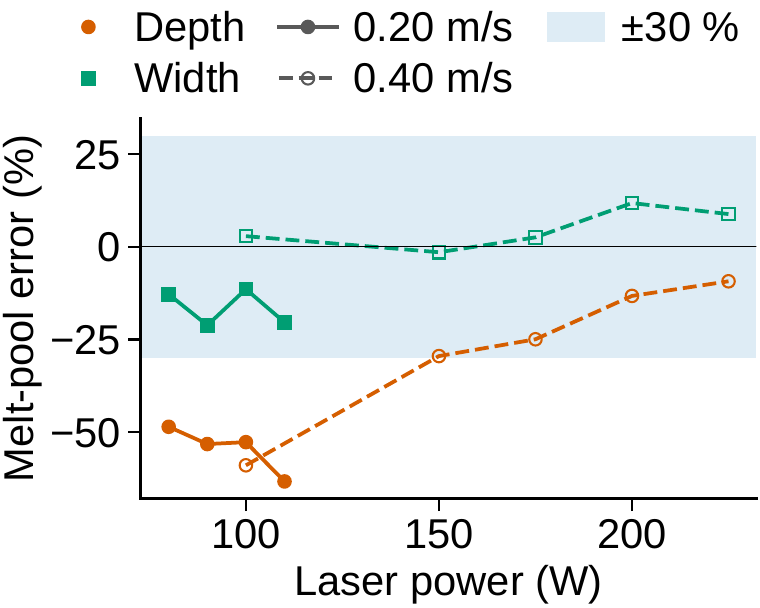}
  \subcaption{Melt-pool error against power.}\label{fig:calib-transition-err}
\end{subfigure}\hfill
\begin{subfigure}[t]{0.49\textwidth}
  \includegraphics[width=\linewidth]{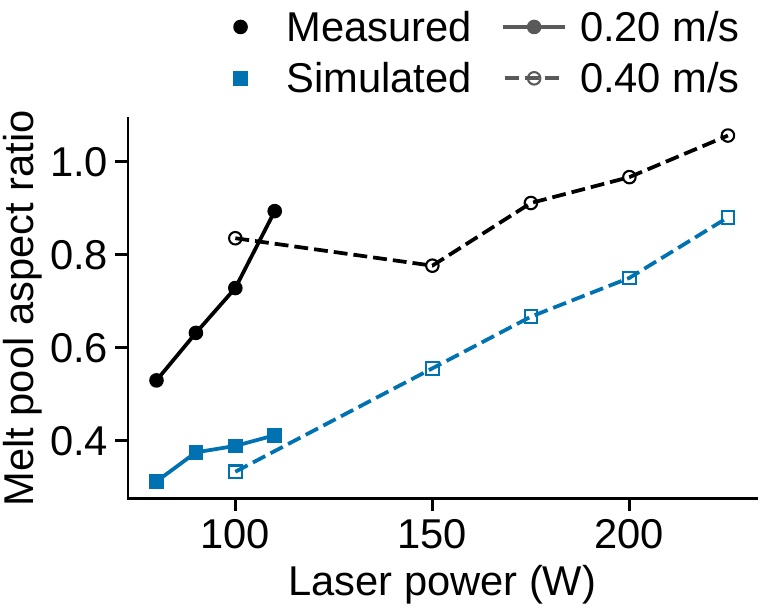}
  \subcaption{Melt-pool aspect ratio against power.}\label{fig:calib-transition-aspect}
\end{subfigure}
\caption{Melt-pool geometry for Ti-6Al-4V at $r_0=\SI{60}{\micro\metre}$,
against metallographic sections~\cite{galiegue_acoustic_2025}. (a) Signed error
against measurement for melt-pool depth and width at both scan speeds, with shading
marking agreement within \SI{30}{\percent}. (b) The corresponding aspect ratio,
melt-pool depth divided by melt-pool width, for both speed series.}
\label{fig:calib-transition}
\end{figure}

Below the keyhole threshold the depth agreement worsens while the width
agreement remains accurate. On the power series of Gali\`egue et
al.~\cite{galiegue_acoustic_2025} at \SI{0.20}{\metre\per\second} and
$r_0 = \SI{60}{\micro\metre}$, our melt-pool depth falls
\SIrange{49}{63}{\percent} short of the measurement across
\SIrange{80}{110}{\watt}, while our melt-pool width stays within
\SI{30}{\percent} of it at every power (\cref{fig:calib-transition}). The two
errors move apart as power rises. The measured pool deepens faster than it
widens, its aspect ratio climbing from \num{0.53} to \num{0.89}, while ours
rises only from \num{0.31} to \num{0.41}. The data source does not state whether
its depths run from the original plate surface or from the top of the
resolidified bead, and ours run from the original surface. Our bead stands \SI{5}{\micro\metre} above the surface at \SI{90}{\watt},
so that convention accounts for about \SI{7}{\percent} of the
\SI{68}{\micro\metre} deficit there and none of the rest.
That series spans only \SIrange{80}{110}{\watt}. A second series from the same
source, plate, spot and observable at \SI{0.40}{\metre\per\second} extends the
comparison to \SI{225}{\watt}, over which the depth error shrinks steadily from
\SI{-59}{\percent} to \SI{-9}{\percent}, ending inside our agreement scale, while
the width error stays small with no systematic sign
(\cref{fig:calib-transition-err}). Speed itself is not the cause. At
\SI{100}{\watt}, the one power the two series share, doubling the speed moves
the depth error from \SI{-53}{\percent} to \SI{-59}{\percent}. What the second
series adds is reach further across the keyhole threshold. Its cavity deepens from
\SIrange{17}{181}{\micro\metre} over the five powers, against a
\SI{60}{\micro\metre} beam radius, and the deficit is
largest where the cavity is shallowest.

That largest deficit falls on a measurement that breaks the usual speed ordering. At \SI{100}{\watt} the source reports \SI{121.8}{\micro\metre} at
\SI{0.40}{\metre\per\second} against \SI{98.7}{\micro\metre} at
\SI{0.25}{\metre\per\second} and \SI{85.9}{\micro\metre} at
\SI{0.30}{\metre\per\second}, with the aspect ratio inverted the same way,
\num{0.835} against \num{0.728} at \SI{0.20}{\metre\per\second}
(\cref{fig:calib-transition-aspect}). Within its own series the depth falls from
\SI{121.8}{\micro\metre} at \SI{100}{\watt} to \SI{111.1}{\micro\metre} at
\SI{125}{\watt}, then rises linearly from \SI{150}{\watt} to \SI{400}{\watt}. Back-extrapolation gives \SI{95.7}{\micro\metre} at \SI{100}{\watt}, putting
the tabulated value about a quarter above its own trend. The source gives no
per-point uncertainty, but reports that only the low-speed conditions were
sectioned and averaged over three locations, the others over one, that depth varies along a track by about ten percent, and that keyhole onset is unstable at low power. It gives no keyhole frequency for this point alone~\cite{galiegue_acoustic_2025}.
We score the tabulated value. Against the fitted trend our depth error there would
be \SI{-48}{\percent} instead of \SI{-59}{\percent}.

Three measurements point to the vapor cavity as the source of the deficit. The
first is the conducted melt shell, the metal melted between the cavity wall and
the melt-pool boundary. Taken as melt-pool depth minus depression depth, the
simulated shell is nearly constant, $41 \pm \SI{2}{\micro\metre}$ at
\SI{0.20}{\metre\per\second} and $34 \pm \SI{3}{\micro\metre}$ at
\SI{0.40}{\metre\per\second} (standard deviation over the simulations at each speed),
while the melt-pool depths themselves vary by a factor of four.
The model therefore melts about the right thickness of metal around whatever
cavity it produces, and its depth errors are cavity errors. The second is melt-pool width, which depends mostly on beam
size~\cite{khairallah_laser_2016} and matches the measurement within \SI{30}{\percent} at every power
in the series. The third is the
cavity itself, which agrees once keyholing is established
(\cref{fig:calib-keyhole}). That leaves the formation of the cavity below the
threshold. Subtracting the \SI{41}{\micro\metre} shell at \SI{0.20}{\metre\per\second} from
the measured melt-pool depth implies a cavity four to six times deeper than ours.

Exact cavity onset is hard to predict here. Evaporation is only just
switching on, so the onset condition in \cref{eq:recoil} matters most, and
recoil grows exponentially with surface temperature, turning a small temperature
error into a large pressure error. A benchmark of laser-induced vapor depression
turned a \SI{1.1}{\percent} interface-temperature error into a
\SI{16.7}{\percent} recoil-pressure error~\cite{much_hybrid_2026}.

\paragraph{Why the deficit seems to be a cavity error.}

\begin{figure}[htbp]
\centering
\begin{subfigure}[t]{0.49\textwidth}
  \includegraphics[width=\linewidth]{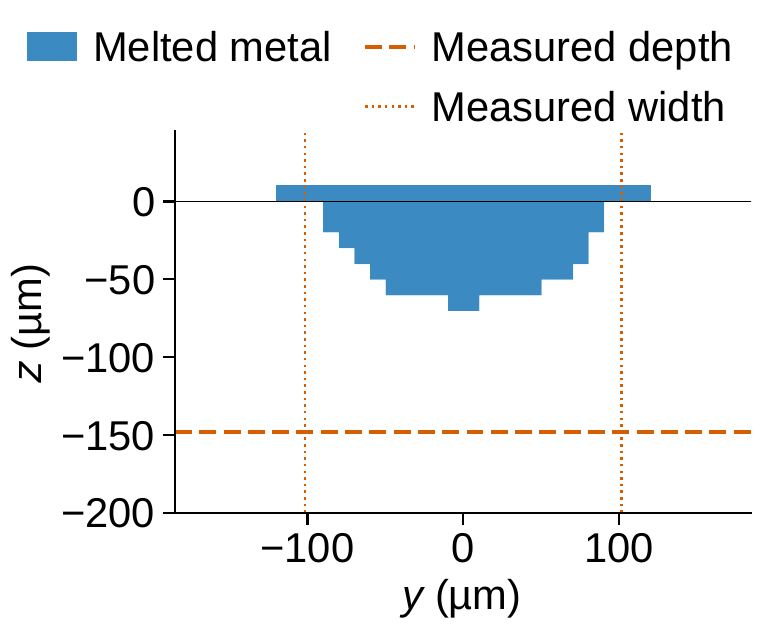}
  \subcaption{Beam radius $r_0 = \SI{60}{\micro\metre}$.}\label{fig:calib-spot-r60}
\end{subfigure}\hfill
\begin{subfigure}[t]{0.49\textwidth}
  \includegraphics[width=\linewidth]{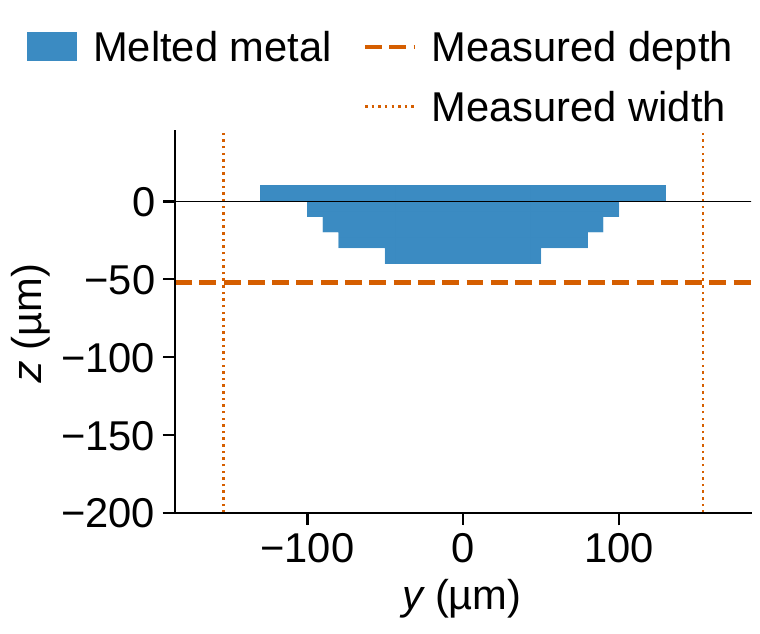}
  \subcaption{Beam radius $r_0 = \SI{130}{\micro\metre}$.}\label{fig:calib-spot-r130}
\end{subfigure}
\caption{Simulated melt cross-sections at \SI{100}{\watt} and
\SI{0.20}{\metre\per\second} for two beam radii, taken as the union over time of
the melted cells in the plane across the track at the beam and shown in physical
orientation, with the substrate surface at $z=0$, for (a) $r_0=\SI{60}{\micro\metre}$
and (b) $r_0=\SI{130}{\micro\metre}$. The dashed line marks the measured melt-pool
depth and the dotted lines the measured width~\cite{galiegue_acoustic_2025}.}
\label{fig:calib-spot}
\end{figure}

We therefore ran a larger spot, spreading the same power over more area so
that no vapor cavity forms. At $r_0 = \SI{130}{\micro\metre}$ the depth error
falls from $-\SI{53}{\percent}$ to $-\SI{23}{\percent}$, about one computational
cell, and the aspect ratio comes close to the measurement, though the width falls short
(\cref{fig:calib-spot}), consistent with a cavity-specific deficit.

Simulated width there falls \SI{35}{\percent} short of measurement. Across the change of spot our pool widens by $1.11\times$ against a measured
$1.52\times$. We appear to under-predict lateral spreading at large spots, with
the same measurement ambiguity as before, since width taken at the $z=0$ surface or at
the top of the re-solidified bead differs by \SIrange{20}{30}{\micro\metre}. This is a
known difficulty, as a recoil calibration matched to one spot size in the same
solver has been reported not to transfer to another~\cite{zenz_critical_2025}.

\paragraph{Other materials.}

Agreement varies by material, and every scored material appears in
\cref{fig:calib-materials}.
CoCr agrees on both depth and width at all three conditions, with a
$-\SI{18}{\percent}$ depth bias common to all three. Of three 304L stationary
spot welds~\cite{hawk_laser_2022}, two fall within \SI{30}{\percent} of
experiment, mean signed error $+\SI{15}{\percent}$. For IN718 we scored seven
AM-Bench challenge conditions spanning \SIrange{245}{325}{\watt},
\SIrange{0.80}{1.20}{\metre\per\second} and three spot sizes. All seven fall
within \SI{30}{\percent} on both depth and width, mean signed depth error
$-\SI{0.4}{\percent}$. 

\paragraph{Depth and coupling through one scan.}

One benchmark experiment also supports comparison in time. The absorptance
record and the radiographs of the AM-Bench Ti-6Al-4V scan were taken
simultaneously on the same shot, the frame numbers indexing the absorptance trace
and the two agreeing on frame time to
\SI{0.3}{\micro\second}~\cite{simonds_ability_2024,simonds_asynchronous_2022}. Cavity depth and absorptance
therefore share one time axis, testing the validated model against a melt pool's
history. Absorptance is tabulated in the record, and we segment cavity depth ourselves from the X-ray frames.

\begin{figure}[htbp]
\centering
\begin{subfigure}[t]{0.49\textwidth}
  \includegraphics[width=\linewidth]{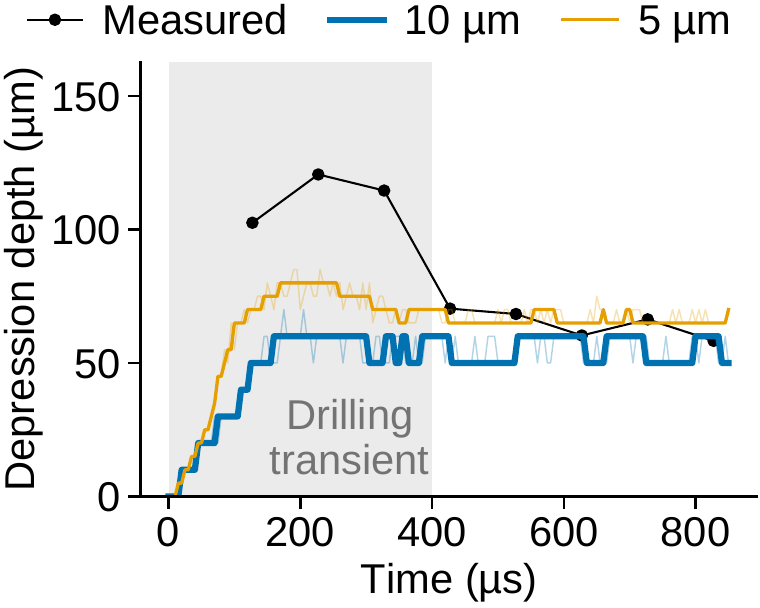}
  \subcaption{Depression depth, scan.}\label{fig:calib-history-depth}
\end{subfigure}\hfill
\begin{subfigure}[t]{0.49\textwidth}
  \includegraphics[width=\linewidth]{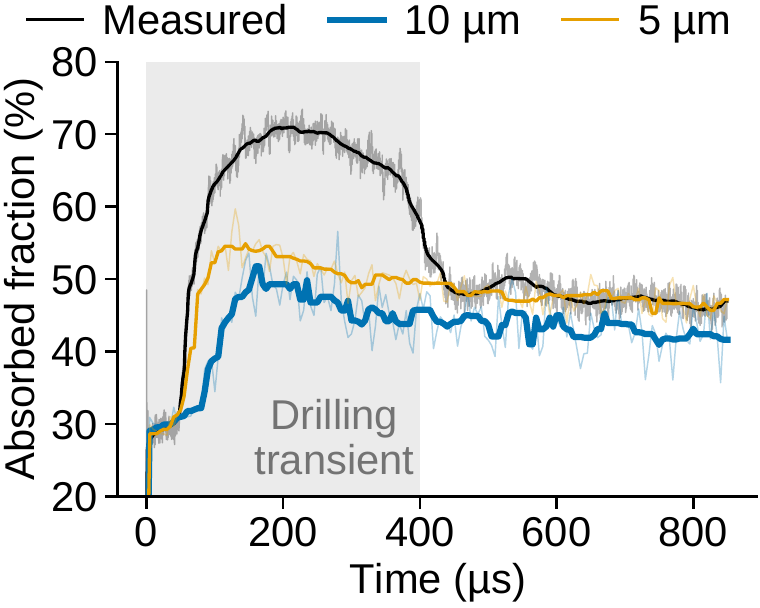}
  \subcaption{Absorbed fraction, scan.}\label{fig:calib-history-abs}
\end{subfigure}

\vspace{0.6em}
\begin{subfigure}[t]{0.49\textwidth}
  \includegraphics[width=\linewidth]{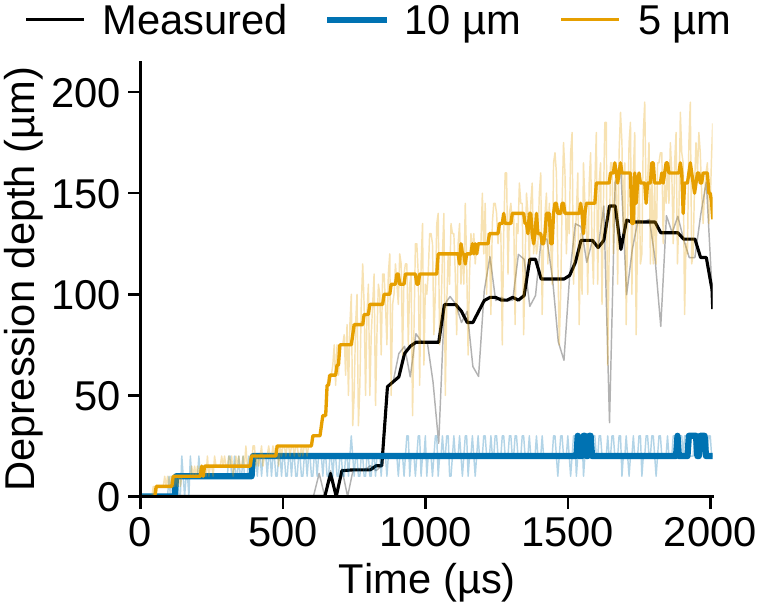}
  \subcaption{Depression depth, stationary spot.}\label{fig:calib-spot-depth}
\end{subfigure}\hfill
\begin{subfigure}[t]{0.49\textwidth}
  \includegraphics[width=\linewidth]{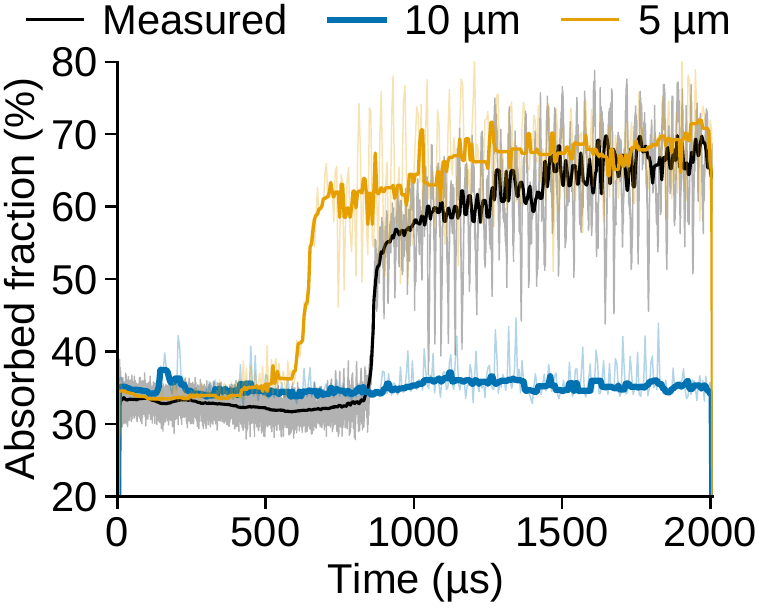}
  \subcaption{Absorbed fraction, stationary spot.}\label{fig:calib-spot-abs}
\end{subfigure}
\caption{Absorptance and depression depth for two configurations of one
measurement record~\cite{simonds_ability_2024,simonds_asynchronous_2022}, measured and simulated on a
common time axis at both cell sizes. (a, b) One Ti-6Al-4V scan at
\SI{195}{\watt} and \SI{0.7}{\metre\per\second}, with shading marking the
drilling transient. (c, d) The stationary
\SI{101.9}{\watt} {spot} from the same record.}
\label{fig:calib-history}
\end{figure}

Measured depth and measured absorbed fraction move together
(\cref{fig:calib-history}). The cavity deepens to about \SI{120}{\micro\metre}
as the absorbed fraction rises to \SI{73.5}{\percent}, and both then relax to a
plateau near \SI{60}{\micro\metre} and \SI{48.5}{\percent}. Those are raw values. The running medians drawn in \cref{fig:calib-history} peak at
\SI{71}{\percent} and settle at \SI{46}{\percent}. A deeper cavity traps
more of the beam through multiple reflections, and this shot shows both
signatures of that at once.

Our model follows the settled keyhole and understates the initial drilling
transient. Before \SI{400}{\micro\second} the measured cavity is about twice as
deep as ours. After that the two agree to about ten percent and they coincide
from \SI{627}{\micro\second} on. Absorbed fraction behaves the same way,
rising from \SI{29}{\percent} on the cold surface to \SI{52}{\percent}
where the measurement reaches \SI{71}{\percent}, then settling at
\SI{43}{\percent} against a measured \SI{46}{\percent}. Both quantities fall short over the same interval and recover at the
same point, which places the shortfall {in} the drilling phase. Halving the
cell deepens the cavity throughout, from \SIrange{23}{36}{\micro\metre} averaged over \SIrange{40}{100}{\micro\second} and from \SIrange{55}{67}{\micro\metre} {over \SIrange{600}{850}{\micro\second}}, while the absorbed-fraction peak
rises by three points. Neither effect closes the drilling gap.

We tested two candidate causes on this simulation, each a single change to the solver input and compared with a new run of the unchanged simulation, with transient depths averaged over \SIrange{120}{400}{\micro\second}. Moving $T_{v1}$ down to \SI{3425}{\kelvin}
deepened the cavity throughout, by \SI{5.9}{\micro\metre} during drilling and
\SI{4.8}{\micro\metre} once settled, a uniform offset where a drilling-specific
cause would have deepened the transient alone. Halving the computational cell size
added \SI{17.3}{\micro\metre} to the transient depth and \SI{10.8}{\micro\metre}
to the settled cavity, the larger of the two and the most concentrated in
drilling, though its effect on overall geometry is negligible
(\cref{fig:calib-xray}).
The geometry scored elsewhere in this section averages over a steady
window, which sits inside the interval where the two agree.

\cref{fig:calib-xray} shows where that agreement holds in space. The trailing
wake matches at every frame, with a rim about \SI{20}{\micro\metre} above the
surface in both, and once settled both cavities reach \SI{60}{\micro\metre} at
the same position. Inside the transient the measured
cavity is \SI{115}{\micro\metre} deep against our \SI{50}{\micro\metre}, a
comparison taken over the whole sequence of frames since the cavity oscillates
continuously, ours over \SIrange{50}{60}{\micro\metre} about a mean of
\SI{55}{\micro\metre}. The refined \SI{5}{\micro \metre} boundary comes slightly
closer to the transient keyhole. At \SI{127}{\micro\second} it sits below the coarse one, still well short of the
measured cavity, while at \SI{627}{\micro\second} the two coincide. Cell size
therefore governs the drilling cavity and stops mattering once the keyhole is
established.

\paragraph{What a finer mesh costs.}

Every simulation in our datasets uses a uniform \SI{10}{\micro\metre} cell, a choice
set by the cost of generating a corpus. Halving the cell to \SI{5}{\micro\metre} multiplied solver cost by
\num{12} on a controlled pair of otherwise identical simulations. Our 150-simulation
free-form-path corpus took \num{703} core-hours, which would be about three days at the ten cores our workstation gives a simulation. The same corpus at \SI{5}{\micro\metre}
would take about five weeks using the same setup. 

Our chosen resolution of \SI{10}{\micro\metre} sits within the range used in
practice. Cells per beam radius span roughly \numrange{2}{45} across recent
laser-melting simulations, from a FLOW-3D\textsuperscript{\textregistered} study resolving a
\SI{43}{\micro\metre} spot with \SI{10}{\micro\metre} cells~\cite{lai_unveiling_2024}
to an adaptively refined computation at \num{45}
cells~\cite{much_improved_2024}. Ours is \num{6.0} at the
\SI{60}{\micro\metre} spot of the transition condition. The range is wide
because the field has no settled convention. A 2020 review found mesh
sensitivity undiscussed in virtually all publications~\cite{cook_simulation_2020},
and while later studies commonly compare two or three cell
sizes~\cite{ebrahimi_numerical_2021,much_hybrid_2026}, they still select a mesh
on cost without an error estimate. The one numeric criterion we
located, thirty-five cells across the melt-pool
width~\cite{ebrahimi_influence_2022}, is a single group's proposal. We carry
about twenty.

\begin{figure}[htbp]
\centering
\begin{subfigure}[t]{0.49\textwidth}
  \includegraphics[width=\linewidth]{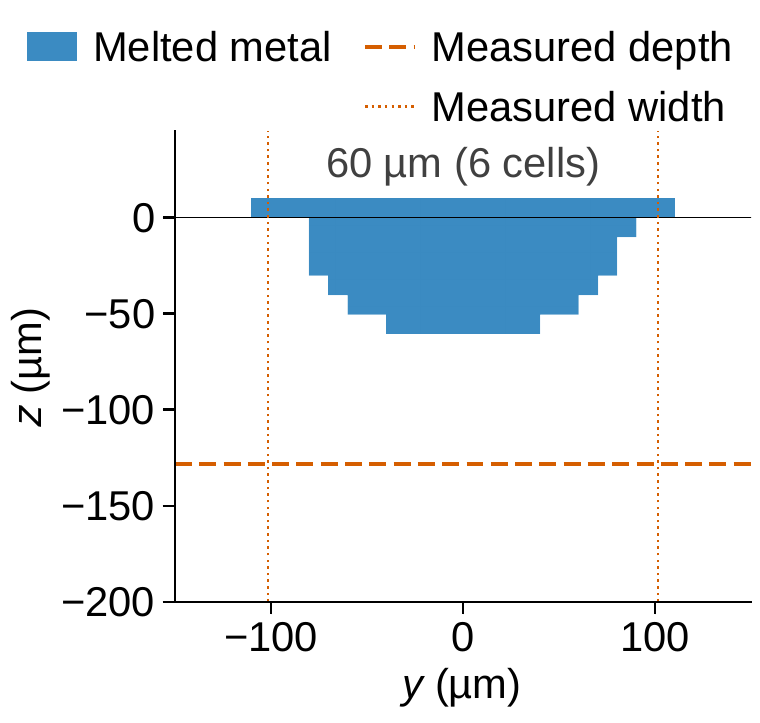}
  \subcaption{\SI{10}{\micro\metre} cells.}\label{fig:calib-mesh-c10}
\end{subfigure}\hfill
\begin{subfigure}[t]{0.49\textwidth}
  \includegraphics[width=\linewidth]{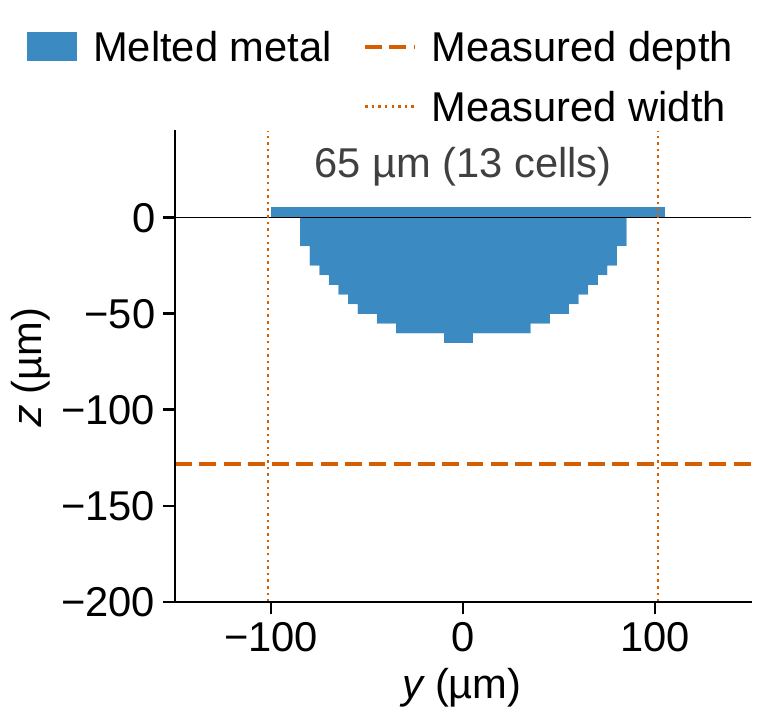}
  \subcaption{\SI{5}{\micro\metre} cells.}\label{fig:calib-mesh-c5}
\end{subfigure}
\caption{Simulated melt cross-sections for the one pair of simulations differing in cell size alone, just below the keyhole threshold, (a) at \SI{10}{\micro\metre} and (b) at
\SI{5}{\micro\metre}. Shading is the union over time of melted cells and the
substrate surface is at $z=0$. Dashed and dotted lines mark the measured melt-pool
depth and width at this condition~\cite{galiegue_acoustic_2025}.}
\label{fig:calib-mesh}
\end{figure}

\paragraph{When a finer mesh is needed.}

We thus tested the effect of resolution on a few simulations. On one simulation just below the keyhole threshold, whose vapor depression is two cells (\SI{20}{\micro\metre}) deep, we re-ran the \SI{90}{\watt} condition of
\cref{fig:calib-transition} at \SI{5}{\micro\metre}, changing the cell size and
nothing else in the rendered solver input. Melt-pool depth moved from
\SI{60}{\micro\metre} to \SI{65}{\micro\metre} against \SI{128}{\micro\metre}
measured, below the resolution floor, and width did not move (\cref{fig:calib-mesh}).

Two further pairs bound how far that carries. The first is the stationary spot
of the same NIST record, where the \SI{2.008}{\milli\second} {beam-on duration} is long
enough to watch a keyhole form (\cref{fig:calib-history}c,d). At
\SI{10}{\micro\metre} the model holds a three-cell depression for the whole
beam-on duration and never drills. At \SI{5}{\micro\metre} it drills, and over the settled
window its cavity averages $139 \pm \SI{28}{\micro\metre}$ against a measured
$115 \pm \SI{25}{\micro\metre}$, each spread the frame-by-frame oscillation of
the keyhole itself, so the two agree to within the fluctuation of either.
Halving the cell decides whether the mechanism appears at all. The absorbed
fraction, which no mesh quantizes, confirms it. At \SI{5}{\micro\metre} it
doubles as the keyhole forms, as the measurement does, and at
\SI{10}{\micro\metre} it never moves.

The second pair bounds it from the other side, on a metallographic condition. Halving the cell on the \SI{100}{\watt},
\SI{0.40}{\metre\per\second} simulation moves neither quantity available there.
Melt-pool depth is \SI{50.0}{\micro\metre} at both cell sizes, \SI{59}{\percent} short of the measurement either way, and the cavity holds at
\SI{17}{\micro\metre}. That simulation is converged at \SI{10}{\micro\metre}, so
its deficit belongs to the model. Regime is what separates the two pairs. The
\SI{100}{\watt} cavity is a shallow depression \SI{17}{\micro\metre} deep
against a \SI{60}{\micro\metre} beam radius, while the stationary spot reaches
\SI{195}{\micro\metre} and is a keyhole in the ordinary sense.

The one discrepancy refinement does not remove is
onset. The simulated keyhole opens at \SI{675}{\micro\second} against a
measured \SI{867}{\micro\second}, which holds across segmentation choices and
which the independently measured absorptance puts at \SI{890}{\micro\second}.
What decides whether a mesh can represent a cavity is its width, because the
multiple reflections that drive a keyhole down need a channel a ray can bounce
along. A stationary spot drills a narrow bore, measured at
\SIrange{24}{42}{\micro\metre} or two to four cells at
\SI{10}{\micro\metre}, while a moving beam drags a broader one, seven to nine
cells across the three \SI{0.40}{\metre\per\second} conditions where one
can be measured.

The cell count has to come from the measurement or from a refined run,
since a mesh too coarse to open a channel reports no channel to measure.

We therefore find the \SI{10}{\micro\metre} mesh adequate for the scanned tracks that make up the
corpus here, but inadequate for a long stationary dwell with a narrow cavity.

\paragraph{Scope of the validated simulation model.}

The validated simulation settings are used for every dataset generated in this study. Results from pre-validation data are marked as such where they
appear. Re-solving the process grid from \cite{benoit_fast_2026} under the new settings changes
melt-pool depth by $-36$ to $-\SI{44}{\percent}$ at five conditions spanning
\SIrange{90}{190}{\watt} and
\SIrange{0.47}{0.95}{\metre\per\second}, and by $-\SI{68}{\percent}$ at a
sixth, the slowest sampled at \SI{0.195}{\metre\per\second}, so models and physical results from the two generations are not directly interchangeable.

\FloatBarrier
\section{Architecture benchmark on the LP-FNO dataset}\label{sec:benchmark}

The dataset of our previous work~\cite{benoit_fast_2026} is \num{83} straight
tracks in Ti-6Al-4V on the pre-validation solver settings, at \SI{10}{\micro\metre}
on a \numproduct{0.9x0.4x0.3}\,\si{\milli\metre} domain, power \SIrange{40}{190}{\watt}
and scan speed \SIrange{0.1}{1.0}{\metre\per\second}, in the quasi-steady formulation of \cref{sec:formulation}. That paper reports its metrics as eight-fold cross-validation
means over the set. We here give an indicative comparison using a split \num{65}\,/\,\num{9}\,/\,\num{9} where every model is scored on the
nine held-out simulations only, since a cross-validated comparison has high computational costs over three architectures. \Cref{tab:benchmark} gives a comparison of metrics. We can see that improved hyper-parameters and richer inputs seem to lead to substantially better metrics compared to our previous model.

\begin{table}[htbp]
\centering
\caption{The three architectures on the pre-validation straight-path set,
nine held-out simulations. One training run per
row, best per column in bold. The last row is the
published eight-fold cross-validation mean of the LP-FNO model from \cite{benoit_fast_2026} on the same set.}
\label{tab:benchmark}
\small

\begin{tabular*}{\textwidth}{@{\extracolsep{\fill}}lccccc}
\toprule
 & \tabhead{T MAE\\(\si{\kelvin})} & \tabhead{T RMSE\\(\si{\kelvin})} & \tabhead{BB MAE\\(\si{\kelvin})} & \tabhead{Rel. err.\\(\%)} & \tabhead{Melt-pool\\IoU} \\
\midrule
FNO   & 6.40 & 21.4 & 23.5 & 1.11 & 0.927 \\
RIGNO & 7.90 & 22.2 & 27.9 & 1.37 & 0.932 \\
{GAOT} & \textbf{6.22} & \textbf{19.0} & \textbf{21.9} & \textbf{1.08} & \textbf{0.935} \\
\midrule
LP-FNO~\cite{benoit_fast_2026}& 15.7 & 33.5 & --- & 2.49 & 0.922 \\
\bottomrule
\end{tabular*}
\end{table}

\section{Details about the AiiDA provenance store}\label{sec:provenance-supp}

Every simulation and model result in this paper resolves to a node in a provenance store
(\Cref{sec:pipeline}).

\paragraph{Contents.} The store is an AiiDA~\cite{huber_aiida_2020}
profile backed by PostgreSQL, with array payloads in a content-addressed object
repository on disk. The published store is a smaller part of it, restricted to the data directly relevant
to this paper. The results tables resolve to \num{45} trained models, the \num{33}
of \cref{tab:arch-steady,tab:paths-accuracy,tab:preheat} and the twelve of
\cref{tab:twoalloy}, and to the evaluations listed row by row in the provenance
log released with them.

\paragraph{What a node holds.} A simulation node stores one simulation as a
single compressed array file holding the eleven exported fields, among them temperature, metal--gas fraction, liquid fraction, velocity and pressure, on the full stored grid at every exported time step, with the process
parameters, mesh description and complete solver manifest. A dataset collection
holds no arrays, referencing its members by identifier
(\cref{fig:prov-roles}), so a simulation is stored once however many collections read
it, and collections compare directly for shared membership. A model node holds
the trained weights and the resolved configuration behind them. An evaluation node holds one batch of the predicted and reference fields it scored and every metric computed from them, not only those tabulated here. In the published store, those fields sit in the restricted-access record.

\begin{figure}[htbp]
\centering
\includegraphics[width=\linewidth]{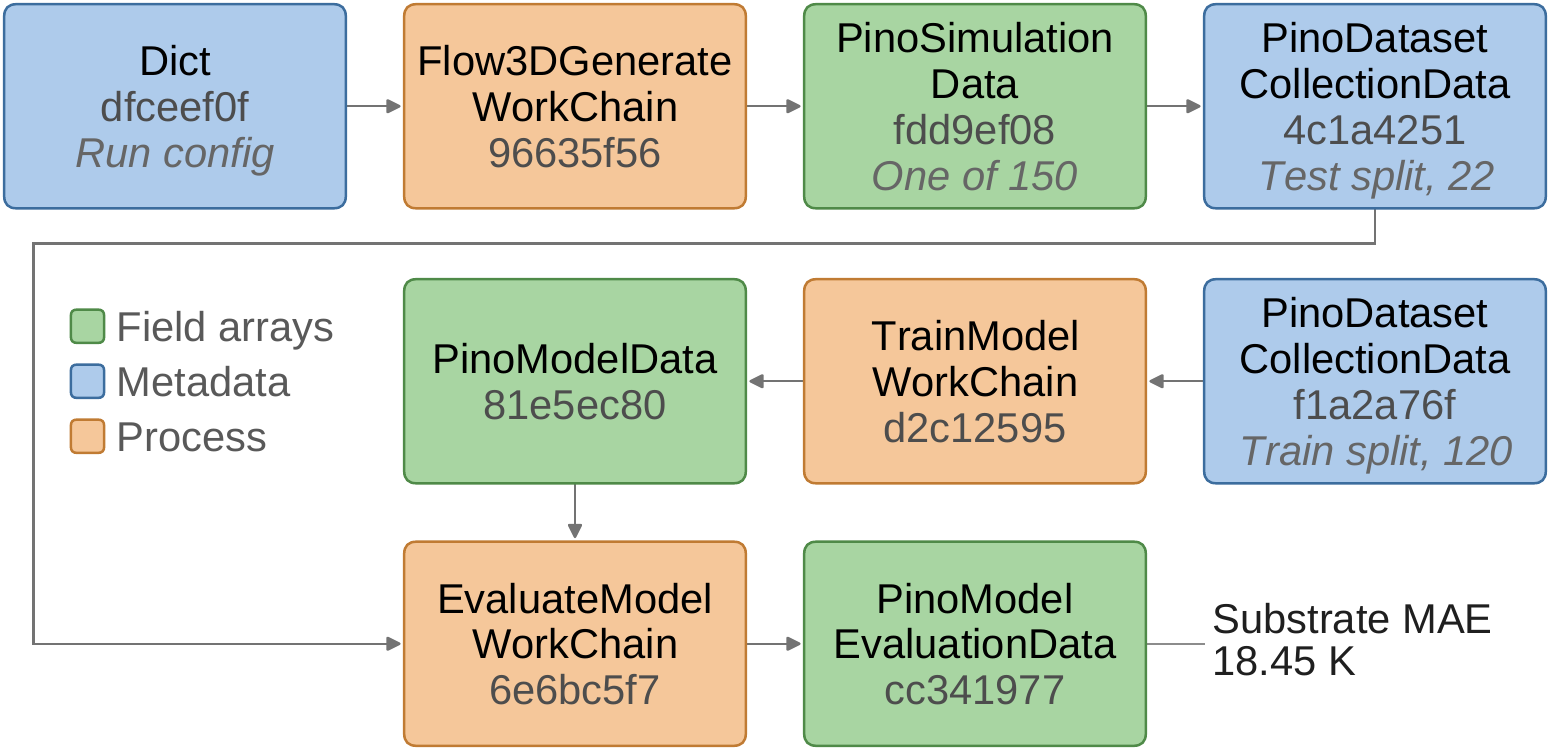}
\caption{The provenance of one reported number, as stored. Boxes are nodes labeled by AiiDA class and the first eight characters of their UUID in the deposited archive, and arrows are stored links. Green nodes hold field arrays, blue nodes metadata only, and orange nodes the processes that
created them. The chain runs from one simulation's run configuration through the solver
job and the dataset splits to the trained model and its evaluation.}
\label{fig:lineage}
\end{figure}

\paragraph{Following one number back.}
\Cref{fig:lineage} draws the ancestry of one reported quantity, the
\SI{18.45}{\kelvin} substrate error of the FNO on the free-form paths in \cref{tab:paths-accuracy}.
The chain is what the database returns. The evaluation resolves to the model and the test split, the model to the training split and
training job, each split to the simulations composed into it, and each simulation to its solver run
and configuration. The same walk can be made from any simulation or model result in this paper.

\paragraph{What the export contains.} The supplementary export is the cited nodes, their member simulations and the sub-graph reachable from them, omitting simulation array payloads for size. Metrics, configurations, manifests and links are kept. Some collections, such as derived ones, record membership as a stored identifier
list instead of graph edges, which the store resolves identically
but a graph export does not follow, so their membership is read from node
attributes.

\paragraph{Keeping it on one machine.} A solved simulation is large. The exported fields of a single \numproduct{90x40x40} track at the cadence used here are typically \SIrange{1.5}{3.9}{\giga\byte} of floating-point arrays. Five measures keep the store small.
Each simulation is stored as one compressed array file, which is lossless and worth a
factor of \numrange{16}{38}, so the median stored time series is about \SI{100}{\mega\byte} and a steady frame under \SI{1}{\mega\byte}. Storing the same arrays as separate objects was measured
at roughly twenty times that and is refused at ingest. The solver's own results
file, which runs from a few gigabytes to \SI{60}{\giga\byte} per simulation, is
deleted as soon as the export is verified by reopening it, and kept when the
export is missing or partial, since it is then the only copy of a simulation that
cost a license seat. Collections hold no arrays of their own
(\cref{fig:prov-roles}), so the splits, the pooled sets and every overlapping
dataset of \cref{tab:datasets} cost nothing beyond their membership lists. A
derived collection stores the transformation and recomputes on read, so a
resolution study adds no copies of the fields it subsamples. And the object
repository is content-addressed, so identical objects are stored once. When space needs to be freed, array payloads can be evicted while the
node, its metadata and its provenance stay in place.

\FloatBarrier%

\section{Cost of the frame the problem is posed in}\label{sec:frame-difficulty}

\Cref{tab:frame-difficulty} scores the models of \cref{tab:arch-steady} in the
two time formulations of \cref{sec:formulation}, on the same \num{13} held-out
simulations of the straight-path split. \emph{Native} means trained
and scored in the same frame. \emph{Laboratory $\rightarrow$ quasi-steady} means the
laboratory model predicts every stored frame, the predictions are averaged over
time through the transform of \cref{sec:formulation}, and the result is scored
against the quasi-steady truth. Each row gives the mean over three training seeds.

For FNO and GAOT, directly predicting in the quasi-steady frame of reference leads to better metrics. Surprisingly, the RIGNO metrics are similar, or even better when predicted in the laboratory frame and then cropped and averaged into the quasi-steady formulation, showing that RIGNO can learn just as well on the direct time-dependent predictions.

\begin{table}[htbp]
\centering
\caption{The models of \cref{tab:arch-steady} scored in the two time formulations of \cref{sec:formulation}, on the
same \num{13} held-out straight-path simulations. \emph{Native} rows show models trained and scored in the same frame. The \emph{laboratory $\rightarrow$ quasi-steady} rows average the laboratory model's per-frame predictions over time and score them against the quasi-steady truth. Mean over three training seeds.}
\label{tab:frame-difficulty}
\small
{%
\begin{tabular*}{\textwidth}{@{\extracolsep{\fill}}lccccc}
\toprule
 & \tabhead{T MAE\\(\si{\kelvin})} & \tabhead{T RMSE\\(\si{\kelvin})} & \tabhead{BB MAE\\(\si{\kelvin})} & \tabhead{Rel. err.\\(\%)} & \tabhead{Melt-pool\\IoU} \\
\midrule
\multicolumn{6}{@{}l}{\emph{FNO}} \\
Quasi-steady, native & \textbf{7.71} & {29.6} & \textbf{15.0} & \textbf{1.25} & \textbf{0.920} \\
Laboratory, native & {10.01} & \textbf{27.3} & {49.1} & {2.25} & {0.863} \\
Laboratory $\rightarrow$ quasi-steady & {17.91} & {40.3} & {53.2} & {2.92} & {0.866} \\
\midrule
\multicolumn{6}{@{}l}{\emph{RIGNO}} \\
Quasi-steady, native & 11.36 & 40.0 & \textbf{23.6} & 1.84 & {0.926} \\
Laboratory, native & \textbf{6.65} & \textbf{21.4} & {30.9} & \textbf{1.49} & {0.926} \\
Laboratory $\rightarrow$ quasi-steady & {11.50} & {28.6} & {32.7} & {1.88} & \textbf{0.935} \\
\midrule
\multicolumn{6}{@{}l}{\emph{GAOT}} \\
Quasi-steady, native & \textbf{9.74} & \textbf{32.3} & \textbf{16.8} & \textbf{1.58} & \textbf{0.915} \\
Laboratory, native & {11.13} & {41.7} & {72.8} & {2.50} & {0.826} \\
Laboratory $\rightarrow$ quasi-steady & {20.01} & {52.2} & {78.9} & {3.27} & {0.841} \\
\bottomrule
\end{tabular*}}
\end{table}
\FloatBarrier

\section{Smoothing the GAOT oscillations}\label{sec:gaot-smoothing}

\begin{figure}[h]
\centering
\begin{subfigure}[t]{0.49\textwidth}\includegraphics[width=\linewidth]{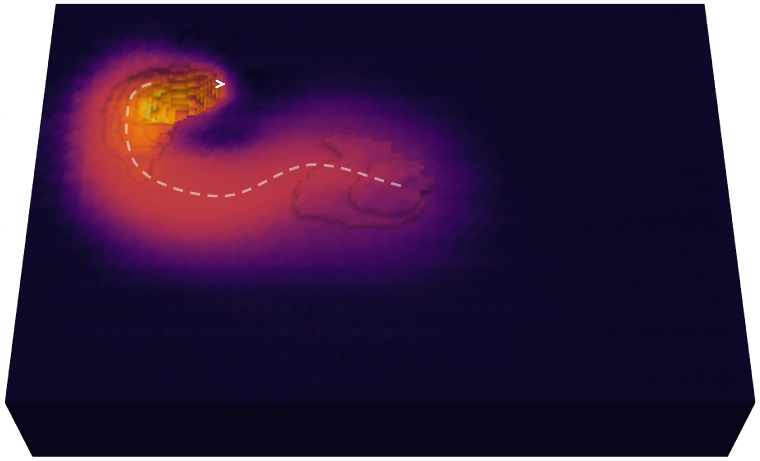}\subcaption{Before smoothing.}\label{fig:gaot-smoothing-before}\end{subfigure}\hfill
\begin{subfigure}[t]{0.49\textwidth}\includegraphics[width=\linewidth]{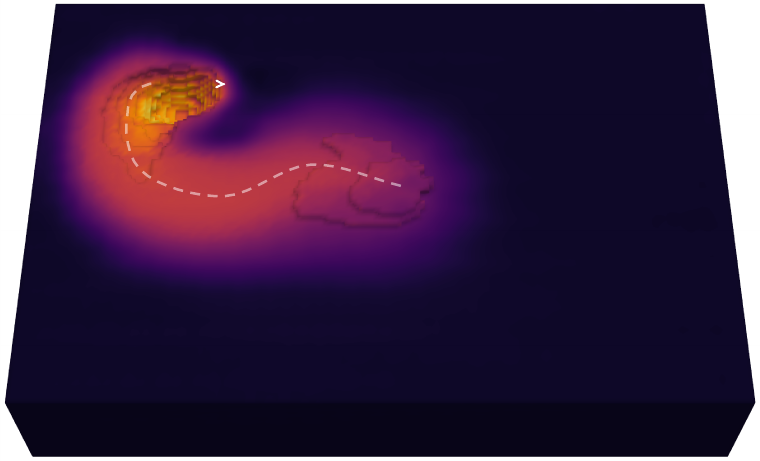}\subcaption{After smoothing.}\label{fig:gaot-smoothing-after}\end{subfigure}

\vspace{0.3em}
\includegraphics[width=0.5\textwidth]{Figs/rp_artifact_cbar.pdf}\hspace{1em}\includegraphics[width=0.2\textwidth]{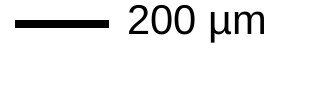}\hspace{1em}\includegraphics[width=0.22\textwidth]{Figs/rp_artifact_track_key.pdf}
\caption{The GAOT prediction of \cref{fig:rp-gaot} before the Gaussian smoothing \subref{fig:gaot-smoothing-before} and after it \subref{fig:gaot-smoothing-after}, rendered as in \cref{fig:rp-artifact} and on the same color scale. Over the frame shown, the relative error goes from \SI{4.50}{\percent} to \SI{4.19}{\percent} and BB MAE from \SI{112.7}{\kelvin} to \SI{109.0}{\kelvin}, and melt-pool IoU stays at \num{0.736}. After smoothing, oscillation artifacts from the prediction are effectively gone.
}
\label{fig:gaot-smoothing}
\end{figure}

The oscillations in the bulk substrate of the GAOT prediction in \cref{fig:rp-gaot} seem to be mostly superficial, in that they add little to the prediction error. To check this, we smooth the predicted temperature of the frame shown there with a Gaussian filter, without retraining the model. The standard deviation of the filter is a quarter of the spacing of the regular \numproduct{44 x 20 x 24} lattice of latent tokens, which gives \SI{8.7}{\micro\metre}, \SI{13.0}{\micro\metre} and \SI{4.2}{\micro\metre} along $x$, $y$ and $z$. The filter acts on the network's temperature before the masking of \cref{sec:formulation}, and its average leaves out the cells with $\alpha < 0.5$. We use the smoothed temperature in full where it lies below \SI{1200}{\kelvin}. Above that we blend it with the unsmoothed prediction. The weight of the smoothed temperature falls linearly to zero at the melt threshold $(T_s+T_l)/2$, which is \SI{1898}{\kelvin} for Ti-6Al-4V. The melt pool therefore changes little. We do not filter the metal volume fraction $\alpha$, which leaves the vapor cavity as predicted. In \cref{fig:gaot-smoothing} the oscillations are largely gone, the relative error and BB MAE fall by about \SI{7}{\percent} and \SI{3}{\percent} of their unsmoothed values, and melt-pool IoU does not change.

\FloatBarrier
\end{appendices}

\bibliographystyleS{sn-nature}
\bibliographyS{references}

\end{document}